\documentclass[useAMS,usenatbib,a4paper, fleqn]{mnras}
\usepackage{ae,aecompl}
\usepackage[dvipsnames]{xcolor}

\usepackage{booktabs}
\title[Observational Bias and YMC Characterisation II.]
{Observational Bias and Young Massive Cluster Characterisation \\II. Can {\it Gaia} accurately observe young clusters and associations?}
\author[Buckner et al.]
{Anne S.M. Buckner\thanks{E-mail:
a.buckner@exeter.ac.uk}$^{1}$, Tim Naylor$^{1}$, Clare L. Dobbs$^{1}$, Steven Rieder$^{2}$,  Thomas J. R. Bending$^{1}$\\
$^1$ School of Physics and Astronomy, University of Exeter, Stocker Road, Exeter, EX4 4QL, UK \\
$^2$ Geneva Observatory, University of Geneva, Chemin Pegasi 51, 1290 Sauverny, Switzerland \\
}

\usepackage{amssymb}
\usepackage{amsmath}
\usepackage[pdftex]{graphicx}
\usepackage{epsfig}
\usepackage{multirow} 
\def\aap{{ A\&A}}

\def\aj{{AJ}}

\def\apj{{ApJ}}

\def\mnras{{MNRAS}}

\def\apjs{{ApJS}}
\def\araa{{ARA\&A}}
\def\apss{{Ap\&SS}}

\usepackage{ulem}

\begin{document}
\label{firstpage}
\date{\today}

\pagerange{\pageref{firstpage}--\pageref{lastpage}} \pubyear{2012}

\maketitle

\begin{abstract}
Observations of clusters suffer from issues such as completeness, projection effects, resolving individual stars and extinction. As such, how accurate are measurements and conclusions are likely to be? Here, we take cluster simulations (Westerlund2- and Orion- type), synthetically observe them to obtain luminosities, accounting for extinction and the inherent limits of Gaia, then place them within the real Gaia DR3 catalogue. We then attempt to rediscover the clusters at distances of between 500 and 4300 pc. We show the spatial and kinematic criteria which are best able to pick out the simulated clusters, maximising completeness and minimising contamination. We then compare the properties of the ‘observed’ clusters with the original simulations. We looked at the degree of clustering, the identification of clusters and subclusters within the datasets, and whether the clusters are expanding or contracting. Even with a high level of incompleteness (e.g. $<2\%$ stellar members identified), similar qualitative conclusions tend to be reached compared to the original dataset, but most quantitative conclusions are likely to be inaccurate. Accurate determination of the number, stellar membership and kinematic properties of subclusters, are the most problematic to correctly determine, particularly at larger distances due to the disappearance of cluster substructure as the data become more incomplete, but also at smaller distances where the misidentification of asterisms as true structure can be problematic. Unsurprisingly, we tend to obtain better quantitative agreement of properties for our more massive Westerlund2-type cluster. We also make optical style images of the clusters over our range of distances.

\end{abstract} 



\section{Introduction}
Star clusters and associations are the principal sites of star formation in galaxies, so studying them is important to understand the conditions stars are formed in, and the surrounding conditions of stars and planets as they evolve. Young massive clusters (YMCs) are of particular interest as they host massive stars, which ionise and shape the surrounding gas (\citealt{2018ARA&A..56...41M}, \citealt{2011MNRAS.414..321D}, \citealt{2020ApJ...905...61P})

Observational studies of clusters are subject to non-trivial biases and uncertainties, including stellar membership identification, data incompleteness and 2D projection effects due to the clusters' 3D orientation in the line-of-sight (LoS; \citealt{2009Ap&SS.324..113A}, \citealt{2012A&A...545A.122P}, \citealt{2022MNRAS.514.4087B}, \citealt{2022A&A...659A..72B}). Detailed studies of high mass star forming regions within the Milky Way are therefore typically limited to within a few kilo-parsecs to reduce the severity of these issues, which means we tend to focus on a few well studied objects (such as the Orion, Carina and Rosette nebulae for example). Nevertheless, it remains difficult to estimate how much impact these biases have on the properties we derive for them. 

The aim of this paper series is to assess the potential accuracy of the properties and physics of YMCs derived from observational-based studies, given the above limitations. In Paper I \citep{2022MNRAS.514.4087B} we considered a cluster formed from two colliding clouds, 
to investigate issues arising from 2D projection effects.
We found that generally correct qualitative conclusions are obtained when viewing a cluster from various orientations, though the accuracy of specific values are unreliable. The properties most likely to be interpreted incorrectly were  
whether a cluster is expanding or contacting, and the stellar membership of substructure. 
Other factors affecting observed measurements involve considering gas and dust along the LOS, and the limitations of the telescope.
In this paper we produce synthetic observations of clusters to examine these remaining factors. 

Arguably the current optimal instrument to study clusters is {\it Gaia} (\citealt{2016A&A...595A...2G,  2018A&A...616A...1G, 2021A&A...649A...1G, 2022arXiv220800211G}), due to the all-sky coverage and high precision measurements. 
The availability of {\it Gaia}’s all-sky position, parallax and proper motions has allowed many photometrically-identified clusters to be definitively confirmed as real or asterisms, their membership lists culled and/or expanded, and the refinement of region properties (\citealt{2018A&A...618A..93C}, \citealt{2019MNRAS.487.2385M}, \citealt{2021MNRAS.504..356D}, \citealt{2021EPJST.230.2177M}, \citealt{2021A&A...650A..67R}, \citealt{2022Univ....8..111C}, \citealt{2022MNRAS.512.4464D},  \citealt{2022A&A...660A..11G}). Large-scale searches making use of {\it Gaia}’s parallax and kinematic data are also proving fruitful, particularly for identifying clusters that have been previously missed due to a lower density contrast with the field (\citealt{2019A&A...624A.126C}, \citealt{2019A&A...627A..35C}, \citealt{2019JKAS...52..145S}, \citealt{2020A&A...635A..45C}, \citealt{2020MNRAS.496.2021F}, \citealt{Kounkel2020}, \citealt{2021MNRAS.502L..90F}, \citealt{Quintana2021}, \citealt{2022A&A...661A.118C}, \citealt{2022A&A...660A...4H}, \citealt{2022ApJS..259...19L}, \citealt{2022arXiv220612170H}). However, infrared photometric searches remain useful as {\it Gaia} is an optical survey so it is vulnerable to extinction (due to both local nebulosity and the foreground ISM). For YMCs whose members are still embedded deep within the natal cloud this limitation is a particular problem as highly incomplete {\it Gaia} membership lists are to be expected, with further complexity introduced to the dataset by fainter stars which suffer from large parameter uncertainties \citep{2022arXiv220800211G}. Therefore despite {\it Gaia}’s unprecedented accuracy, with incomplete membership lists potentially containing significant uncertainties and interloping field stars, how reliable are the properties we derive for clusters using it?

Here we attempt to answer this question by placing a simulated cluster within the real {\it Gaia} EDR3 catalogue at various distances. We attempt to (i) distinguish the cluster's stellar members and non-member stars and (ii) recover the cluster's spatial distribution, structure and kinematic properties. We also make synthetic images of our clusters as if they were observed, again including all the field stars within our field of view. \citet{Kounkel2018} also use synthetic clusters to test how well they can characterise structure in their observations of the Orion star-forming complex. In contrast to our approach of using simulated clusters, they begin with assumptions about the properties of the populations of field and cluster stars, and draw samples from them.

This paper is structured as follows. Section 2 details our cluster simulations, and Section 3 our methodology to convert them into realistic {\it Gaia} observations. In Section 4 we show optical images of the cluster observations and we describe how stellar members are identified in Section 5. Section 6 details our analysis methods. Our results are presented in Section 7 and discussed in Section 8.


\section{Cluster data from simulations}
We take data from two simulations from \citet{dobbs2022}, labelled M1R1FB and M5R1FB in that paper. The simulations follow the formation of clusters and associations in a region of spiral arm. They both simulate the same region, but the initial mass of gas is $10^5$ M$_{\odot}$ in M1R1FB, and $5 \times 10^5$ M$_{\odot}$ in M5R1FB. Both include feedback in the form of ionisation and supernovae, but for the timeframes used in this paper, few if any supernovae have occurred. 

The evolution of the two regions, and the types of clusters formed, are quite different in the two simulations. In the lower gas mass simulation, M1R1FB, a loose association consisting of multiple groups of stars forms. This association is not dissimilar in terms of mass, physical size, and substucture to the Orion star-forming region. By contrast M1R5FB produces stellar clusters which are gravitationally bound and more similar to YMCs.

For the analysis presented here, we take the times from M1R1FB and M5R1FB when the older stars have ages around 3\,Myr, and a further time frame from M1R1FB when the older stars have ages around 5.5\,Myr. In reality, star formation will be ongoing but we use a single age parameter to simplify our comparisons. The M5R1FB simulation was not run for as long, so we cannot compare both at the later time. 

The simulations contain both sink as gas particles. The sinks typically represent groups of stars rather than individual stars and we use the grouped star formation tool GRETA\footnote{\url{https://github.com/kyliow/greta}} (GRoupEd sTAr formation; \citealt{liow_grouped_2021}) to convert them to stars, sampling from a Kroupa IMF \citep{kroupa_2001} with a mass range of 0.1-100$M_{\odot}$, then placed randomly within a 10\,pc radius of each sink particle, resulting in an approximately uniform distribution \citep{rieder_ekster_2021}.
A value of 10\,pc was chosen as it is representative of the distance between sink particles in the simulation. The gas particles are used to calculate the line-of-sight extinction due to the natal nebulosity for each star (Section\,\ref{sect_gsim}). 

For a clearer comparison with observations, from now on we refer to the M1R1FB simulation as `Orion-type' 
and the M5R1FB simulation as `Wd2-type'. M5R1FB is comparable to the more massive young clusters in the Milky Way, e.g. Westerlund 1, Westerlund 2, NGC\,3603. In terms of morphology M5R1FB contains two clear clusters and likewise Westerlund 2 has two groups of stars. Hence we relate the region in M5R1FB to Westerlund 2 and name this Wd2-type. Thus we can determine how reliably regions, or clusters, like these are observed, and how well we could observe them at larger distances from the Sun.

Table\,\ref{Tab_clusters} summarises the stellar populations for the Orion-type simulation at the two different times (where the appendices `-3' and `-5.5' in the names represent the ages), and for the Wd2-type simulation.

The stars in each simulation have full 6D phase space (3D positions and velocities) data for each star, and membership is absolute – that is all stars are genuine members of the original datasets and there is no non-member contamination. 

\begin{table}
\caption{List of simulations with the age of the stars, total stellar mass and number of stars.  \label{Tab_clusters} 
} \centering                                      
\begin{tabular}{c | c | c| c}          
\hline\hline                        
Name & Age & Mass  & $\#$ No. stars\\    
& [Myrs] & [$10^3 M_\odot$]&  \\ 
\hline                                   
    Orion-type-3 & 3 & 0.89 &  1635 \\    
    Orion-type-5.5 & 5.5 & 2.30 &  5105 \\   
    Wd2-type & 3 & 30.14 &  50696 \\    
\hline                                             
\end{tabular}
\end{table} 

\section{Placing the simulated datasets in the Gaia catalogue}

The steps we use to place our simulated clusters in the real {\it Gaia} DR3 survey\footnote{\url{https://gea.esac.esa.int/archive/}}\citep{2022arXiv220800211G} are as follows:
\begin{itemize}
\item We generate observational properties of the stars from the simulated data using a {\it Gaia} simulator.
\item Assign uncertainties to the simulated stars through random sampling of stars with comparable magnitude in the real {\it Gaia} catalogue.
\item Use Gaussian sampling to perturb the data values, about these uncertainties 
\item The simulated stars are then combined with the {\it Gaia} catalogue.
\item Stars which would not be detected or resolved by {\it Gaia} are removed.
\end{itemize}
These steps are described fully in the following sections. We repeat these procedures assuming the simulated stars lie at distances between 0.5-4.0\,kpc in 0.5\,kpc increments and at 4.3\,kpc, the original distance the stars would be in the galaxy simulation (i.e. 9 total observations per 
dataset), centered at coordinates ($l=270^o, b=0^o$). For similar real objects, we note that the distance to Orion is $\sim$400\,pc \citep{Jeffries2007,Menten2007}, whilst the distance to Westerlund 2 is uncertain but estimated to be around 3--4\,kpc \citep{Carraro2013,Hur2015}.

\begin{figure*}
 \centering
 \includegraphics[width=0.68\textwidth]{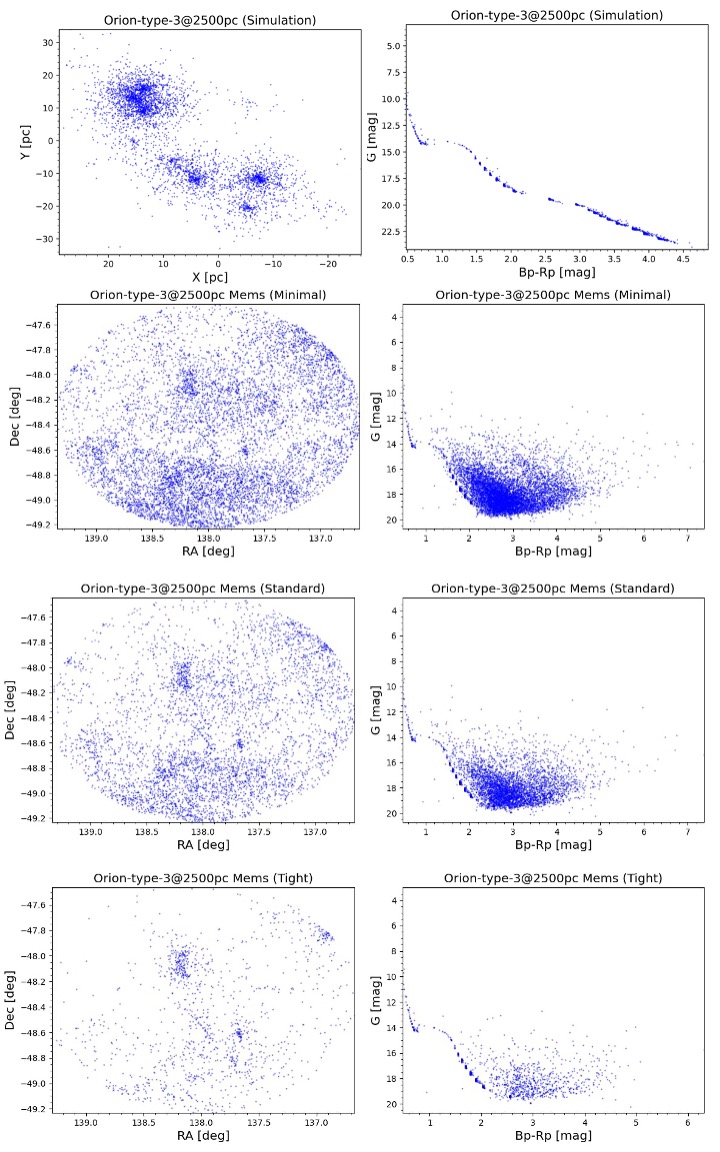}
 
    \caption{These panels show the various stages in creating a synthetic observation and selecting stellar members, as demonstrated for Orion-type-3 placed at a distance of 2500\,pc. Position and colour-magnitude diagrams are plotted for (Top row:) the SPH simulation, (Second row:) members selected using the Minimal criteria, (Third row:) members-only selected using the Standard criteria, (Bottom row:) members-only selected using the Tight criteria. 
    The clustering of members in the colour-magnitude diagrams is because their properties are taken from the model stars with the closest masses and ages to those given by the simulation \citep[for details see][]{2019MNRAS.485.3124K}.   (See section \ref{sect_members} for selection details.)}
    \label{Fig_stages} 
\end{figure*}

We show the positions and colour magnitude diagrams for the simulated dataset from Orion-type-3 before and after these steps in the top two panels of Figure~\ref{Fig_stages}, where the simulated dataset is assumed to lie 2.5\,kpc away. 

\begin{figure*}
\centering
     \includegraphics[width=0.3\textwidth]{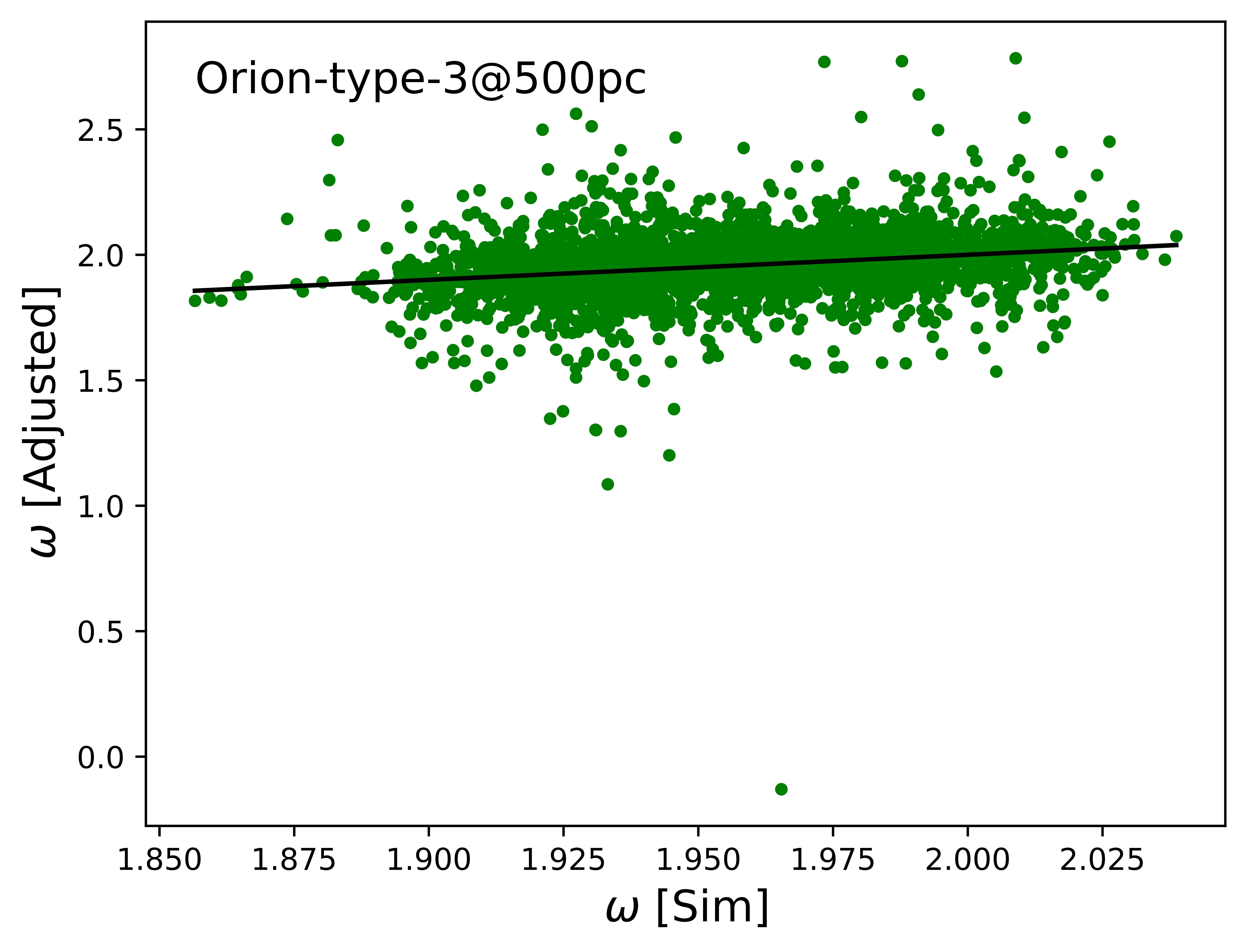}
    \includegraphics[width=0.3\textwidth]{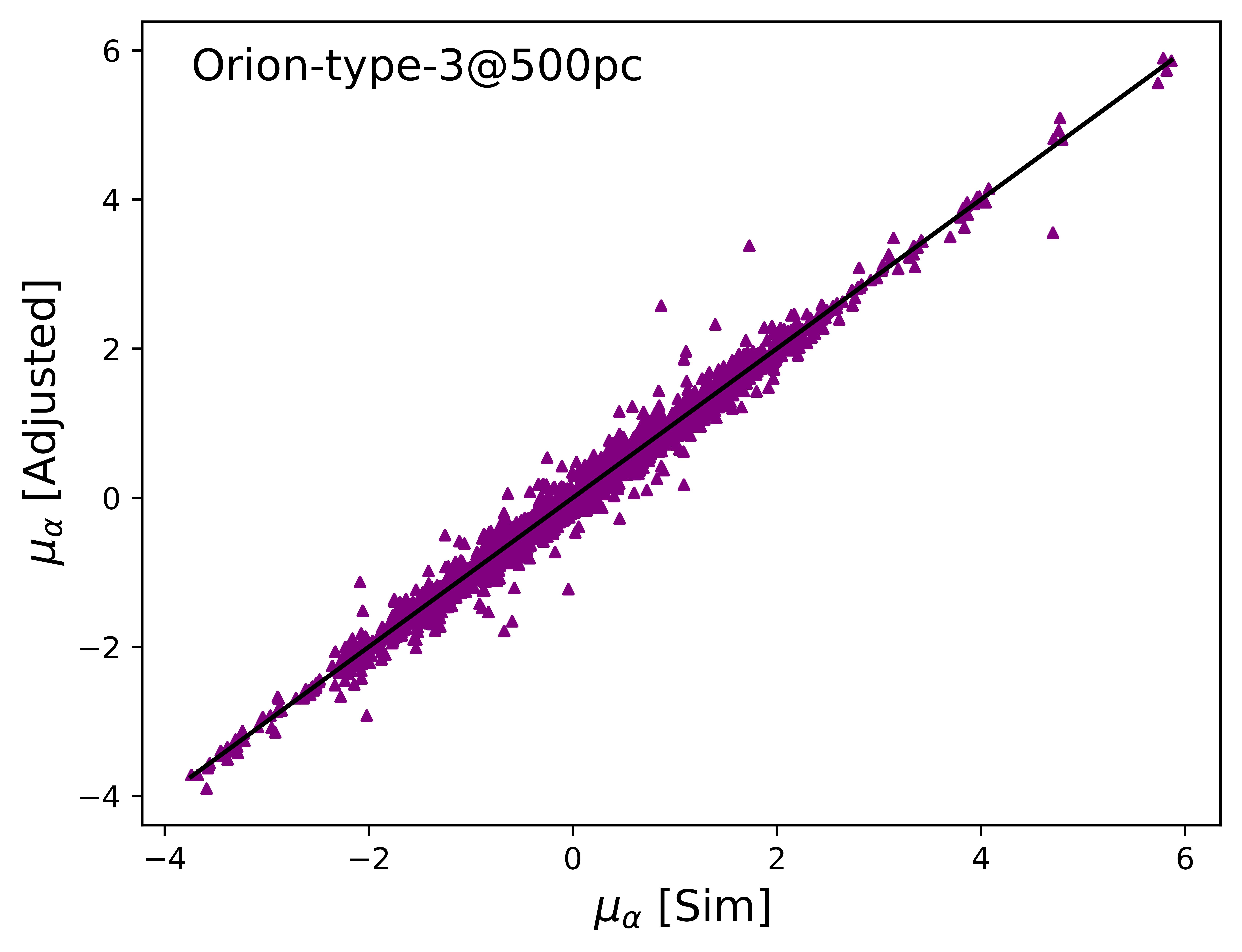}
     \includegraphics[width=0.3\textwidth]{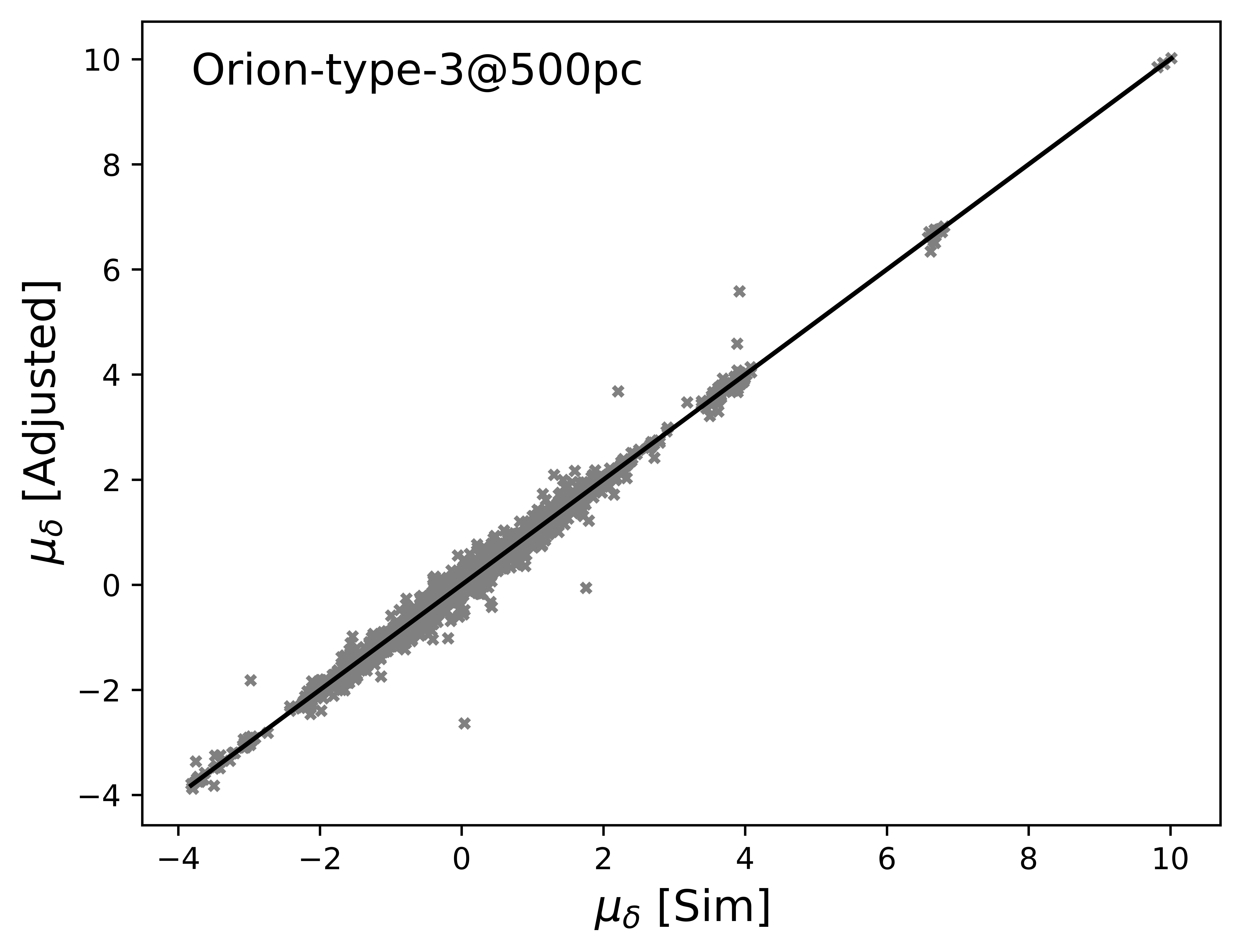}
     
     \includegraphics[width=0.3\textwidth]{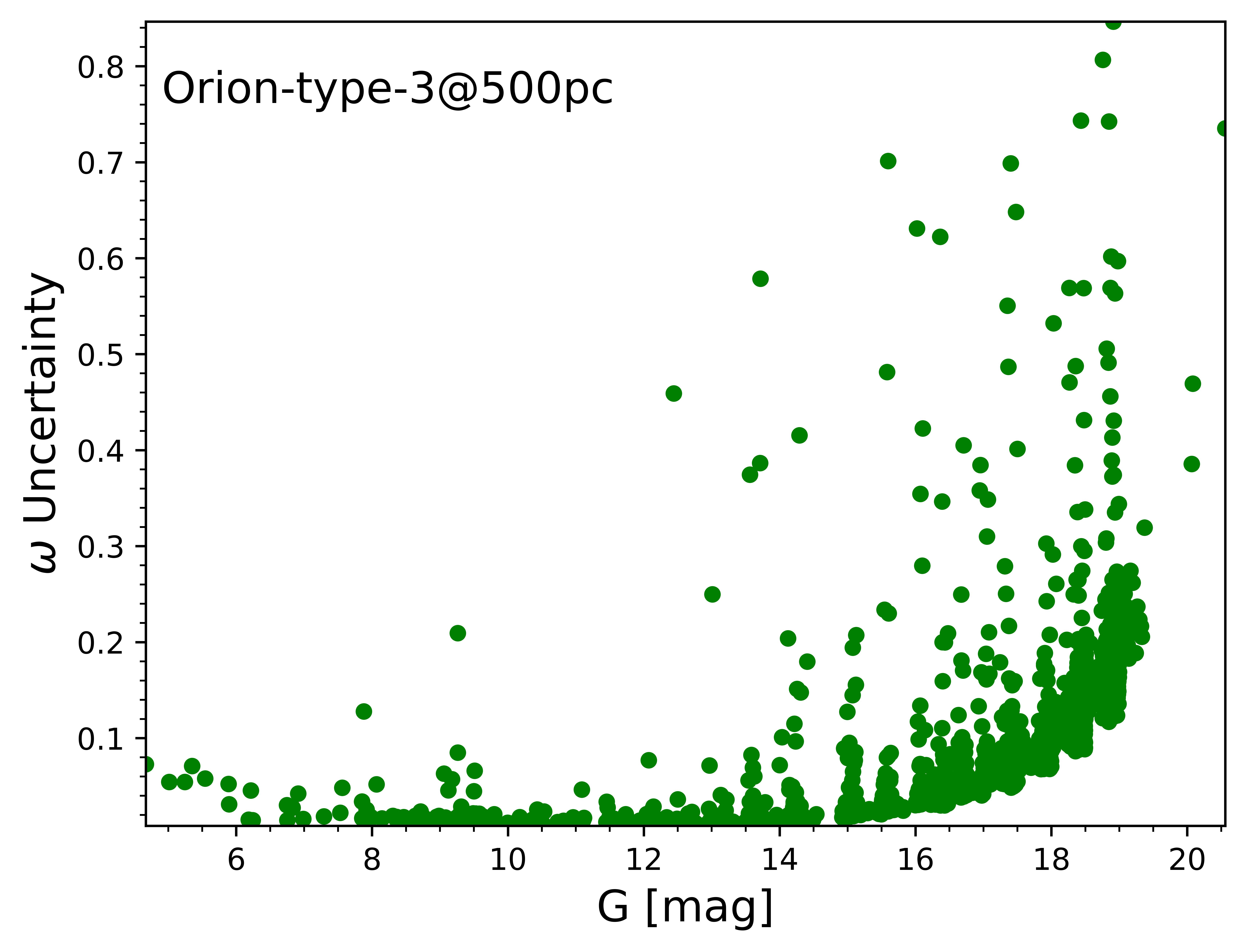}
    \includegraphics[width=0.3\textwidth]{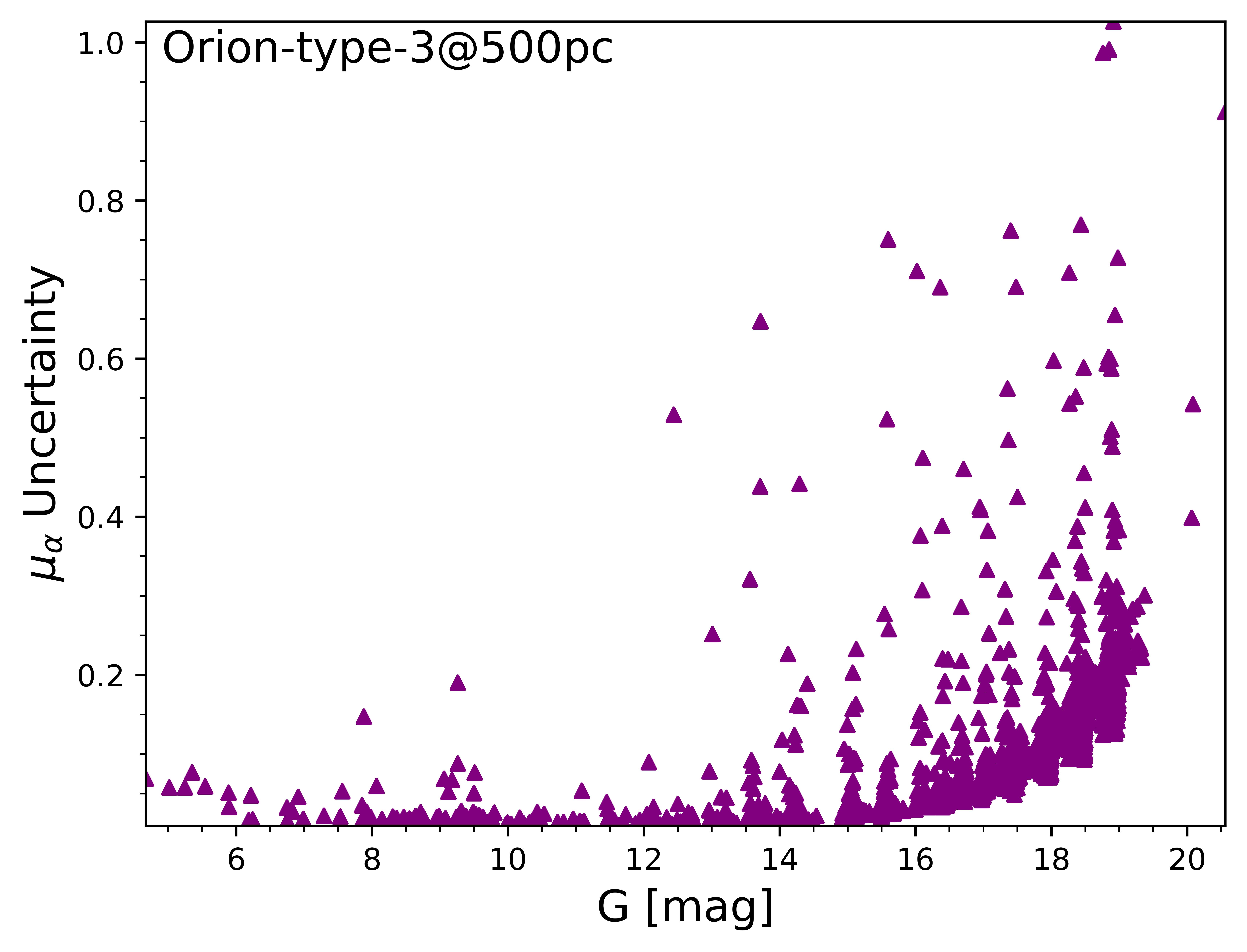}
     \includegraphics[width=0.3\textwidth]{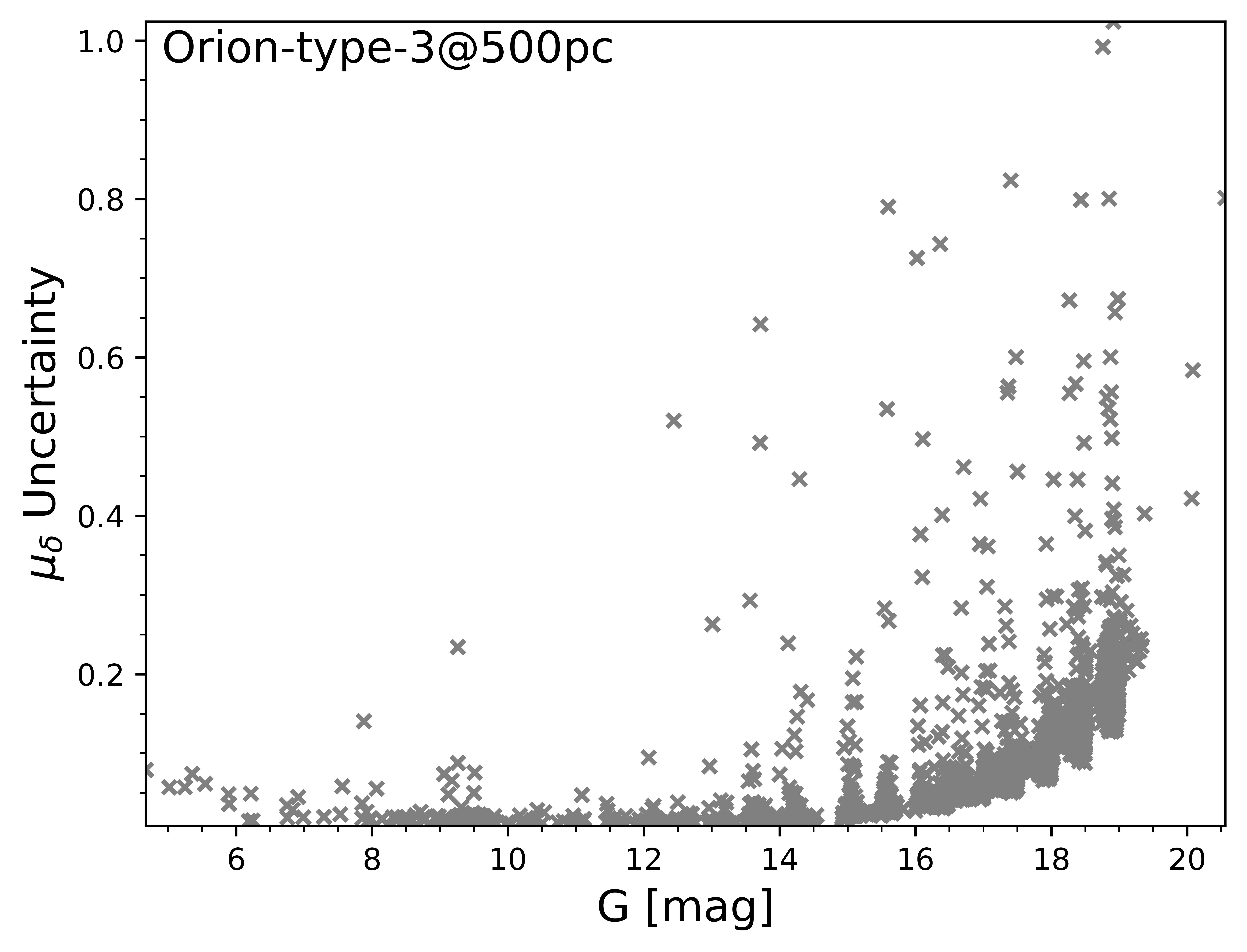}
     
    \includegraphics[width=0.3\textwidth]{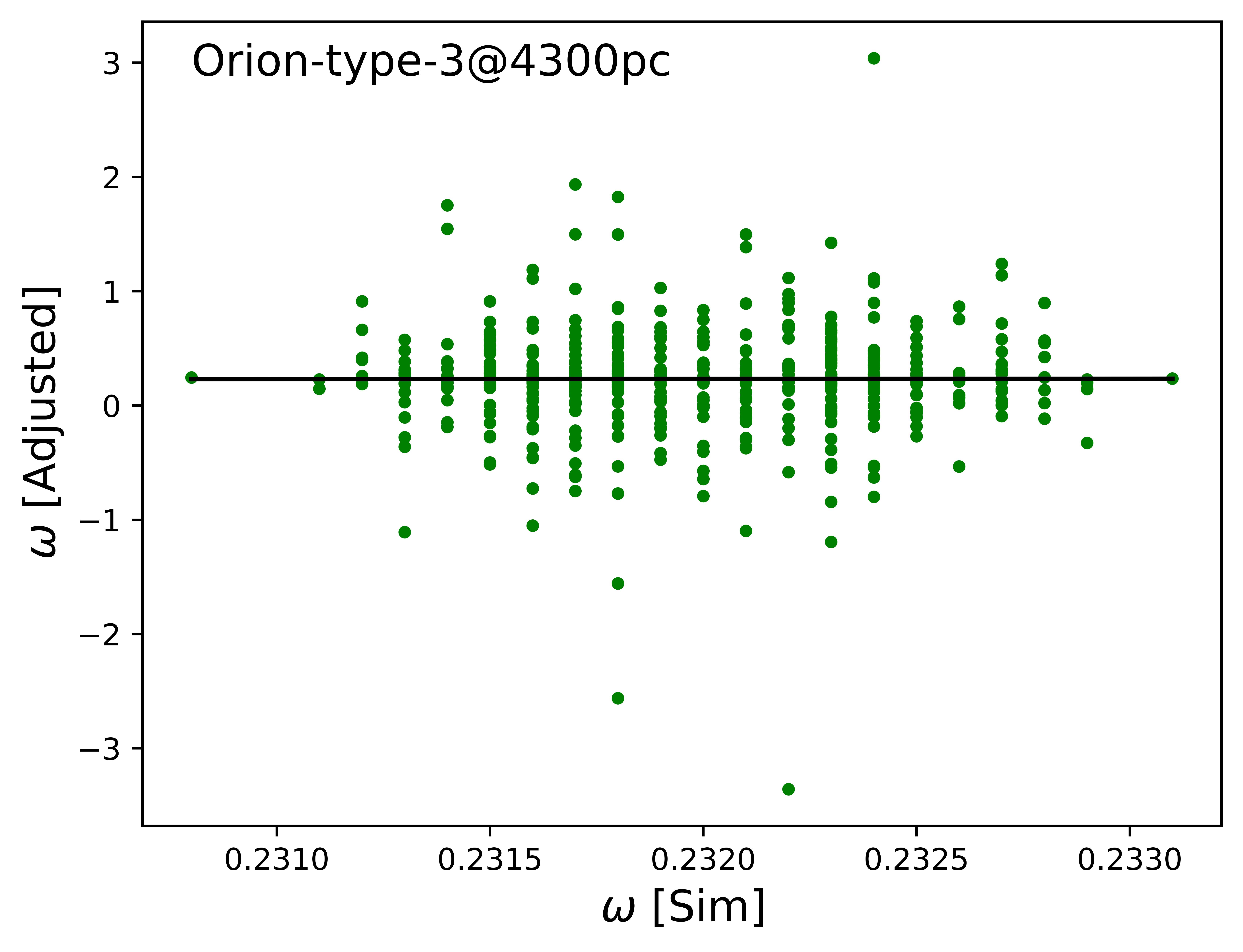}
    \includegraphics[width=0.3\textwidth]{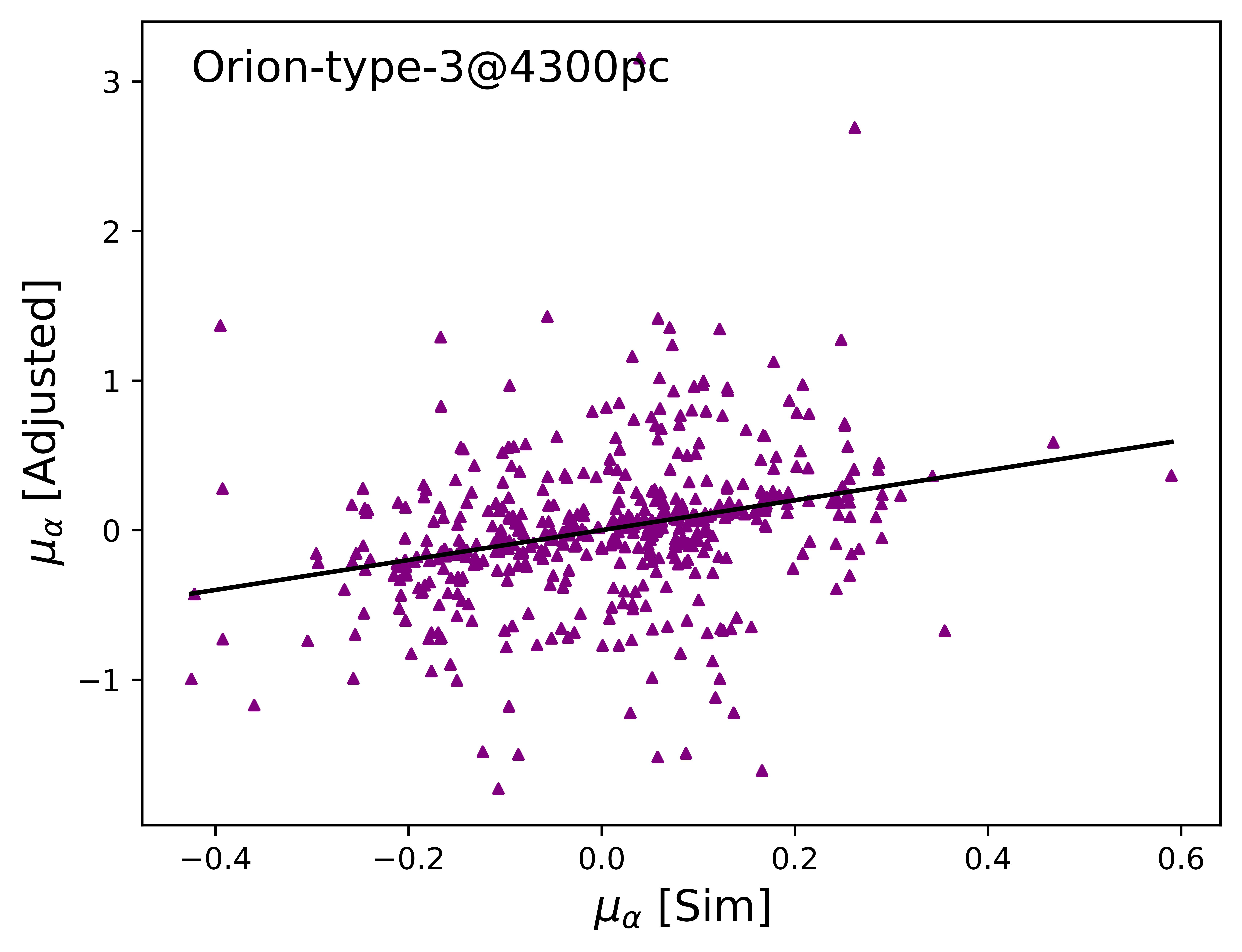}
     \includegraphics[width=0.3\textwidth]{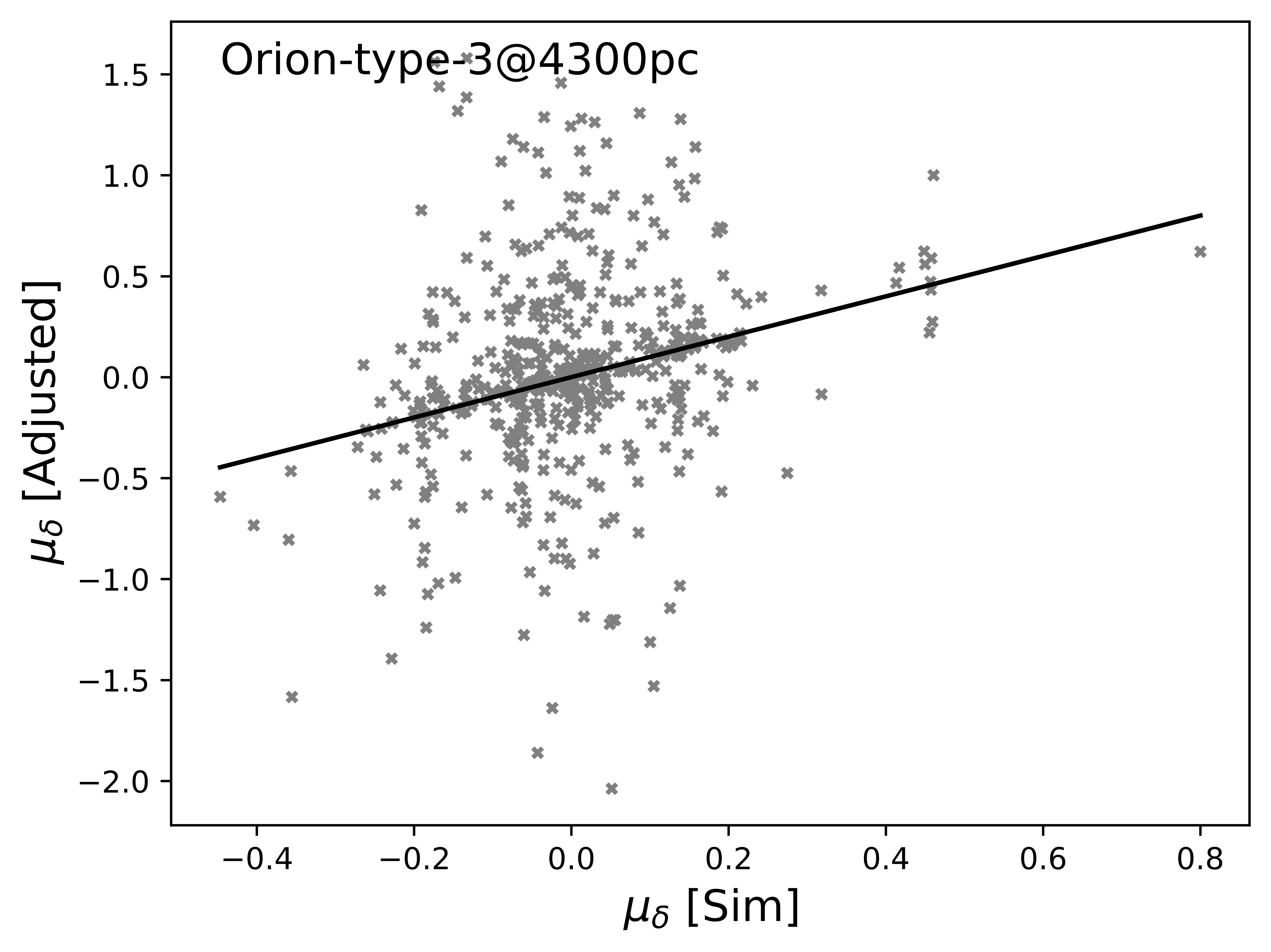}
     
    \includegraphics[width=0.3\textwidth]{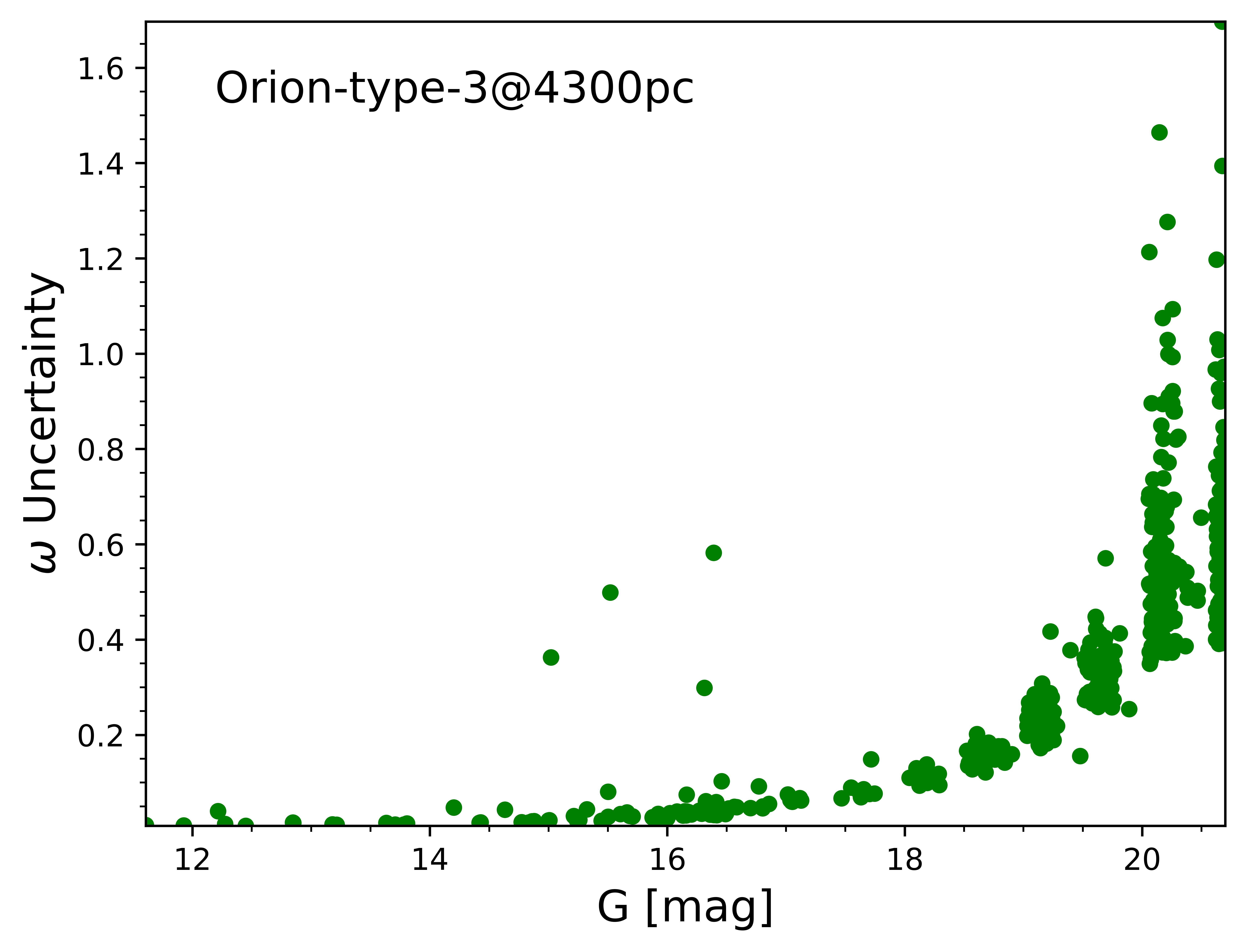}
    \includegraphics[width=0.3\textwidth]{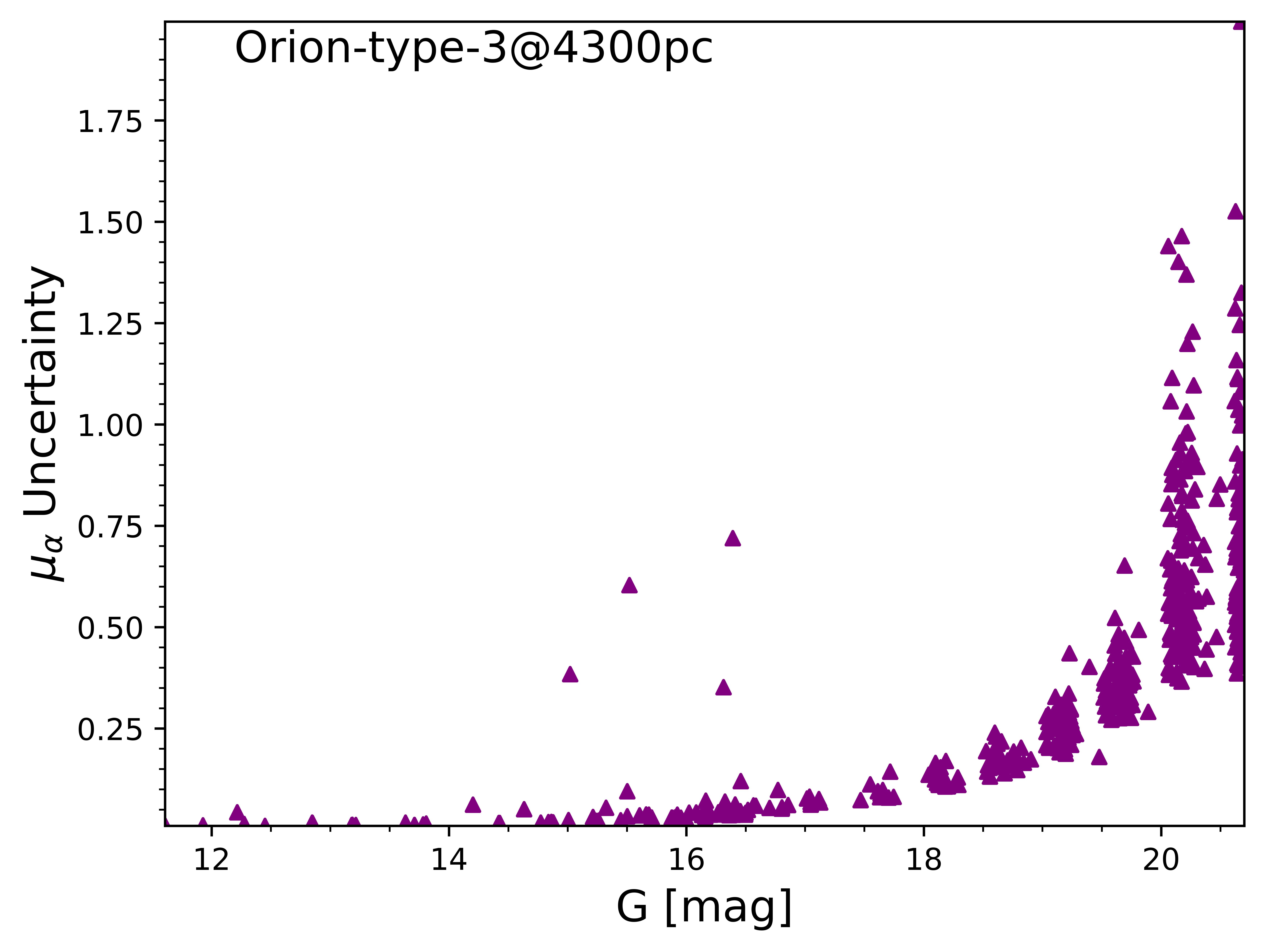}
     \includegraphics[width=0.3\textwidth]{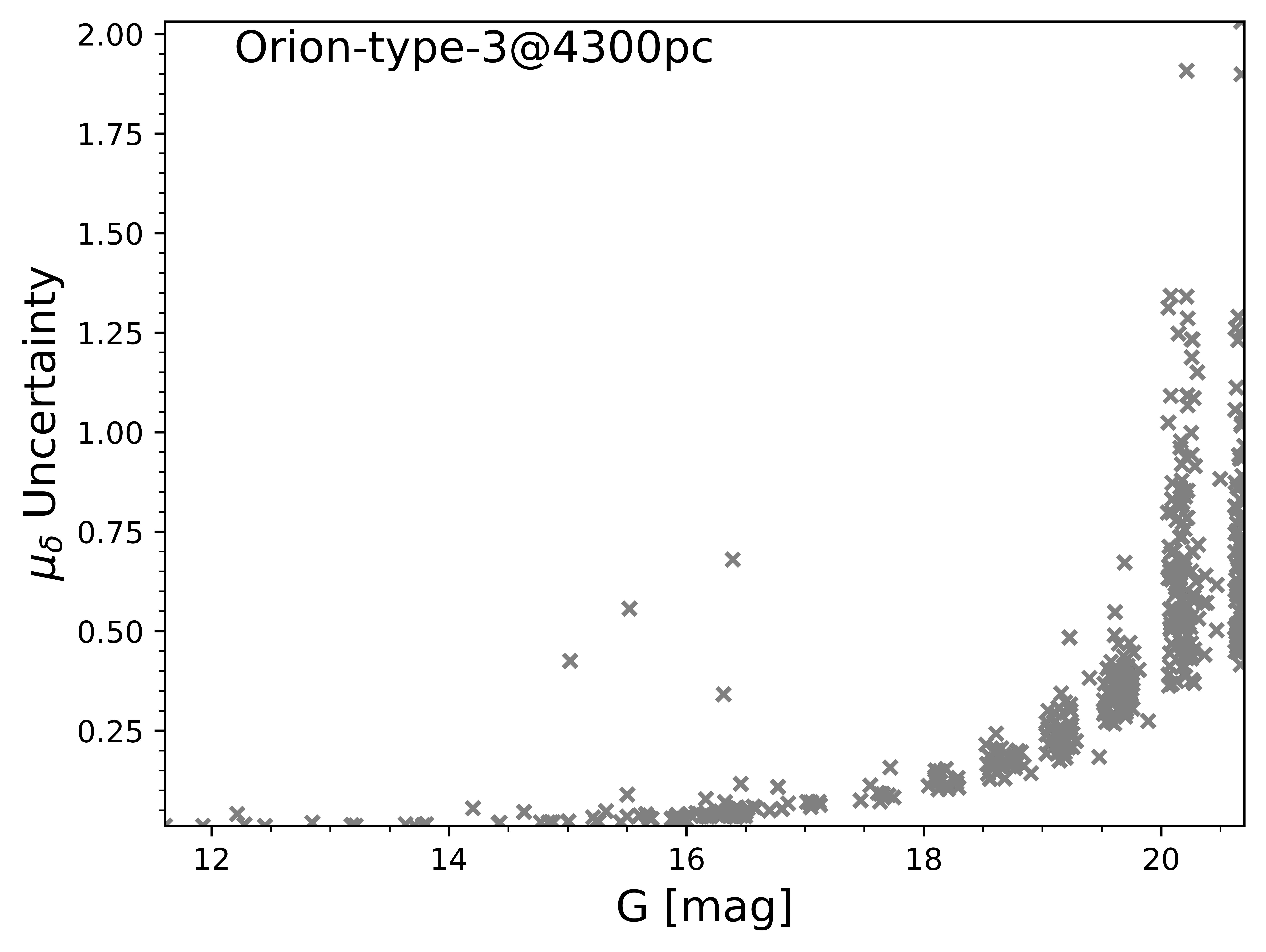}
     
    \caption{The parallax and proper motion uncertainties, and adjusted values, shown a functions of $G$-magnitude and original value from the simulations for stars in Orion-type-3 within the magnitude limit of {\it Gaia} at (rows 1\&2) 500\,pc (rows 3\&4) 4300\,pc. The solid black line shows a 1:1 relationship.}.
    \label{Fig_num_mems} 
\end{figure*}

\subsection{{\it Gaia} Simulator}\label{sect_gsim}

To generate {\it Gaia} observations of the simulated 
stars we use the {\it Gaia} Simulator\footnote{\url{https://github.com/zkhorrami/gaiaSimulations}} tool of synthetic observation generator MYOSOTIS\footnote{\url{https://github.com/zkhorrami/MYOSOTIS}} (Make Your Own Synthetic ObservaTIonS; \citealt{2019MNRAS.485.3124K}).  The simulator works by taking the relative positions, ages and masses of the stars and, using stellar evolutionary and atmosphere models, derives their magnitudes ($G$, $G_{\rm BP}$, $G_{\rm RP}$), parallaxes and proper motions. 
Distance, Line-of-Sight (LoS) and variable extinction - due to the relative depth of stars within the simulated gas cloud along the LoS - are taken into consideration when performing the calculations. 

We calculate the extinction of the stars by considering (i) the column densities of the simulations' natal gas cloud, and (ii) the presence of dust and gas in the Milky Way's Disc, along the line of sight to our cloud.  For the latter we use the canonical value of $A_V=0.7$mag kpc$^{-1}$ \citep{2010MNRAS.409.1281F}. To derive stellar fluxes we used the Dmodel extinction model with $R_V = 3.1$, a solar metallicity (Z=0.015), and turned on the ‘OBtreatment’ option for high-mass stars ($T_{\rm eff}>15 $\,kK) to ensure proper Spectral Energy Distributions (SEDs) were selected for these stars. 

For each synthetic observation, we removed members whose apparent magnitude was calculated to be outside the sensitivity limit of {\it Gaia} (3\,mag $\le G \le$ 20.7\,mag).

\subsection{Introducing Uncertainties}\label{sect_errors}

Values of proper motions, parallaxes and radial velocities assigned to stars by the {\it Gaia} simulator are exact, being scaled from simulation parameters. In reality, however, the values for stars in the {\it Gaia} catalogues represent the most probable value and include an uncertainty. The values for our clusters are therefore too precise and require perturbing to simulate the effects of observational errors 

We began by creating a sample which consisted of sources drawn from the DR3 survey in the LoS of the clusters ($l=270^o,b=0^o$) and radius that encompassed all the synthetic observations ($8.9^o$). No quality constraints were imposed at this stage as all sources detected by {\it Gaia} were required, including those with parameters whose solution was a poor fit to the astrometric model, not available, and/or with large uncertainties. The stars' synthetic observation values from the {\it Gaia} Simulator were then adjusted as follows: 

\begin{enumerate}
\item The DR3 sample is binned by $G$-band magnitude, in increments of 0.5\,mag.
\item Each simulated star $i$, is randomly assigned a renormalised unit weight error (RUWE), and the uncertainties in parallax and proper motion of a DR3 star in the same magnitude bin.
\item A value is randomly drawn from a Gaussian distribution whose mean is equal to the simulated parallax value of star $i$ and whose standard deviation of the associated uncertainty assigned in step (ii). This is the adjusted parallax value for star $i$.

\item Steps iii)-iv) are repeated to obtain star $i$’s adjusted proper motion values.   
\end{enumerate}

Figure\,\ref{Fig_num_mems} shows an example of the uncertainties obtained  as a function of magnitude for one of our datasets, and also the offset spread for stellar parallax and proper motion.  Due to the limited availability of radial velocity measurements in {\it Gaia} to draw from, we do not derive uncertainties for these values.

\subsection{Adding Field Stars}\label{sect_contam}

Many stars along the LoS will be field stars rather than members of a given cluster, so to see how our regions would appear as part of an observed dataset, we added field stars. This was achieved by merging 
the synthetic dataset obtained at the end of (Section\,\ref{sect_errors}) with the DR3 data.
No quality constraints were applied as all field stars detected by {\it Gaia} should be included at this stage. 

\begin{figure}
\centering
   \includegraphics[width=0.42\textwidth]{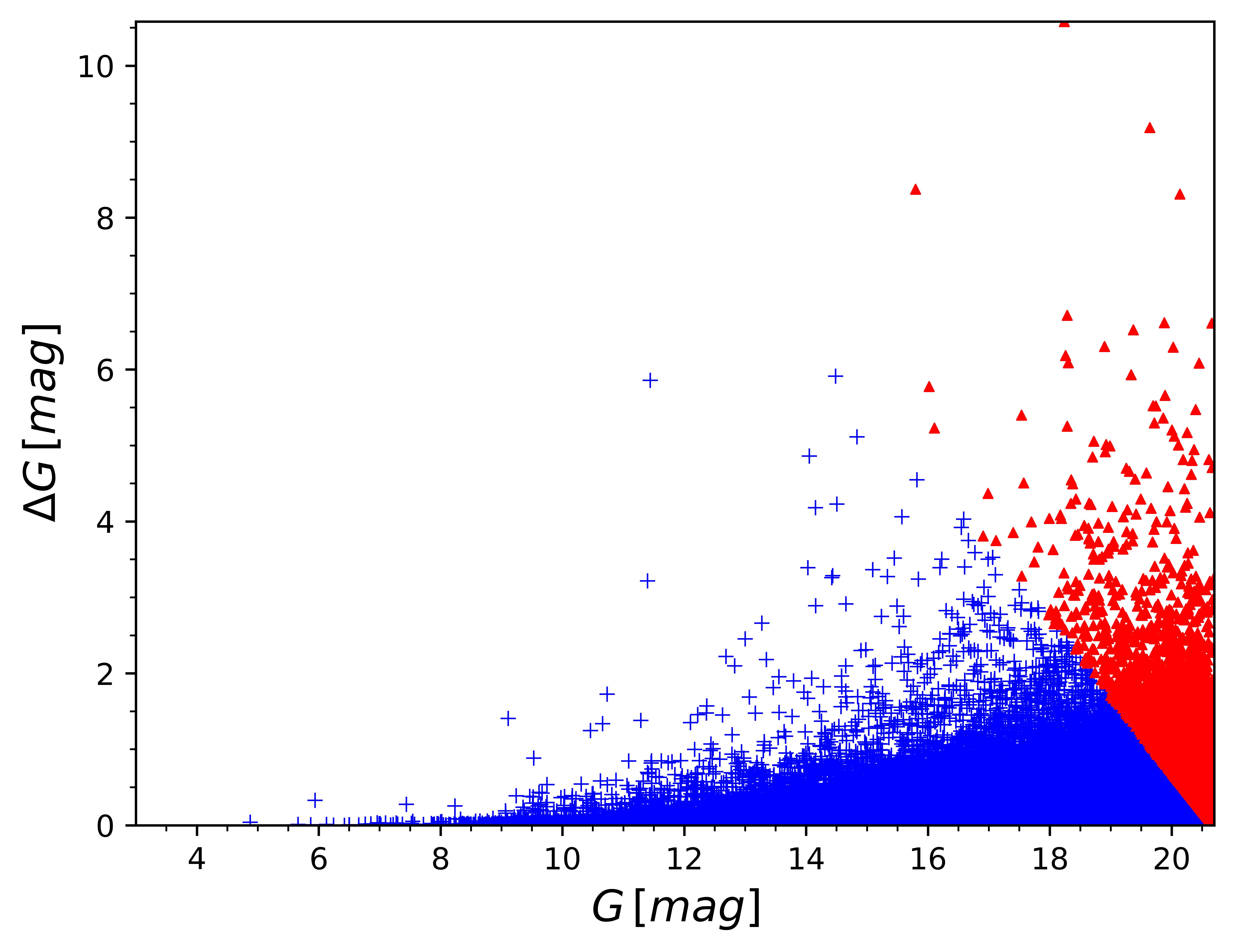}
   \caption{This figure shows the $G$-band magnitudes of stars in our DR3 sample identified as within or background to the SPH cloud of Orion-type-3, when the cluster is placed at a distance of 500\,pc. On the $x$-axis are stars' real $G$-band magnitudes (as listed in the DR3 catalogue), and on the y-axis is the difference between the real and artificially increased $G$-band magnitudes (due to the presence of the cloud in their LoS). Blue crosses represent stars which are still seen by {\it Gaia} after the increase, and red triangles are stars which are now fall below {\it Gaia}'s magnitude limit.}  \label{Fig_background} 
\end{figure}

Parallax measurements were used to identify field stars which were foreground, background and within the dimensions of the simulated cluster’s natal cloud at each proposed distance. Stars within the cloud boundary were kept to represent an expected older star population, which is not included in our simulation data. The line-of-sight $V$-band extinction due to the presence of the cloud was calculated for those identified as being within or background to the cloud, from column densities using a modified routine of the {\it Gaia} Simulator. These were then converted to a $G$-band extinction using the extinction law $A_G/A_V= 0.789 $ \citep{2019ApJ...877..116W} and the stars' catalogued $G$-band magnitudes were adjusted accordingly. Those now fainter than $G=20.7$\,mag were removed from the sample. No modification to the catalogues parameters of stars identified as foreground, or which had no or negative DR3 parallax measurements, were made. Figure\,\ref{Fig_background} shows an example of the changes in $G$-band magnitudes obtained for the DR3 survey samples.

We note that although {\it Gaia} provides the highest precision stellar parallax measurements to date, there are still non-negligible uncertainties on stars' true $z$-axis (line of sight distance) positions. This is because the parallax and associated errors of the {\it Gaia} catalogues represent the mean and standard deviation in a Gaussian distribution of potential true parallaxes, meaning there is a small probability that a star has a true parallax significantly larger or smaller than the given value. Thus, it is expected that some stars we have identified as foreground, within or behind the cloud have been inadvertently mislabelled. 

We justify taking the parallax measures at face value rather than use a posterior probability distribution (like, e.g. \citealt{2021AJ....161..147B}) as the number of stars mislabelled is expected to be small, and statistically insignificant w.r.t to populating our parameter space with a plausible field star distribution. Furthermore, we chose not to assign distances  to stars which had no or negative parallaxes as parallax, rather than distance, will be used to identify cluster members (Section\,\ref{sect_members}) i.e. these stars would be excluded during that stage regardless. 

Figure~\ref{Fig_stages} shows the positions and colour magnitude diagram for one of our datasets, Orion-type-3, before (top) and after (second panel) the field stars are added, assuming a distance of 2.5\,kpc. As would be expected, the field stars engulf the simulated region, however it is still substantial enough to be visible both in position space and as a sequence in the colour magnitude diagram.

\subsection{Resolution and Quality Limitations}\label{sect_resolve}
Having merged our simulated datasets
with real stars from {\it Gaia}, we now remove stars which would not be observed due to resolution and/or quality limitations.  

{\it Gaia}’s ability to resolve two visually close stars as separate sources varies as a function of angular separation, $s$, and $G$-band magnitude contrast. Using close pairs from the full EDR3 catalogue, \citet{2021A&A...649A...6G} empirically defined this function as:

\begin{equation}
\\\\\\ s_{\rm min}= 0.532728 + 0.075526 \cdot \Delta\,G + 0.014981 \cdot (\Delta\,G)^2 \\
\end{equation}
where $s_{\rm min}$ is the minimum angular separation in arcseconds between two stars needed by {\it Gaia} to be resolved as separate sources, and $\Delta\,G$ is the difference in their $G$-band magnitudes. 
Cases where $s<s_{\rm min}$ will include astrometric binaries which are detected as a single source, which will not fit a single-star astrometric solution resulting in a poor fit of the astrometric model and hence would be given a high RUWE by the {\it Gaia} reduction pipeline \citep{2019ApJ...877..116W}.

We combined our synthetic observations (Section\,\ref{sect_errors}) with the DR3 field star selections (Section\,\ref{sect_contam}) to create a 
synthetic plus field catalogue. To identify unresolved cluster stars, the angular separation between all stars in the sample was calculated. When $s<s_{\rm min}$ was found, both stars were assigned a RUWE$\,>\,$1.4. We did not combine the fluxes of unresolved pairs as the high RUWE values resulted in their removal from the sample in the next quality-based selection cut. 
Of course not all such stars would be detected as astrometric binaries, and although removing them all does not affect our results presented in Section \ref{sec:results}, it does mean our CMDs in Figure \ref{Fig_stages} lack the scatter often seen in {\it Gaia} data at faint magnitudes caused by unresolved physical and line-of-sight binaries.
For all remaining synthetic observation stars we assume the  RUWE assigned in Sect\,\ref{sect_errors} ($\,\approx\,$1.0 for most stars), and for the remaining DR3 stars an unchanged RUWE from the listed value. Finally, a selection cut of RUWE$\,<\,$1.4 was made for all stars (synthetic and DR3 sources).

\subsection{Kinematic Data}\label{sect_pm}

Corrections to the proper motions of cluster stars in real observed catalogues are necessary to compensate for (i) perspective contraction, caused by members radial motions and (ii) perspective point convergence, caused by members sharing a common motion (see e.g. \citealt{2020A&A...636A..80B}). As the {\it Gaia} simulator derives proper motion values by simply scaling the simulated X and Y velocities in the local standard of rest, these perspective effects are not present in the synthetic observations and no correction is needed for it. 
We add the additional velocity to the stars owing to the galactic rotation curve at each observed distance, assuming a disc rotation of 220\,km\,s$^{-1}$.

The proper motions of DR3 field stars comprise of two components, which are their true perpendicular motion to the observer’s line of sight (which includes the galactic rotation curve) and a component of velocity owing to solar rotation, i.e.

\begin{equation}
\\\\\\ \mu =\frac{U_{\rm p}+ V_0\,\cos{l}}{r}\\
\end{equation}
where $\mu$ is the proper motion, $U_{\rm p}$ perpendicular velocity and $r$ distance of a star, $l$ galactic longitude, and $V_0$ is the Solar rotation velocity. As our LoS is along $l=270^o$ (Sect\,\ref{sect_gsim}) the term $V_0\,\cos{l}$ is zero, so no correction to the proper motions of DR3 stars are needed.

We consider {\it Gaia} spacecraft’s scanning law, which causes small systematic errors in the proper motion measurements that increase with decreasing angular scale. Following the approach of \citet{2020A&A...636A..80B}, we use the values of these errors given in Table\,7 of \citealt{2021A&A...649A...2L} to exclude all stars from our samples (simulated+real) with a proper motion within uncertainties smaller than the systematic error associated with the angular dimension of each synthetic observation.

To convert between mas\,yr$^{-1}$ to km\,s$^{-1}$ at each cluster distance we use transformation factor $\kappa$, which has a value of 4.7405 at 1000\,pc.

\section{Optical images of the cluster catalogue }
Although {\it Gaia} is not suited to producing images of star clusters, we wished to see how our datasets would appear visually, and (for example) whether the simulated clusters are readily apparent.
We use the {\tt AMUSE-Fresco} software \citep{steven_rieder_2023_7701608, 2018araa.book.....P} to make optical images.
We first calculate the flux of the stars in the Johnson-Cousins $V$, $R$ and $I$ bands from the {\it Gaia} $G$, $G_{\rm BP}$ and $G_{\rm RP}$ magnitudes.
These magnitudes have already been adjusted for extinction from the gas.
{\tt Fresco} then creates an image for each of these bands, using the positions of the stars (RA, Dec. mapped to $x$, $y$).
This image is then convolved with a point-spread function with a core, halo and diffraction spikes to simulate an image as it would be observed from a ground-based telescope.
We create the point-spread function using a \citet{1969A&A.....3..455M} function with a full-width-half maximum of an arcsecond, and then add diffraction spikes as narrow two-dimensional Gaussians aligned along the cardinal directions.
The resulting $V$, $R$ and $I$ images are used as blue, green and red channels in the final image.
Finally, we add a background image showing the column density of the gas to the red channel.

\begin{figure*}
\centering
     \includegraphics[width=1.0\textwidth]{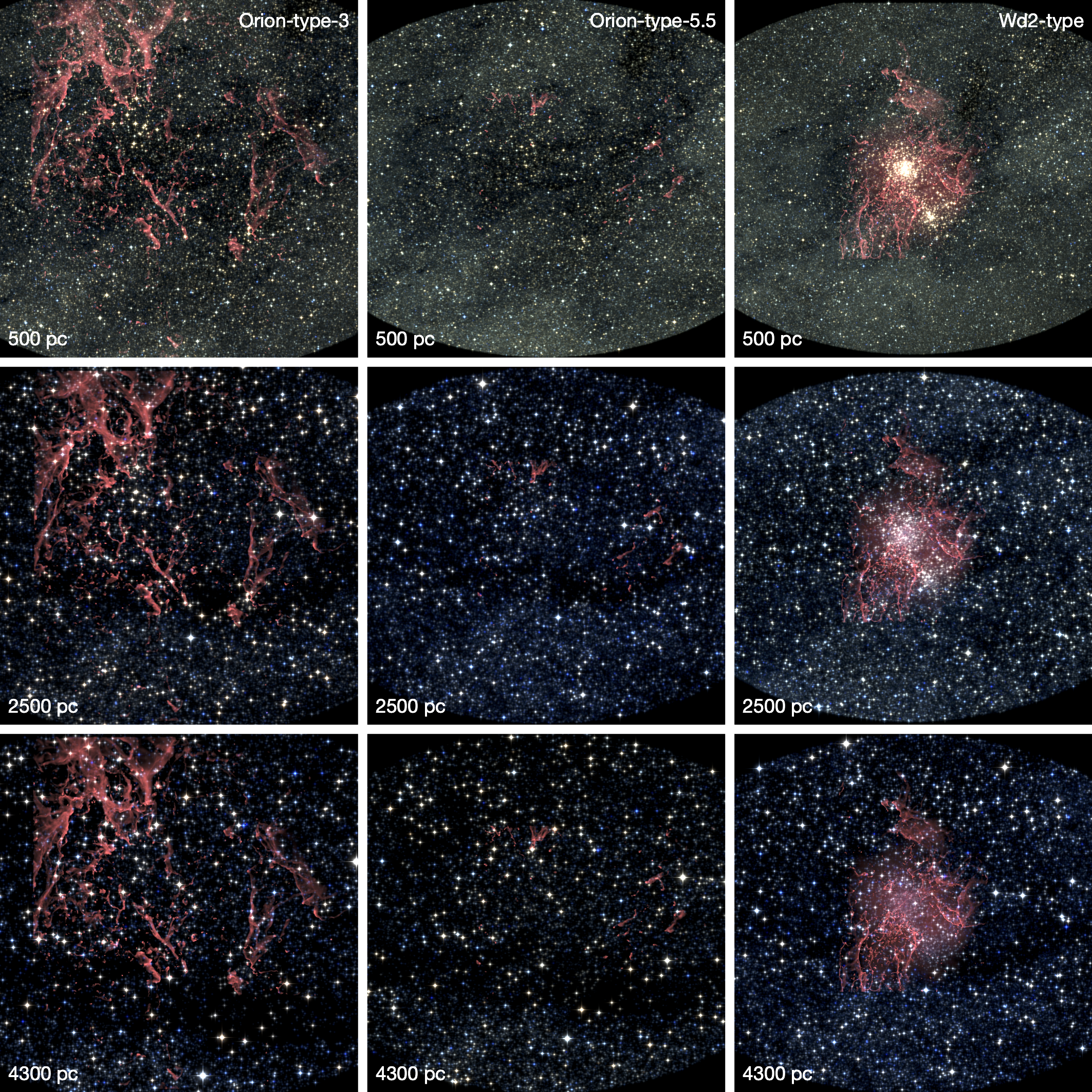}

    \caption{The clusters are shown as would be observed using an optical telescope, Orion-type-3 (left), Orion-type-5.5 (centre) and Wd2-type (right). The Orion-type cluster is visible at 500 pc, but difficult to identify by eye at 4300 pc, and for the later time.
    The Wd2-type cluster is readily observable at all distances. Gas is shown in red.}
    \label{Fig_observedimages} 
\end{figure*}

We show the optical images of the clusters in Figure~\ref{Fig_observedimages}. The Orion-type cluster is visible as an overdensity of stars, 
but is difficult to see at the larger distances of 2500 and 4300 pc, and for the later time. The Wd2-type cluster is easily visible at all distances, which is not surprising as this clusters is so massive, and is also associated with substantial gas.

\begin{figure*}
\centering
     \includegraphics[width=0.33\textwidth]{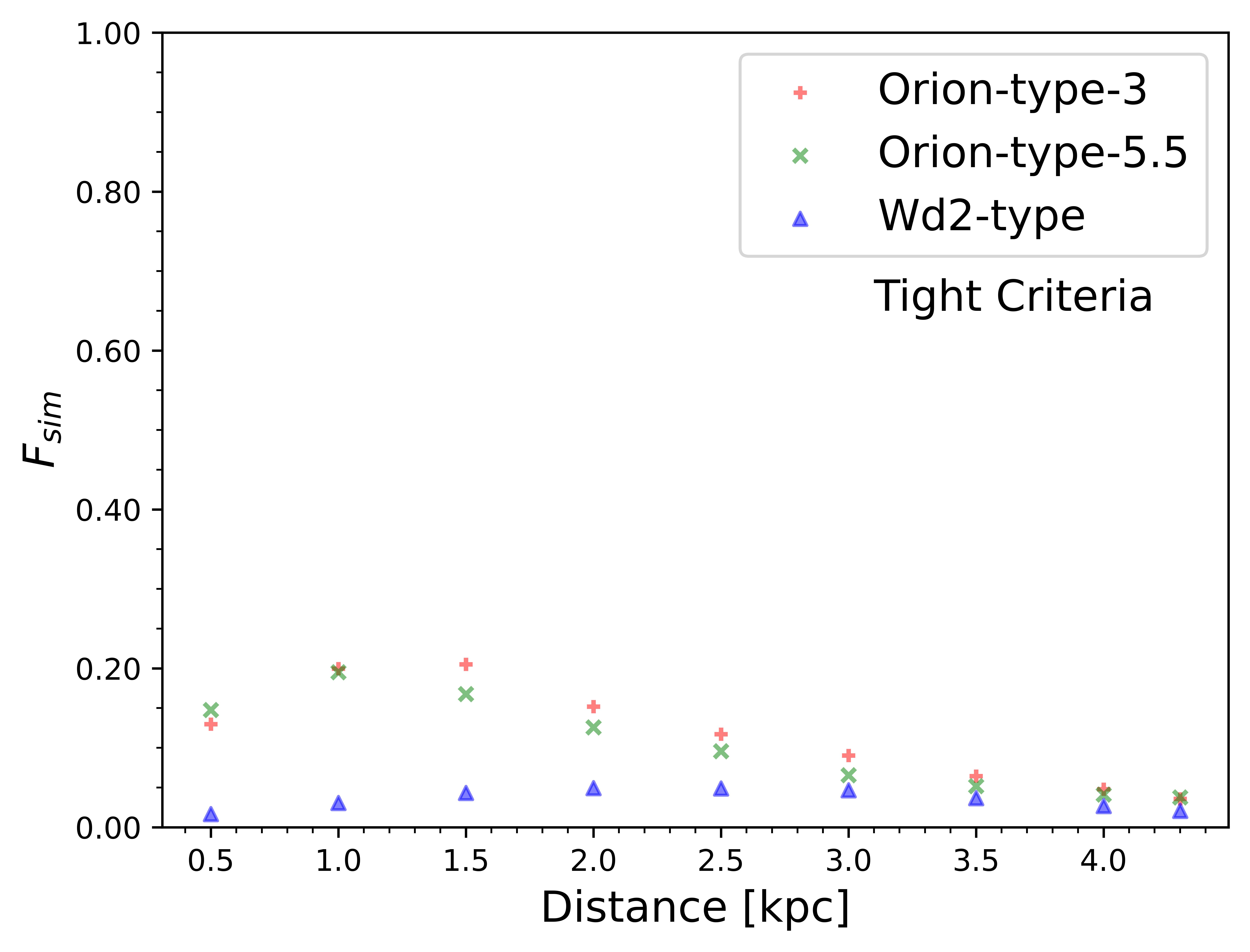}
    \includegraphics[width=0.325\textwidth]{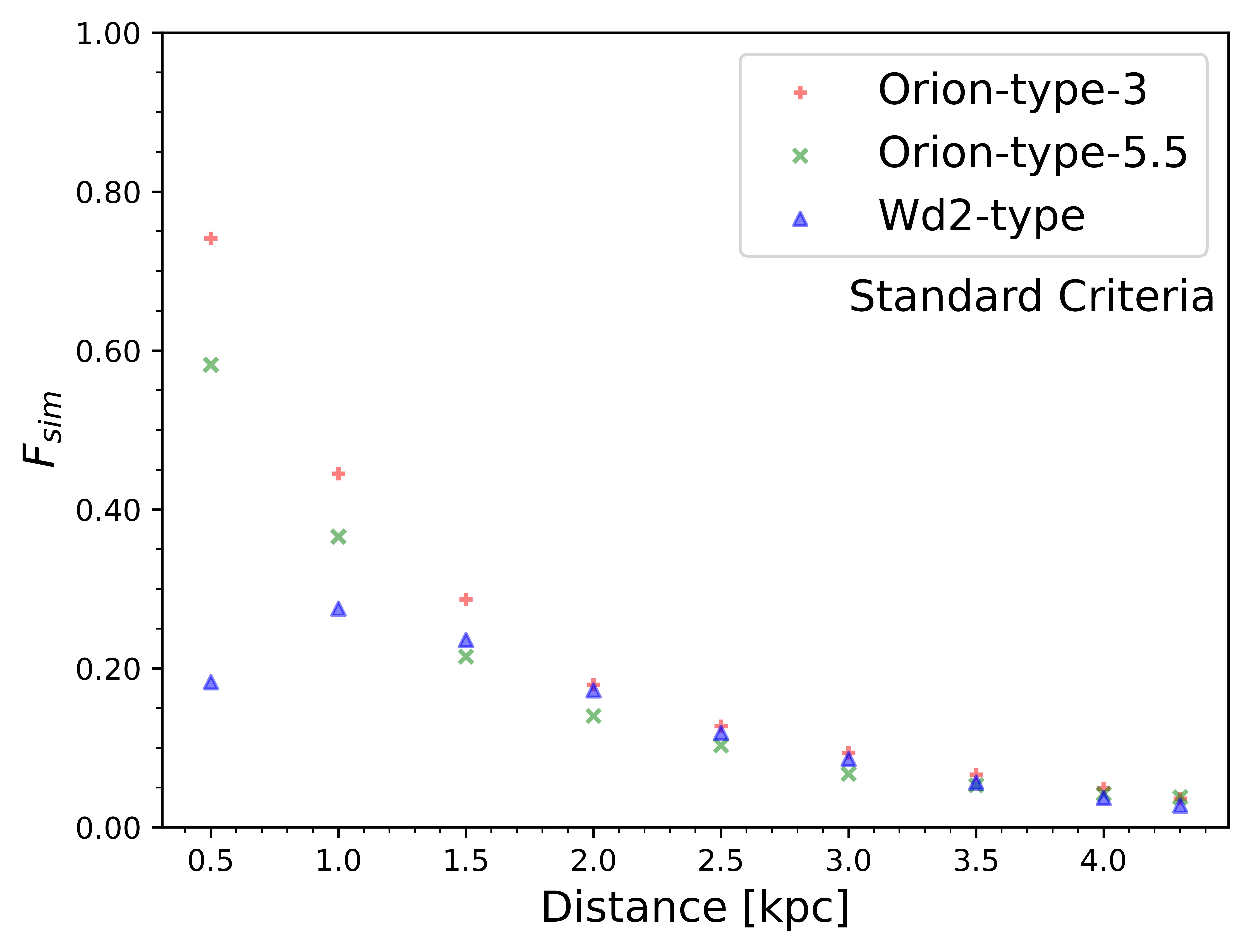}
     \includegraphics[width=0.32\textwidth]{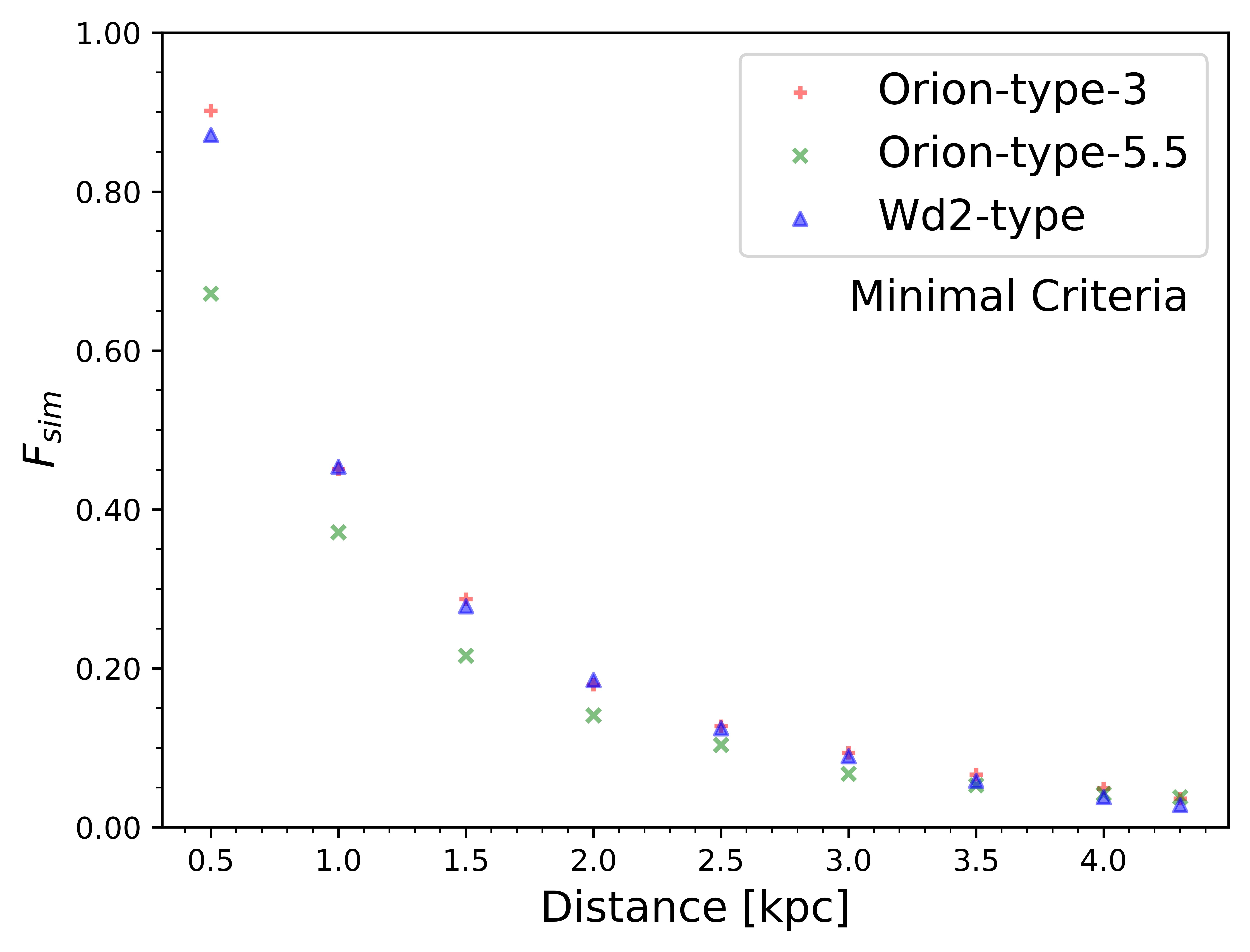}

     \includegraphics[width=0.33\textwidth]{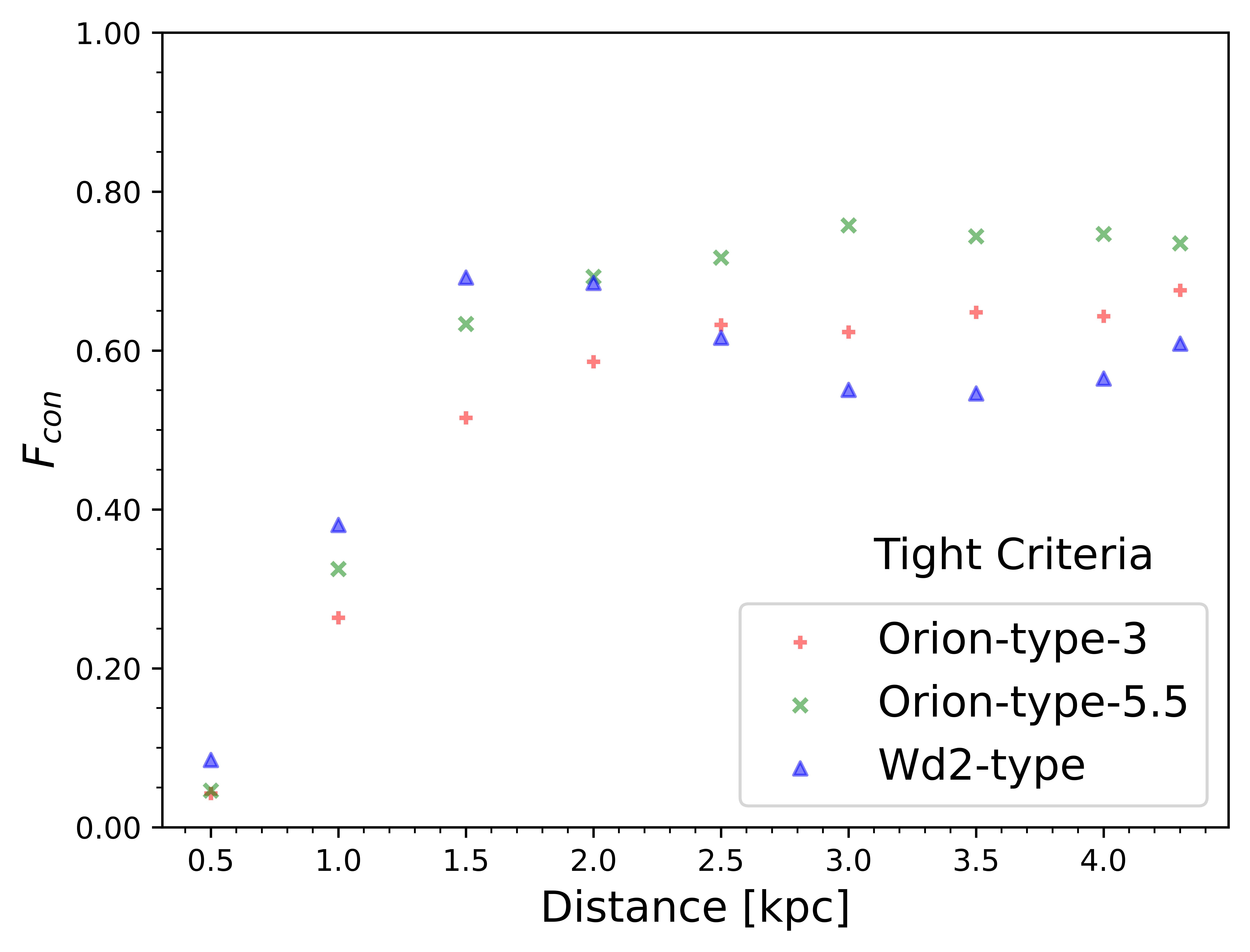}
    \includegraphics[width=0.325\textwidth]{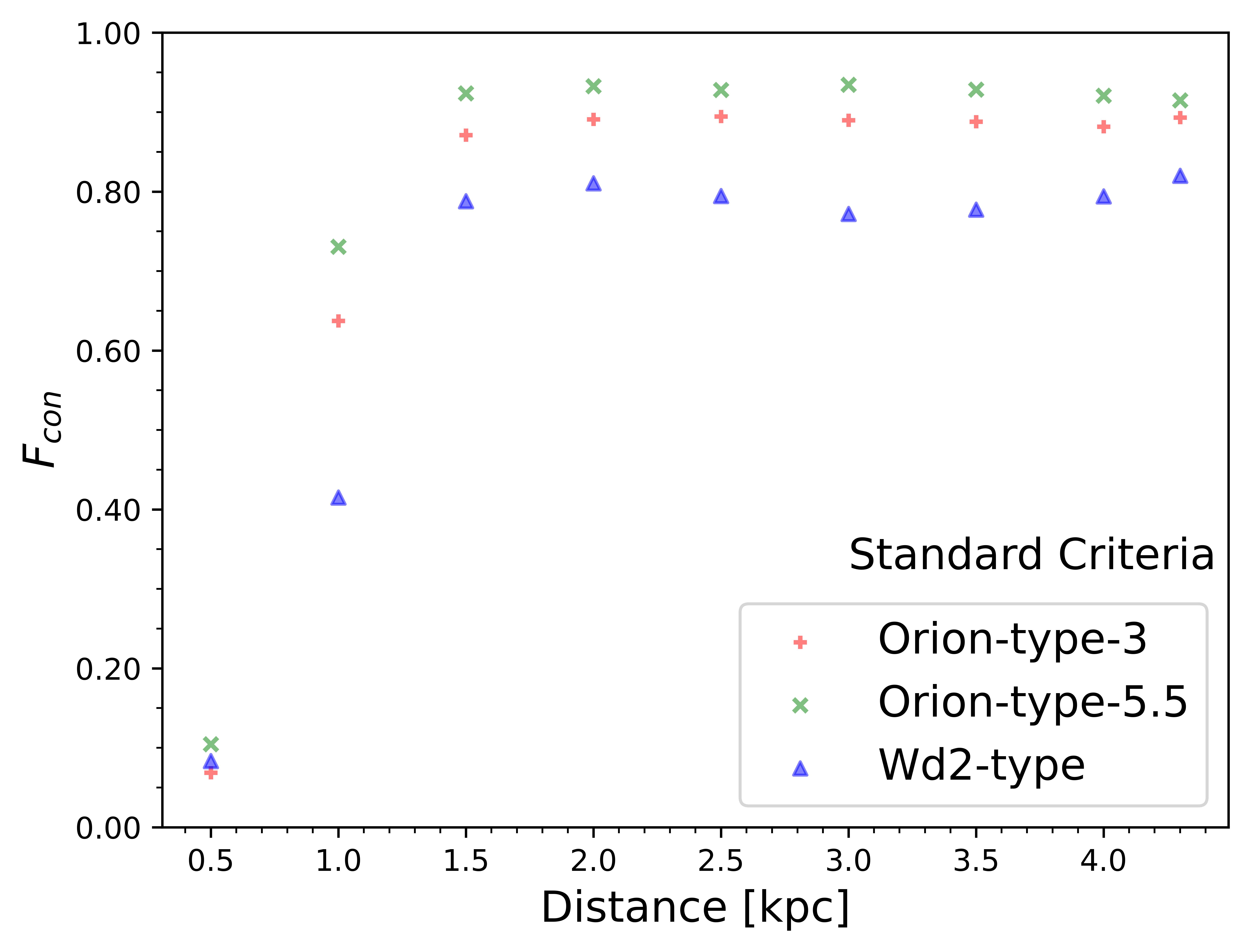}
     \includegraphics[width=0.32\textwidth]{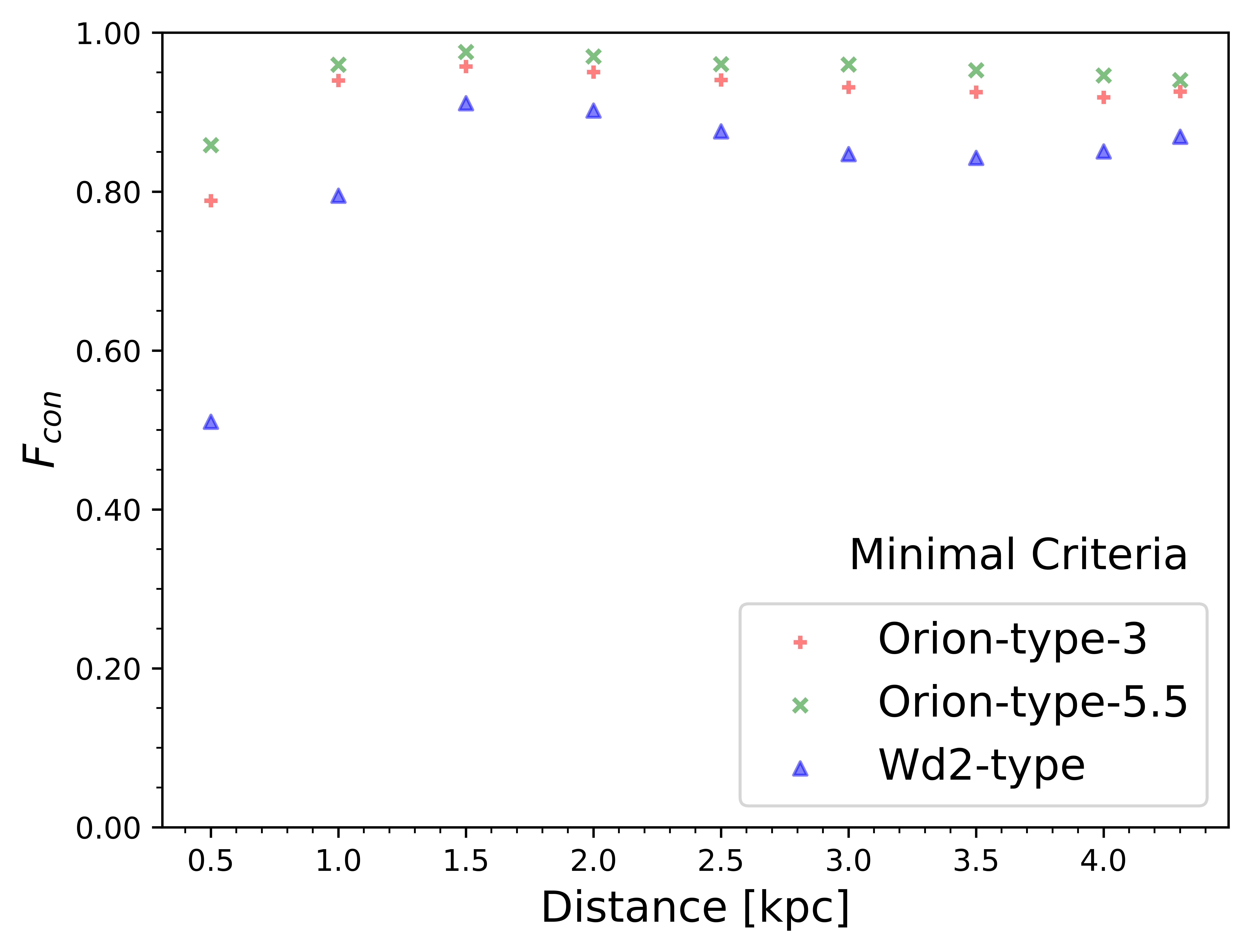}
    \caption{The panels show (Top Row:) fraction, $F_{\text{sim}}$, of observable simulation stars correctly identified as members using the different membership criteria  and (Bottom Row:) contamination of membership lists by non-member field stars as given by the number of incorrectly assigned members as a fraction of the total assigned members for clusters.
    For the minimal criteria up to around $80\%$ of stars in the cluster are wrongly attributed, but the tight criteria is too restrictive particularly at small distances.}
    \label{Fig_seensim} 
\end{figure*}

\section{Stellar Membership}\label{sect_members}
So far our catalogue includes all observable stars (real Gaia and simulated) in the field of view of the simulated datasets. We now identify members of our clusters, or associations, from the catalogue as an observer might do assuming that, as our regions are quite substantial, they would already be known and, for example, listed in an earlier cluster catalogue such as DAML02\footnote{\url{ https://wilton.unifei.edu.br/ocdb/}}. Therefore, rather than search the samples for potential clusters, we attempt to identify members of clusters around their `known' 5D ($l,b,\omega,\mu_{\alpha*},\mu_{\delta}$) position. 

Three different membership criteria were explored.
\begin{enumerate}
\item \textit{Tight} - have proper motions within 0.5\,mas/yr, and parallax within 0.3\,mas, of catalogued values.\\
\item \textit{Standard} - have proper motions within 2\,mas/yr, and parallax within 0.3\,mas, of catalogued values. \\
\item \textit{Minimal} - have a parallax within 0.3\,mas of catalogued values (no proper motion constraints).
\end{enumerate}
 For all three sets of criteria, stars had to be within the catalogued angular radius, about the central longitude and latitude coordinates. The radii used were the maximum angular distance of a simulation star from the cluster centre, plus $10\%$. 

 The criteria loosely follow the prescription of \citet{2018A&A...618A..93C} who adopted these in conjunction with their unsupervised membership assignment code, UPMASK, to identify cluster members in the {\it Gaia} DR2 catalogue. The authors applied the Standard criteria to most clusters, with Tight and Minimal criteria used for compact and dispersed clusters in proper motion space clusters respectively. They concluded applying the criteria in this way struck a good balance between keeping field contamination to a minimum but the criteria remained greater than the apparent dispersion of members for those clusters. 

While we do not use UPMASK to identify cluster members, these criteria are a reasonable starting place to compile a membership list. Advantageously, in this work we know {\it a priori} which stars in our membership lists are true members and which are field contamination, so can quantify the effectiveness of each set of criteria for use in future observational studies.  To this end, we applied all three sets of criteria to the clusters in our catalogue, regardless of their distribution in the proper motion space. 

\citet{2018A&A...618A..93C} also excluded stars fainter than $G$=18.
We experimented with this cut, but found that for our clusters this significantly reduced the number of members correctly identified for only marginal decreases in the contamination.
Hence we imposed no cut in magnitude.

An example of stars selected as members for the Orion-type-3 model using the minimal, standard and tight criteria is shown in Figure~\ref{Fig_stages} (lower panels). As shown, the minimal criteria do a poor job of selecting stellar members, including many stars which are clearly part of the field population. The standard and tight criteria recover stellar members with more success, both in position space and in the colour magnitude diagram, but over-exclude stars, with the tight criterion making the greatest cuts. 

In Fig.~\ref{Fig_seensim} we explore this in a more quantitative way using two metrics.
The first is the number of actual members selected as members expressed as a fraction of the number of observable by Gaia ($F_{\rm sim}$, upper panel). Of course the members selected could include many stars which are not actual members, so our second metric is the fraction of the membership list which is actually non-members ($F_{\rm con}$, lower panel).
Our ideal criteria would capture a very large number of members ({\it{i.e.}} a large $F_{\rm sim}$) with a very low contamination ($F_{\rm con}$).

We can see that for the Orion-type models, the tight criteria achieves the best balance, with $F_{\rm sim}$ typically between 5-20\% and $F_{\rm con}$ below 75\%.
Removing the proper motion cuts entirely for the Orion-type models (right-hand panels) results in unacceptably high contamination (around 95\%), whilst loosening them (middle panels) increases the contamination from around 70\% to 95\%, with no notable gain in true members at distances greater than 1.5~kpc.

The tight criteria also works best for the Wd2-type models, except when the cluster is closer than 1.5~kpc.
This is because the cluster has a relatively large velocity dispersion, because of its large mass, and hence many members are lost in the proper motion cuts.
This is demonstrated in the upper right panel of 
Fig.~\ref{Fig_seensim}, where removing the proper motion cuts allows us to recover most members.
However, this gain is bought at the cost of an increased contamination rate (bottom right panel), showing that there is no good solution for a  nearby Wd2-type cluster; its high velocity dispersion makes proper motion selection problematical. This mirrors the conclusion \citet{2018A&A...618A..93C} reached from real data that even for modest clusters at distances closer than the ones considered here, it is advantageous to drop proper motion cuts.

\begin{figure*}
\centering
     \includegraphics[width=0.33\textwidth]{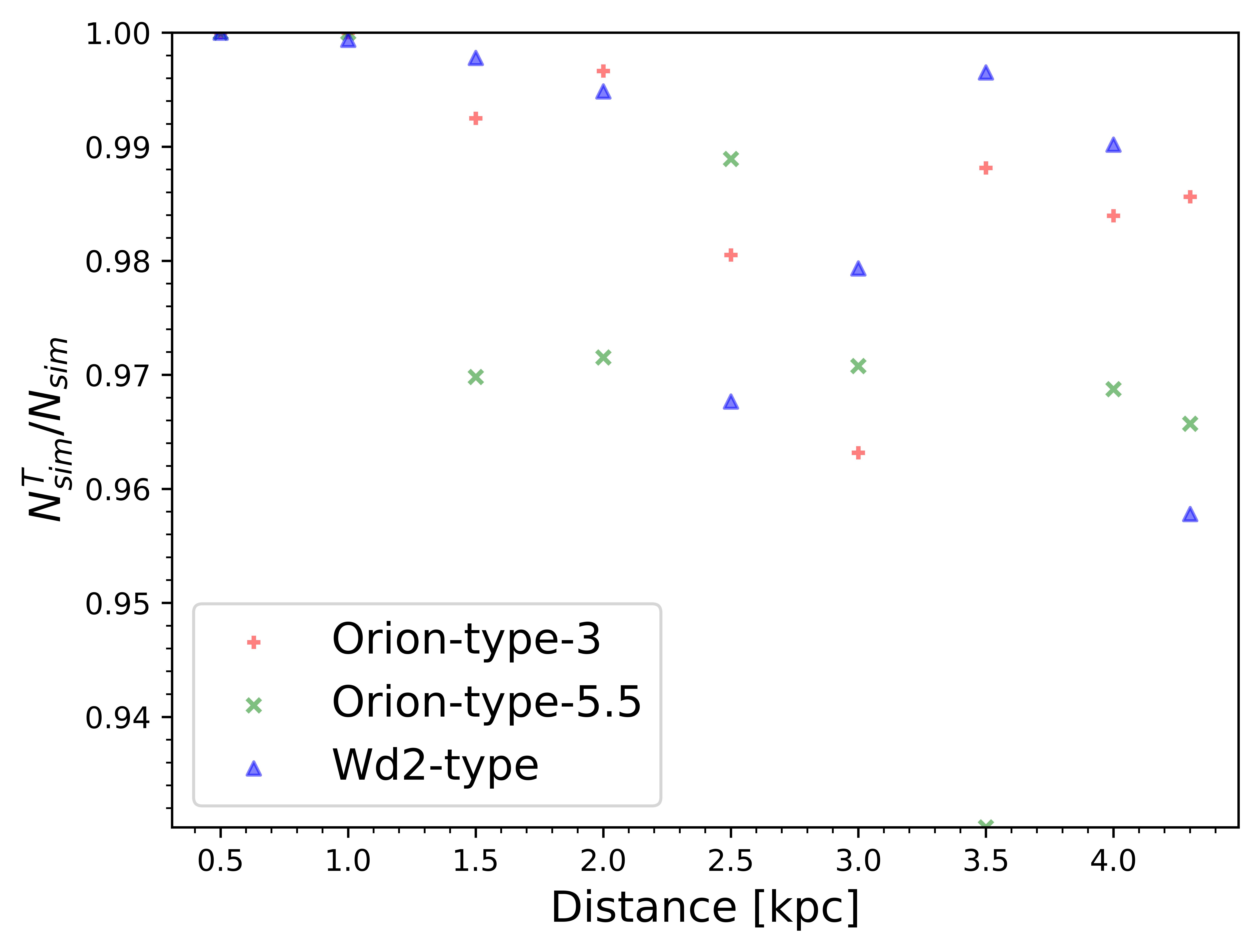}
    \includegraphics[width=0.325\textwidth]{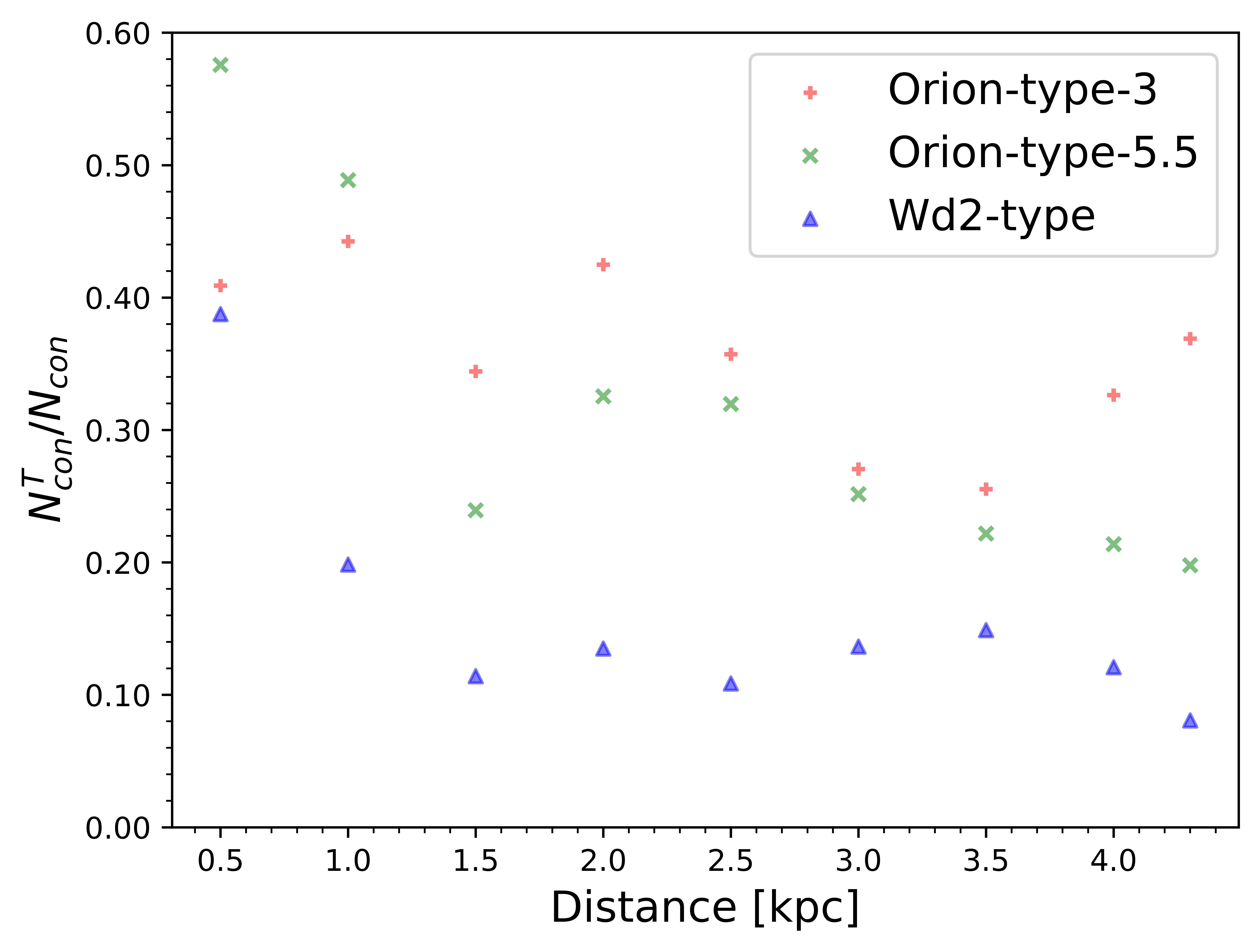}

    \caption{Plots show (Left:) fraction of observable simulation stars correctly identified using the Tight criteria, $N_{\text{sim}}$, still included as members after the position cut $N^{T}_{\text{sim}}$,  (Right:) fraction of contamination of membership lists by non-member field using the Tight criteria, $N_{\text{con}}$, still included as members after the position cut assigned members for clusters $N^{T}_{\text{con}}$.
    By applying the position cuts, based on visible structure in the Ra/Dec plots, we have been able to reduce the contamination by up to $91.2\%$ with minimal impact on the identification of true members.}
    \label{Fig_ConSeensim_trim} 
\end{figure*}

\begin{figure*}
\centering
     \includegraphics[width=0.33\textwidth]{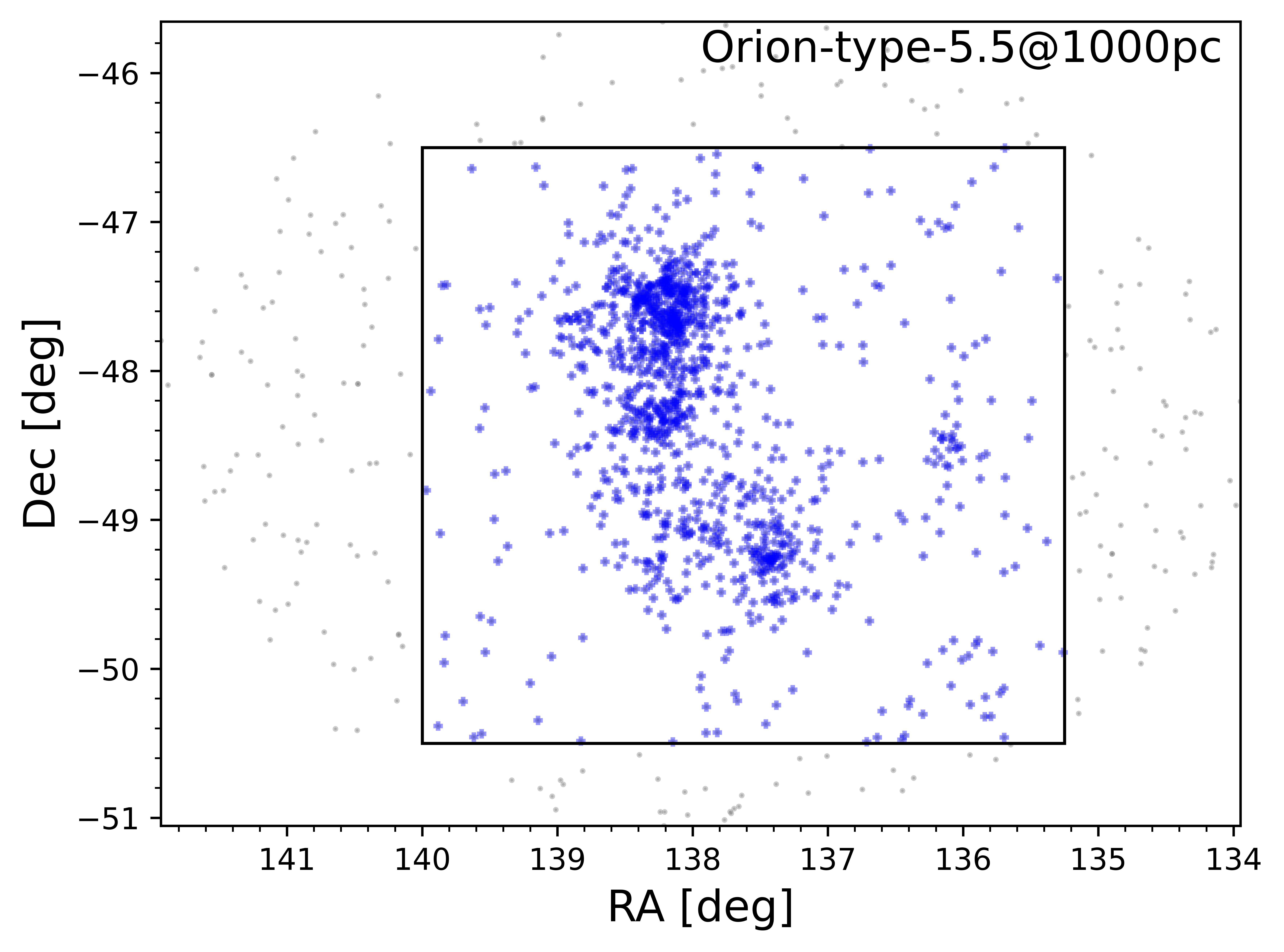}
    \includegraphics[width=0.325\textwidth]{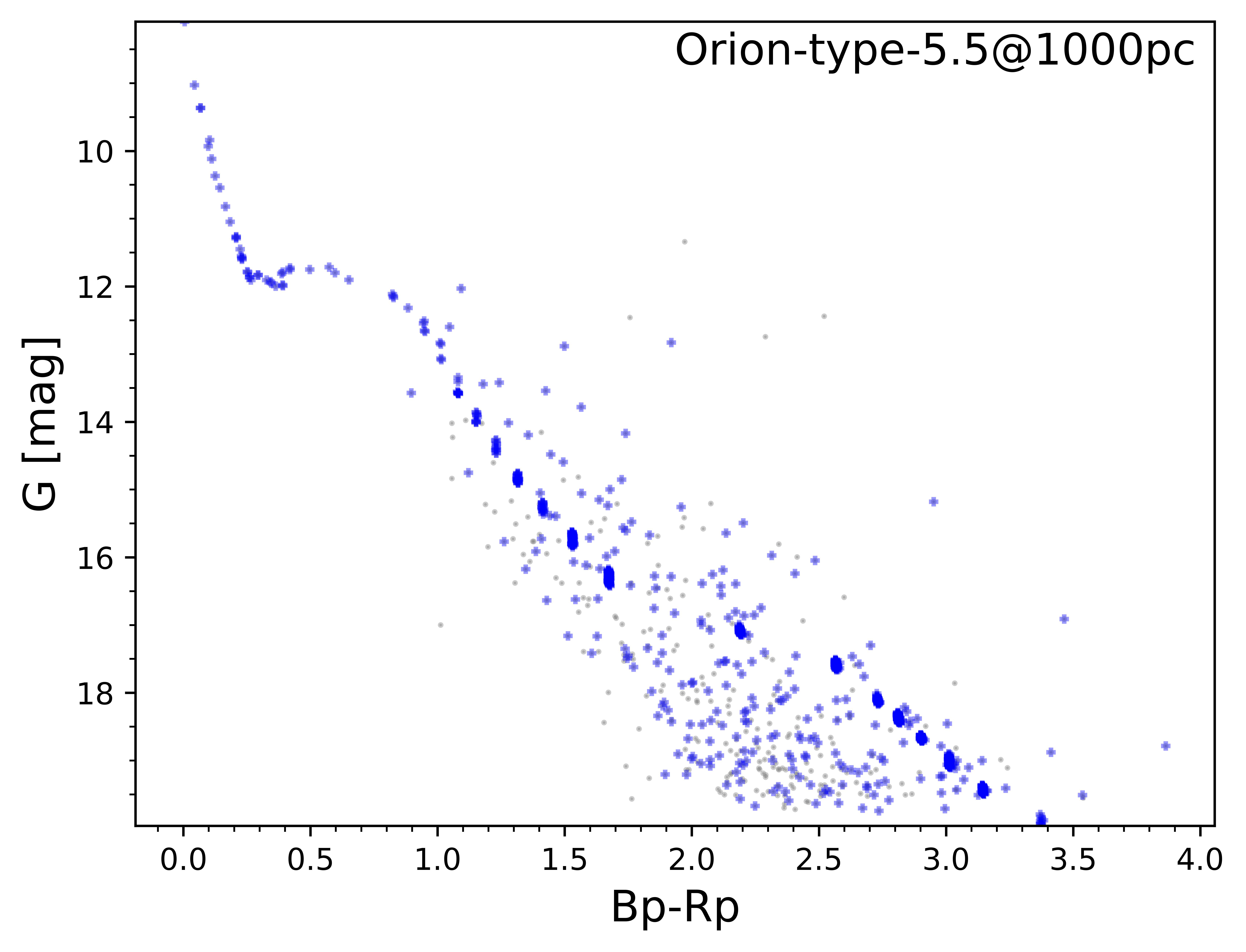}

    \caption{Example of the (Left:) position cuts made to members identified using the Tight criteria and (Right:) the resulting CMD. Blue crosses and grey dots represent members included and excluded by the trim, respectively.}
    \label{Fig_example_trim} 
\end{figure*}

Although the tight criteria, on balance, provided the more favourable outcome for our clusters the contamination rate was surprisingly high. Plotting the positions of `members' found using the tight criteria showed in nearly all cases clear structure at the centre (the real cluster) surrounded by an extended dispersed population, demonstrating clear evidence of contamination in the sample. Mindful that observers do not have {\it a priori} knowledge of the true appearance of a cluster, we attempted to reduce the contamination by making cuts in the RA/Dec space to select regions with apparent structure present (Figures \ref{Fig_ConSeensim_trim} \& \ref{Fig_example_trim}). This approach worked quite well, reducing the contamination between $42.4\%$ - $91.9\%$  at a cost of typically less than a $5\%$ loss in true members. 

\citet{2018A&A...618A..93C} designed what we have called the standard criteria to be sufficiently wide to include most members, whilst eliminating a large number of non-members.
With the advantage of a ``ground truth" we can see that the tight criteria with additional cuts in the position space after member selection produces relatively little contamination, and using this set of criteria the CMD and position plots seem reasonable, at least for the clusters we have simulated, and so we will use them for the remainder of the paper. We note that when placing our clusters outside the mid-plane (where stellar density is significantly lower) the standard criteria produced the best balance of members/contamination and convincing plots. Therefore we recommend careful consideration is given by observers to the galactic location and velocity dispersion of their cluster, as well as inspections of the CMD/position plots trialling both the tight and standard criteria, when considering which constraints works best, and if further constraints are needed, in their specific case.

\section{Analysis Methods}
We analyse the spatial and dynamic properties of the clusters using the same techniques as in Paper I \citep{2022MNRAS.514.4087B}. We outline the techniques below, but for a full description see Paper I.

We assess the spatial association of cluster members using the 2+D statistical clustering tool INDICATE \footnote{\url{https://github.com/abuckner89/INDICATE}} (INdex to Define Inherent Clustering And TEndencies; \citealt{2019A&A...622A.184B}). INDICATE is a local statistic which quantifies each star's degree of association with other stars, given by an index $I$. This index is a ratio of the expected number of neighbours, $N$, for a star $j$ if it was not spatially clustered and its actual number of members, $N_r$. Following \citet{2020A&A...636A..80B} we set $N=5$, such that
\begin{equation}
\\ I_{5,j}=\frac{N_r}{5}
\end{equation}

The index is unit-less with a maximum value of $\frac{S-1}{5}$, where $S$ is the total number of stars in the sample. It is calibrated against random distributions to identify significant values, with values greater than 3$\sigma$ above the mean value for random distributions representative of spatial clustering and higher values denoting progressively greater degrees of spatial association. 

We also use the $\mathcal{Q}$ parameter \citep{2004MNRAS.348..589C} to quantify spatial structure. Unlike INDICATE this parameter provides a global, rather than local, assessment of the stellar spatial distribution based on the 2D positions of members.  Values of $\mathcal{Q}<0.8$, $\mathcal{Q}\approx0.8$ and $\mathcal{Q}>0.8$ denote the cluster has a fractal, random or radial density gradient configuration.

We identify discrete clusters within the data using the Hierarchical Density-Based Spatial Clustering of ApplicatioNs (HDBSCAN*; \citealt{10.1007/978-3-642-37456-2_14}) algorithm. 
HDBSCAN* is similar to DBSCAN, except that for DBSCAN there is a constant search radius $\epsilon$ which is used to determine whether stars within this distance are part of the same cluster. With HDBSCAN*, $\epsilon$ is variable enabling clusters of differing densities to be found. The only input parameter required was ``$min\,samples$". This provides a measure of how conservative the clustering is, and we determined the appropriate value on a cluster-by-cluster basis by checking that the clusters agreed with (i) those identified of visual inspection and (ii) the spatial distribution described by INDICATE.

In terms of the dynamics, we determine the median 2D velocity w.r.t. the system centre, $v^*_{\rm{out}}$, of each cluster identified with HDBSCAN*, where 

\begin{equation}\label{Eq_vout}
 \\ v^{*}_{\rm out}=\vec{v}_{*}\cdot \hat{r}
\end{equation}
and $\hat{r}$ is the 2D outward component of velocity. $v^*_{\rm{out}}$ will be positive if the cluster is expanding, negative if the cluster is contracting, and zero if static.
\begin{figure*}
\centering

    \includegraphics[width=0.33\textwidth]{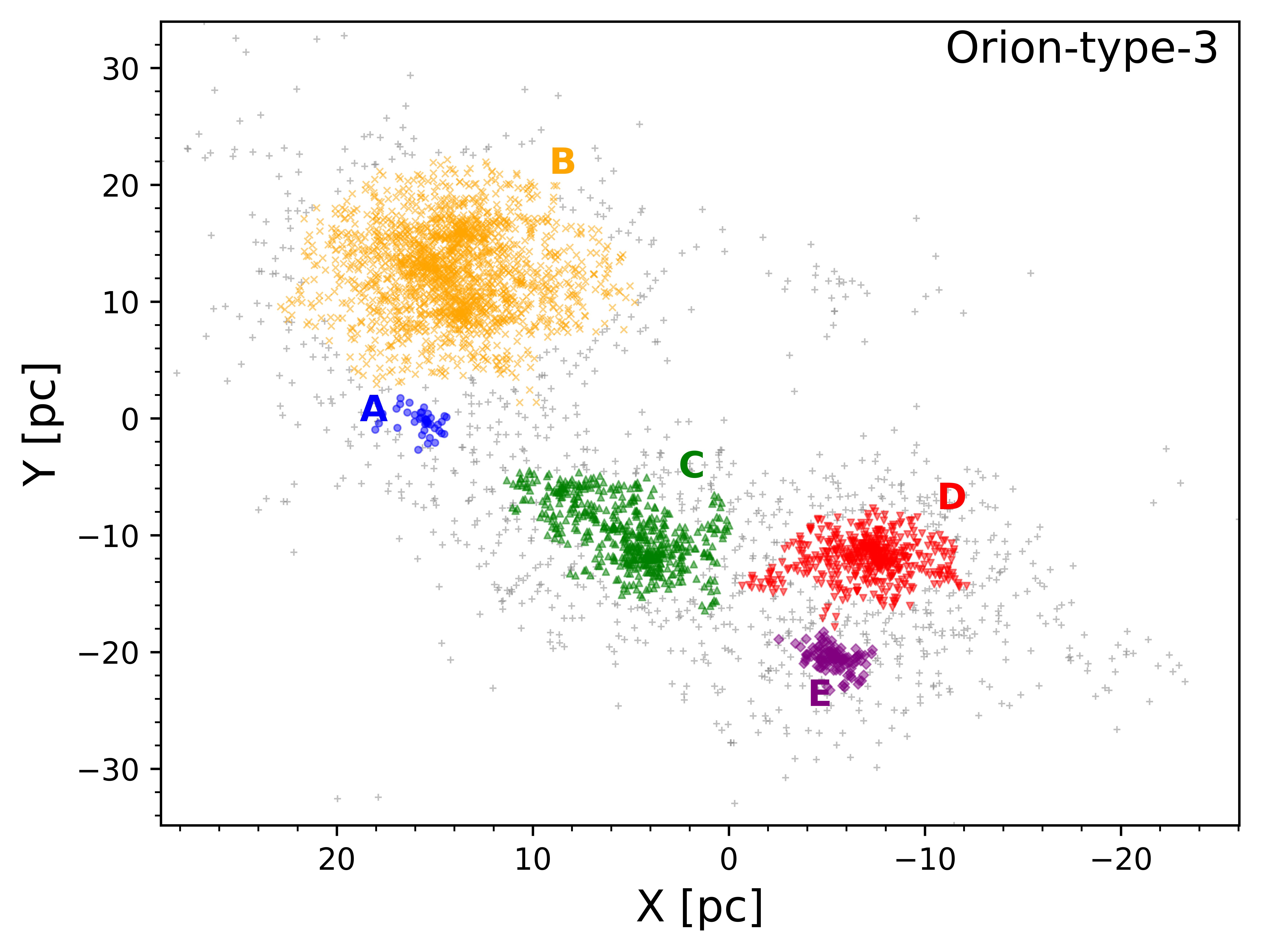}
   \includegraphics[width=0.33\textwidth]{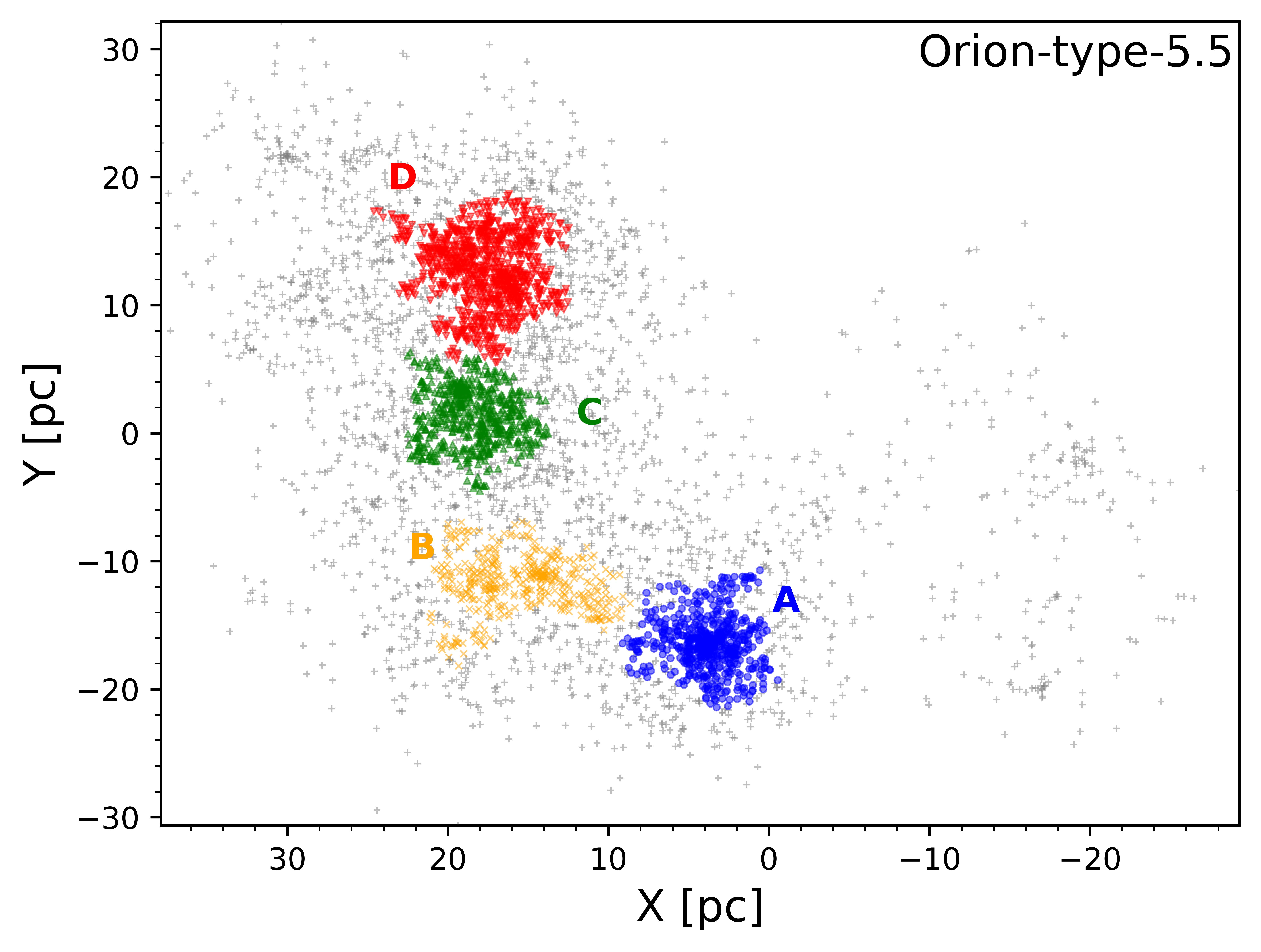}
   \includegraphics[width=0.33\textwidth]{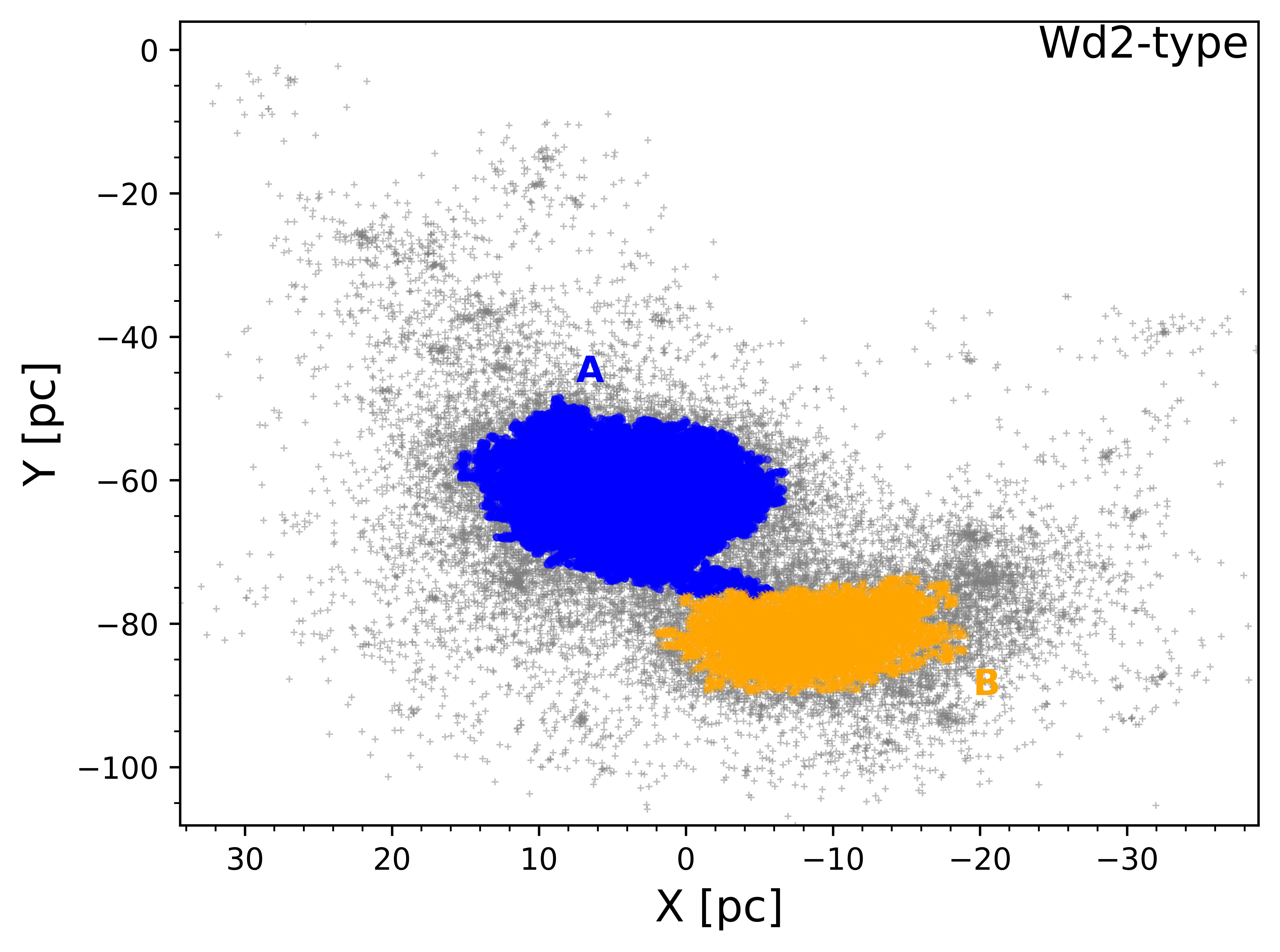}

    \includegraphics[width=0.33\textwidth]{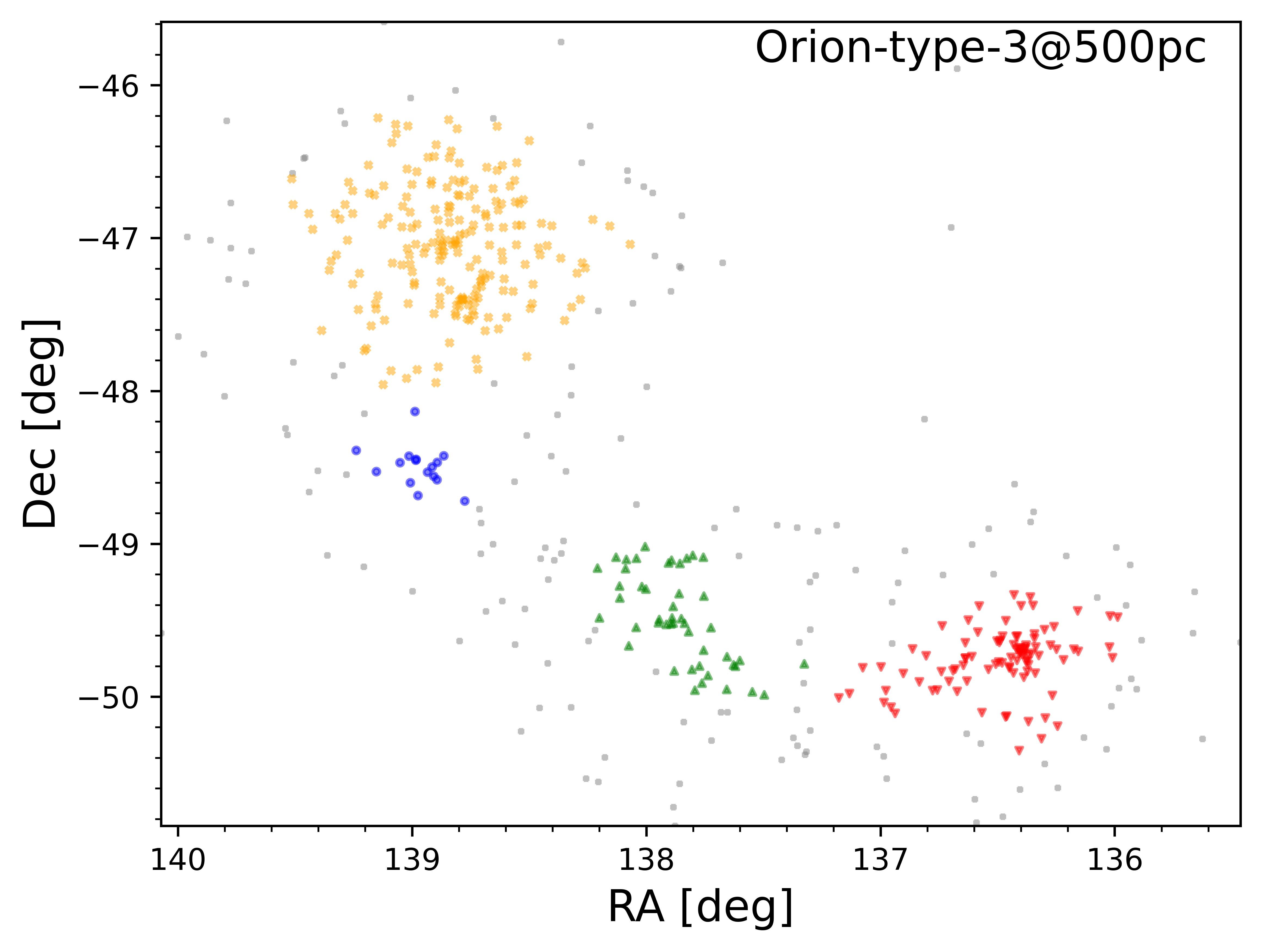}
     \includegraphics[width=0.33\textwidth]{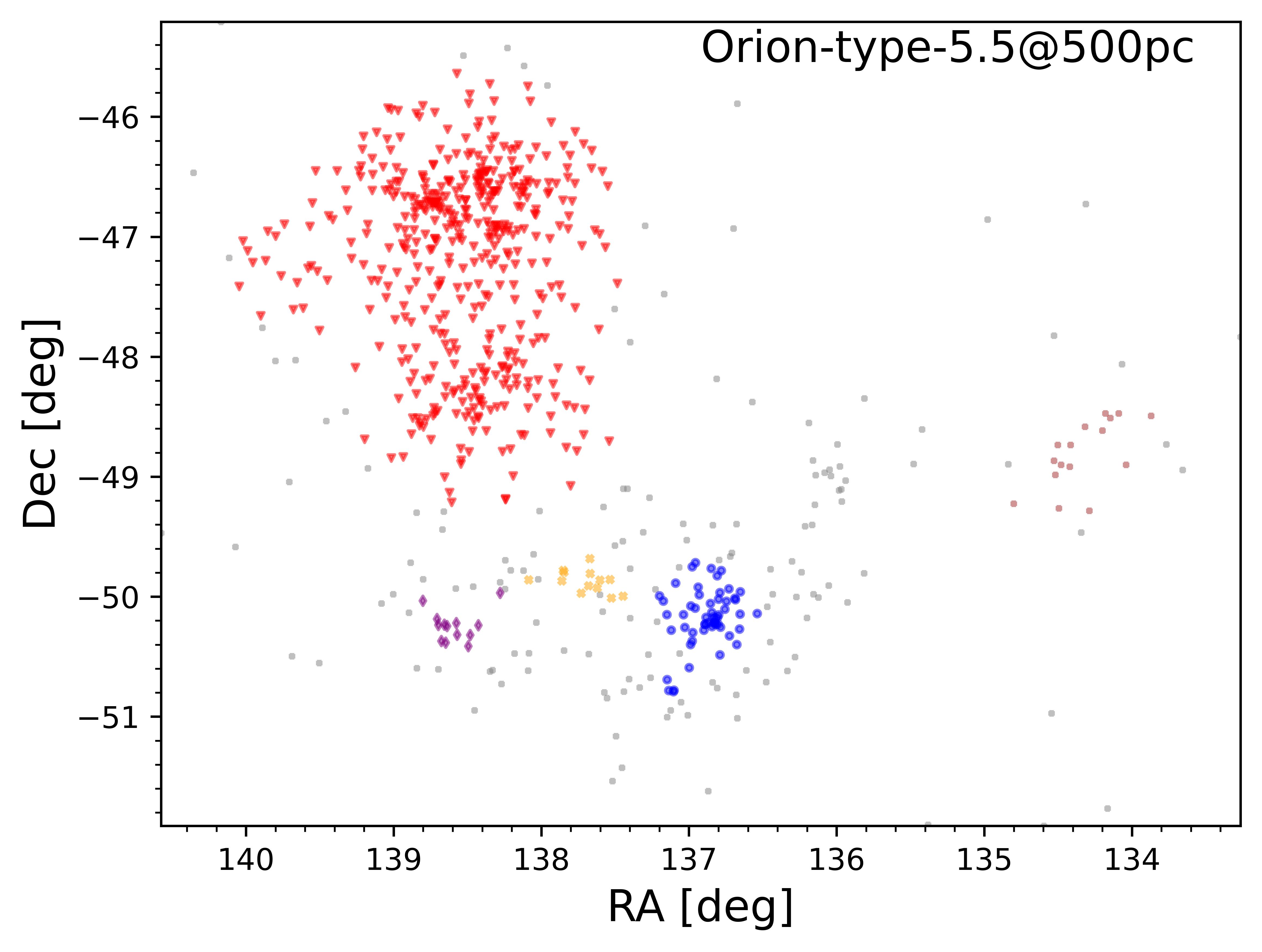}
     \includegraphics[width=0.33\textwidth]  {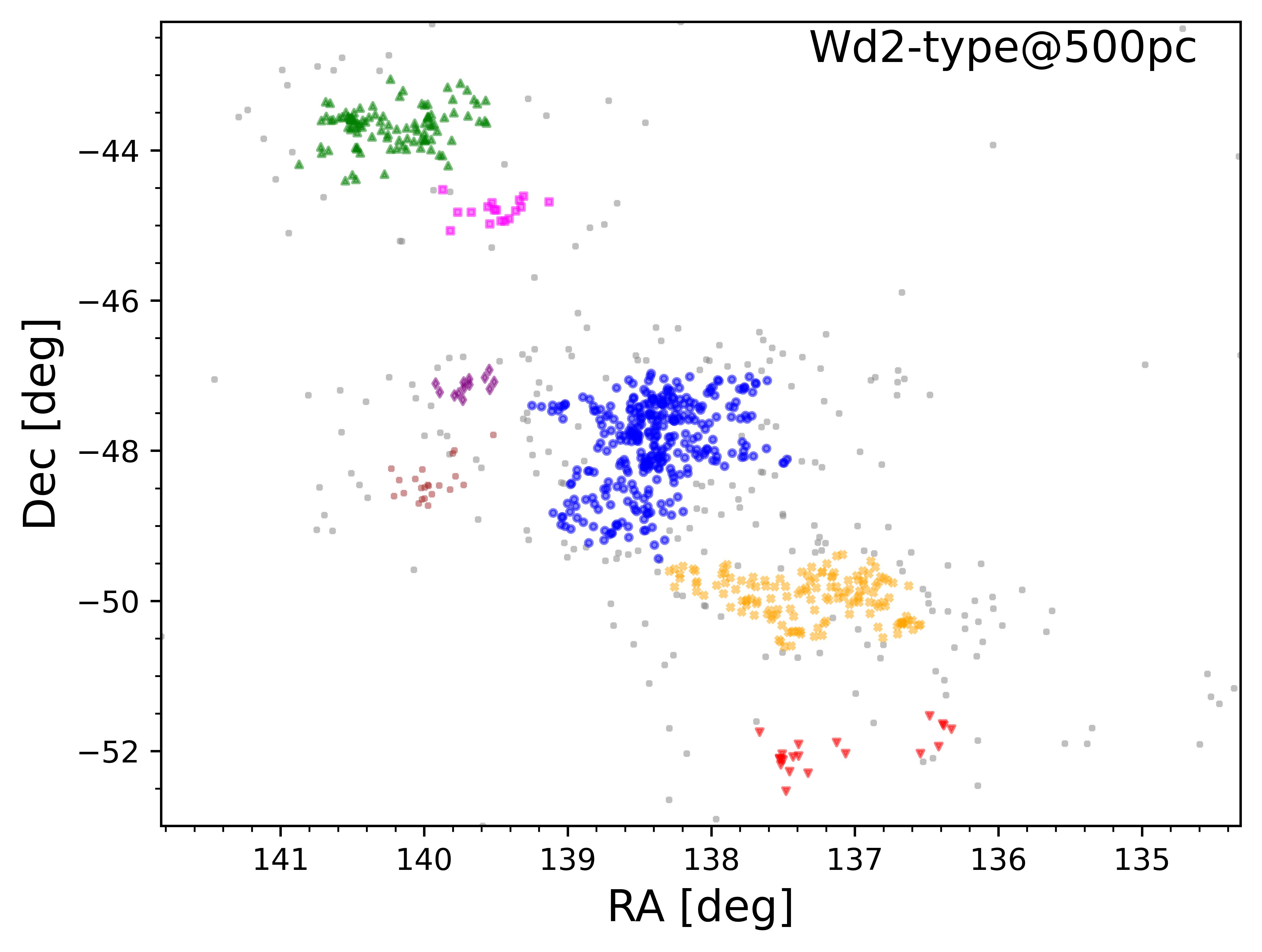}

    \includegraphics[width=0.32\textwidth]{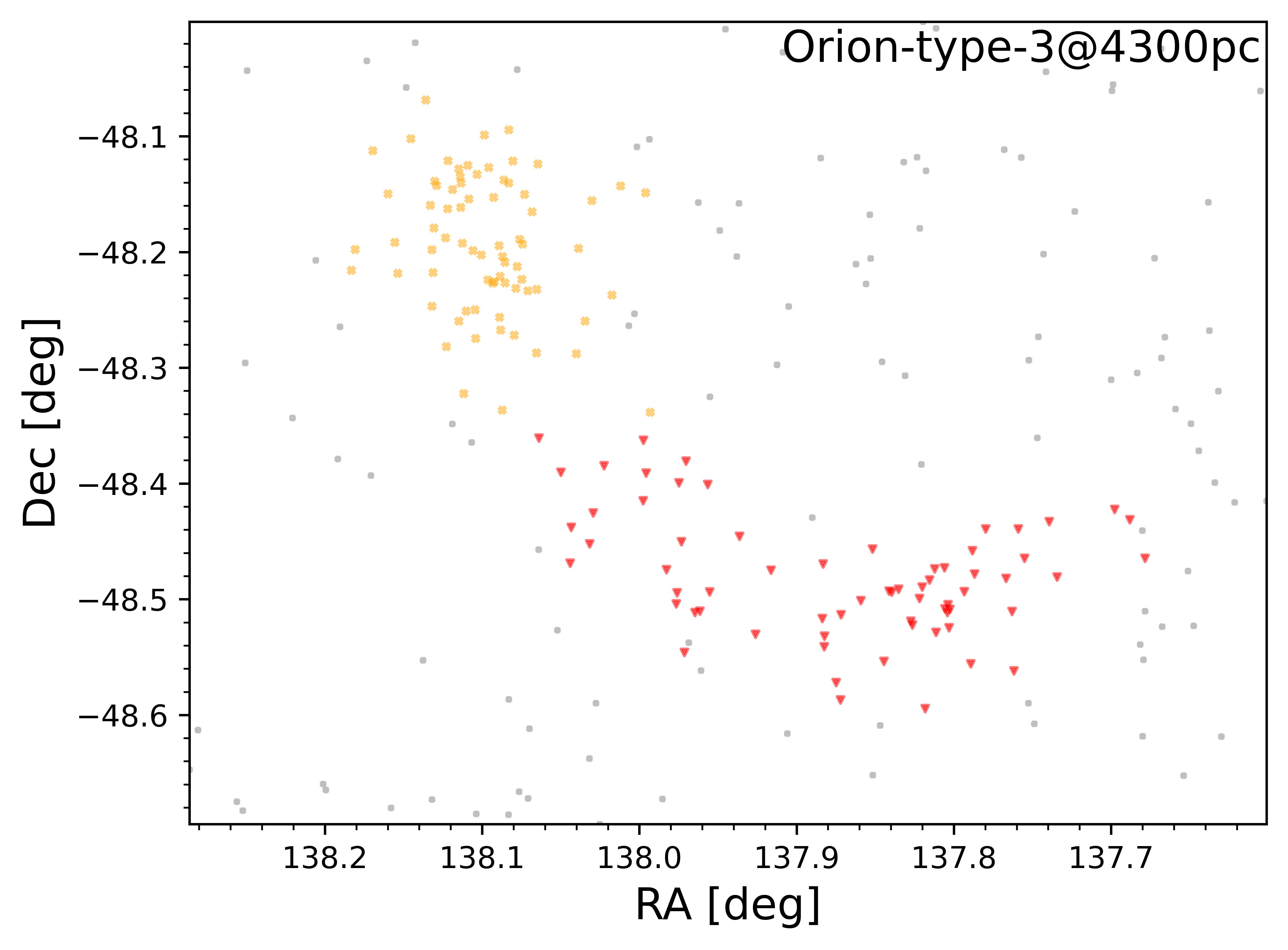}
     \includegraphics[width=0.32\textwidth]{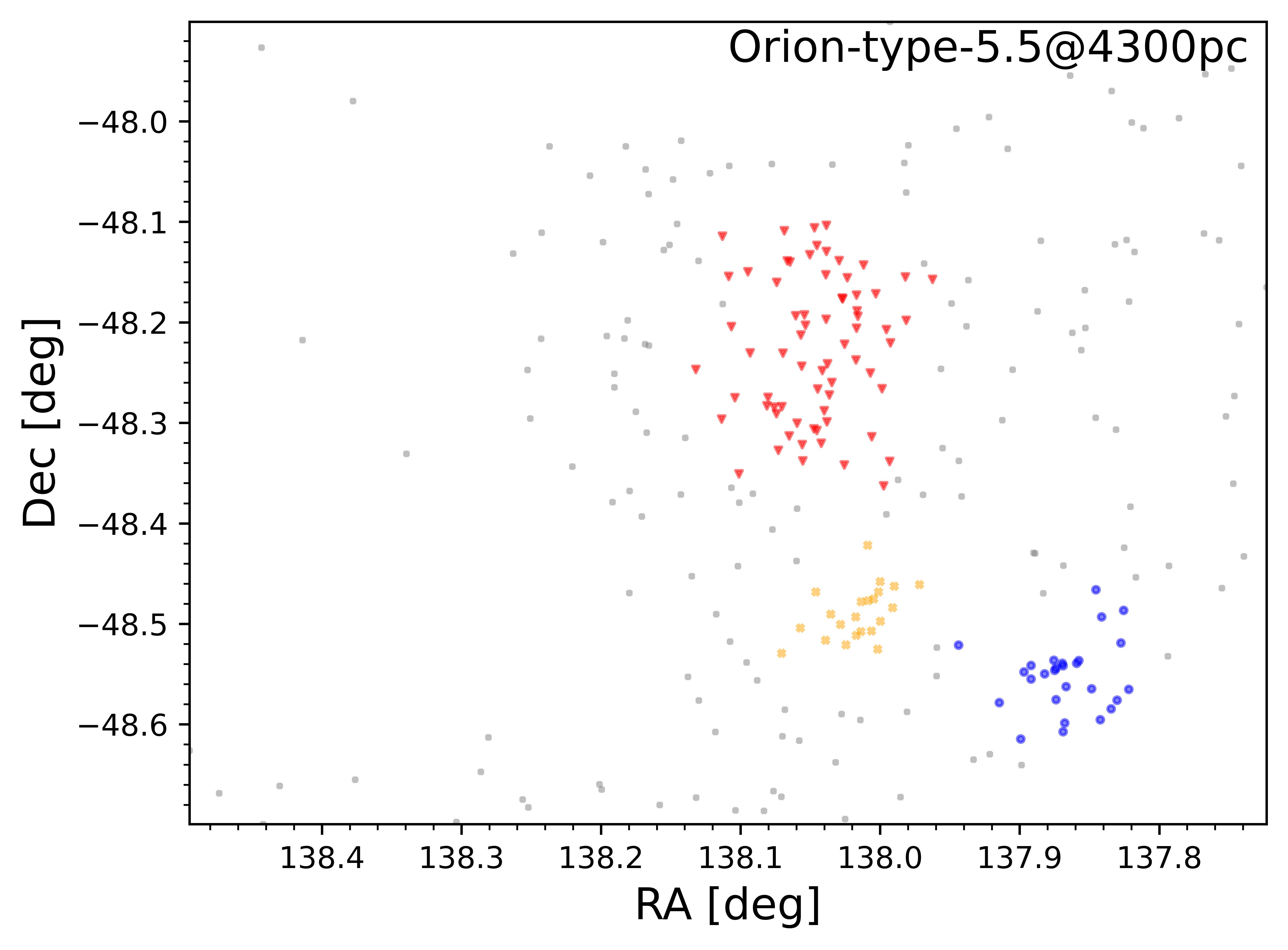}
     \includegraphics[width=0.33\textwidth]{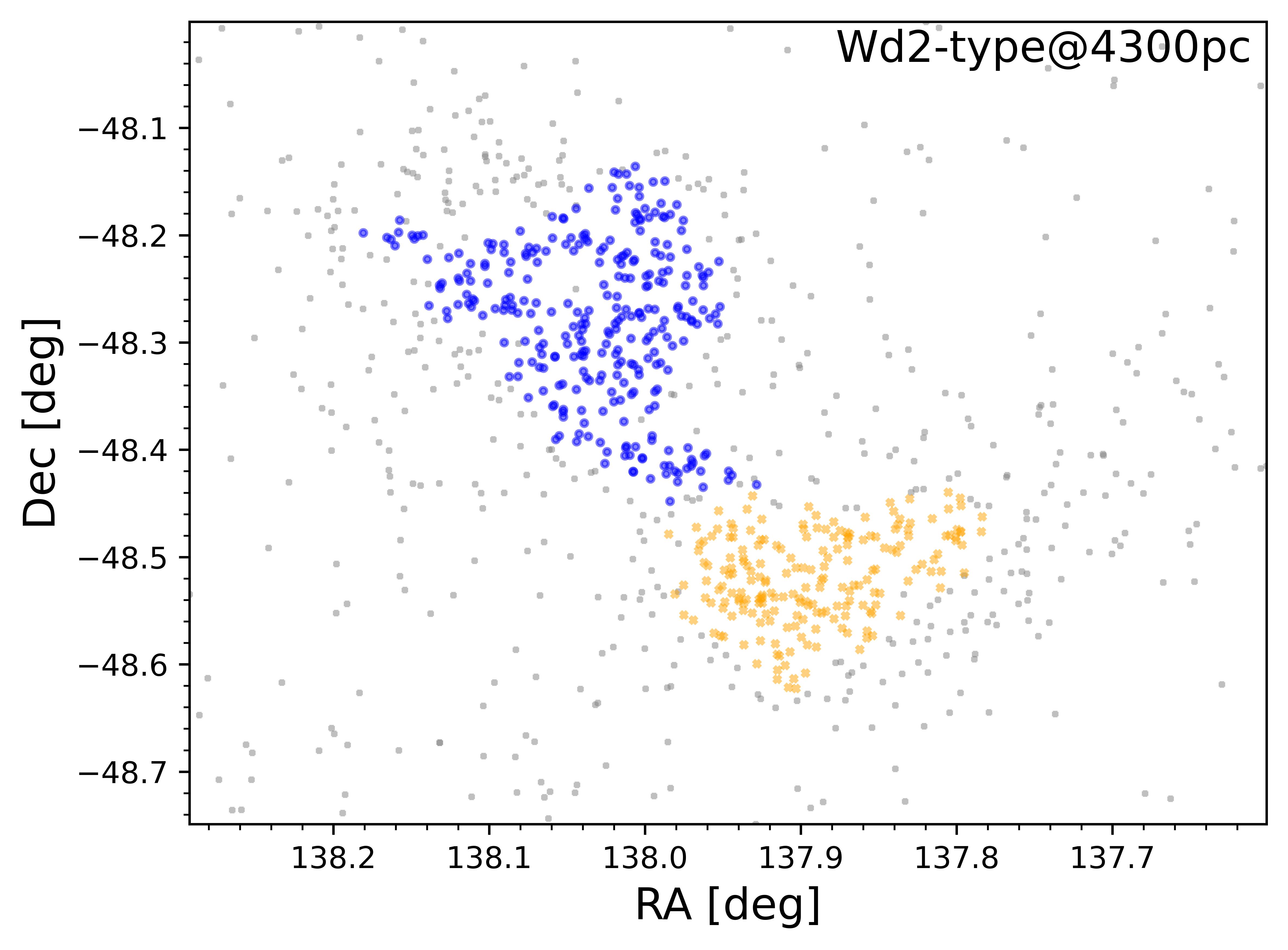}

   \caption{Stellar subclusters found for the simulations and observations (smallest and greatest distance) with HDBSCAN*. Members are denoted by same colours and symbols. Grey crosses represent stars not found to be in a subcluster.}  \label{Fig_sims_hdbscan} 
\end{figure*}


\begin{figure*}
\centering
    \includegraphics[width=0.32\textwidth]{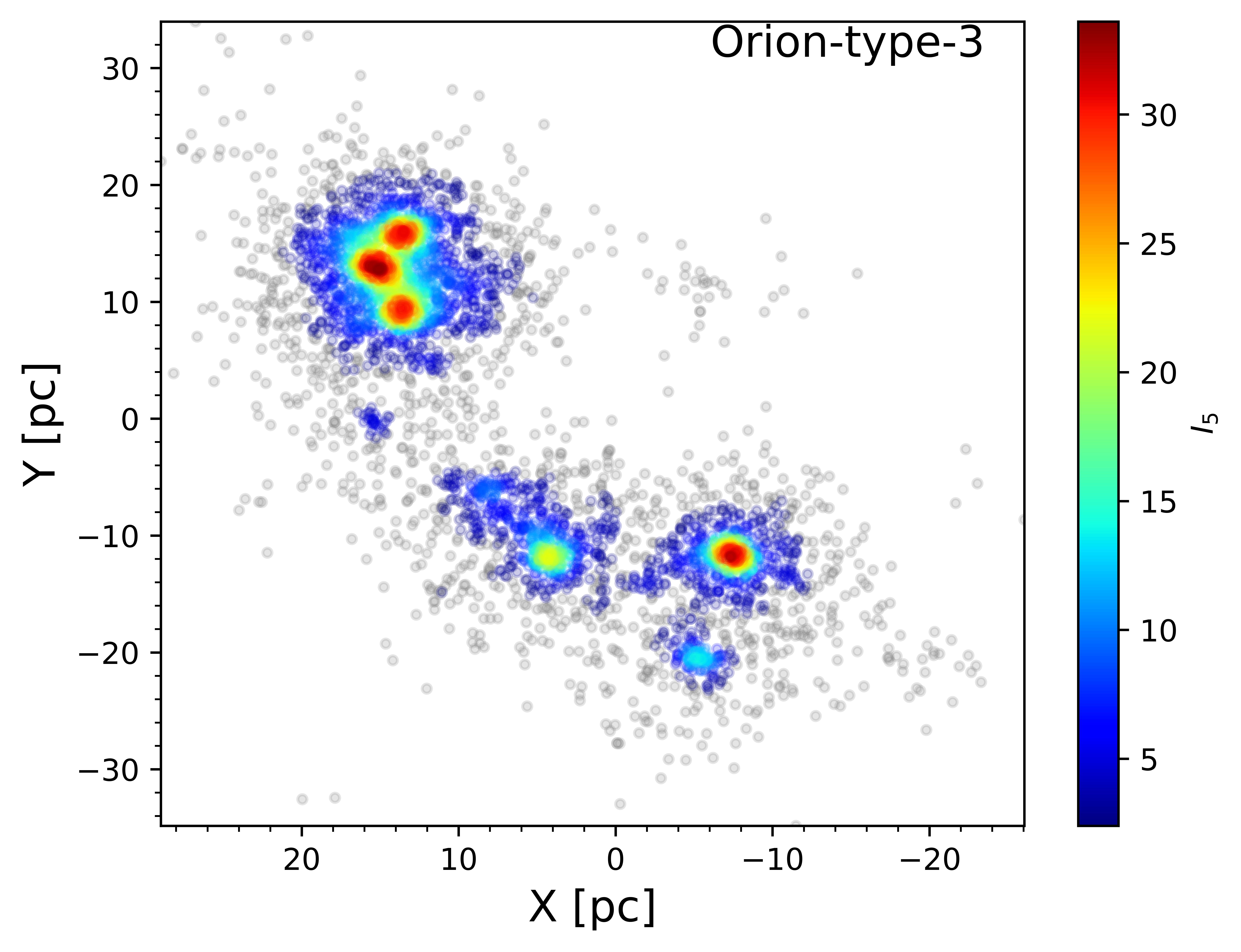}
     \includegraphics[width=0.32\textwidth]{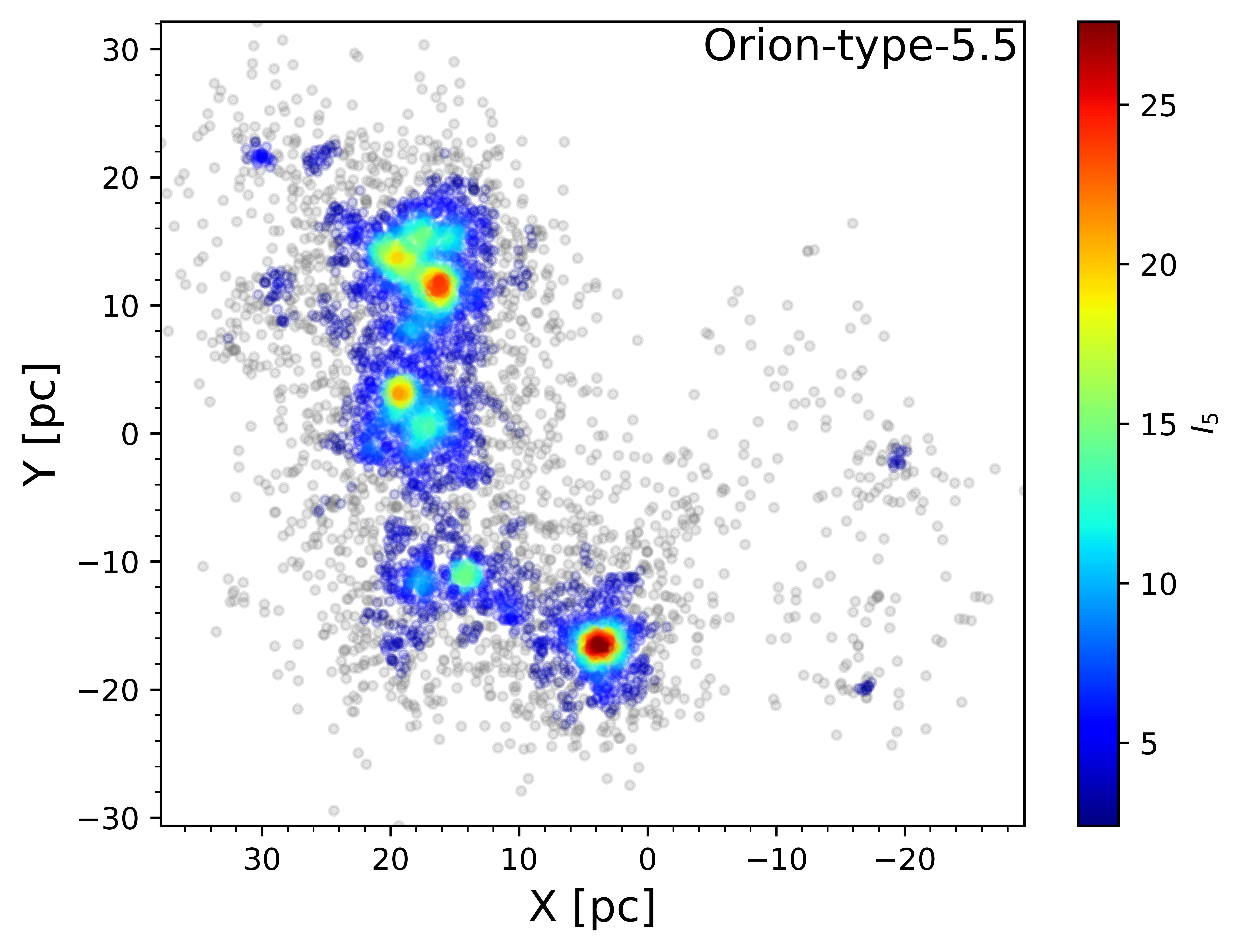}
     \includegraphics[width=0.33\textwidth]{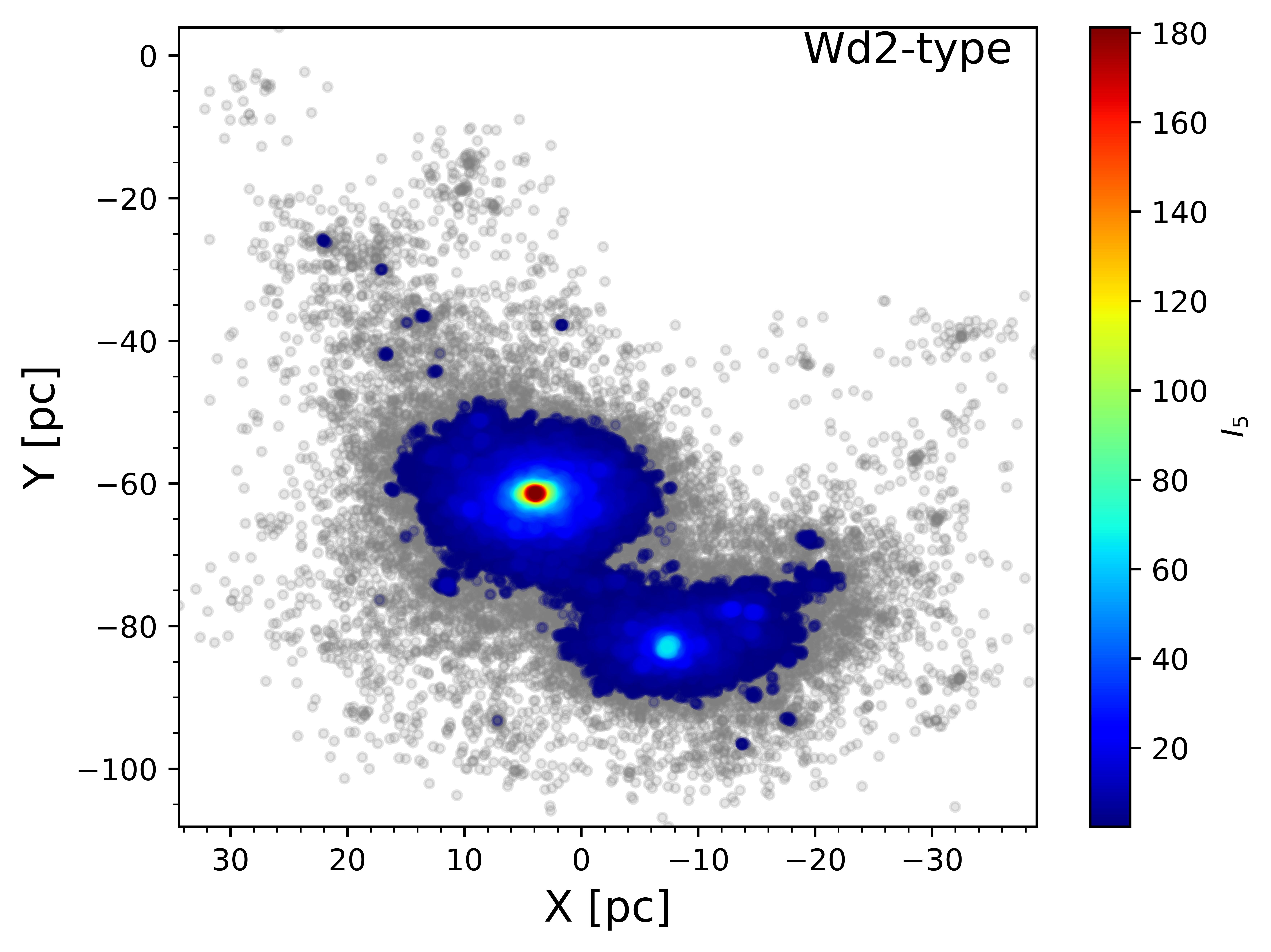}

    \includegraphics[width=0.32\textwidth]{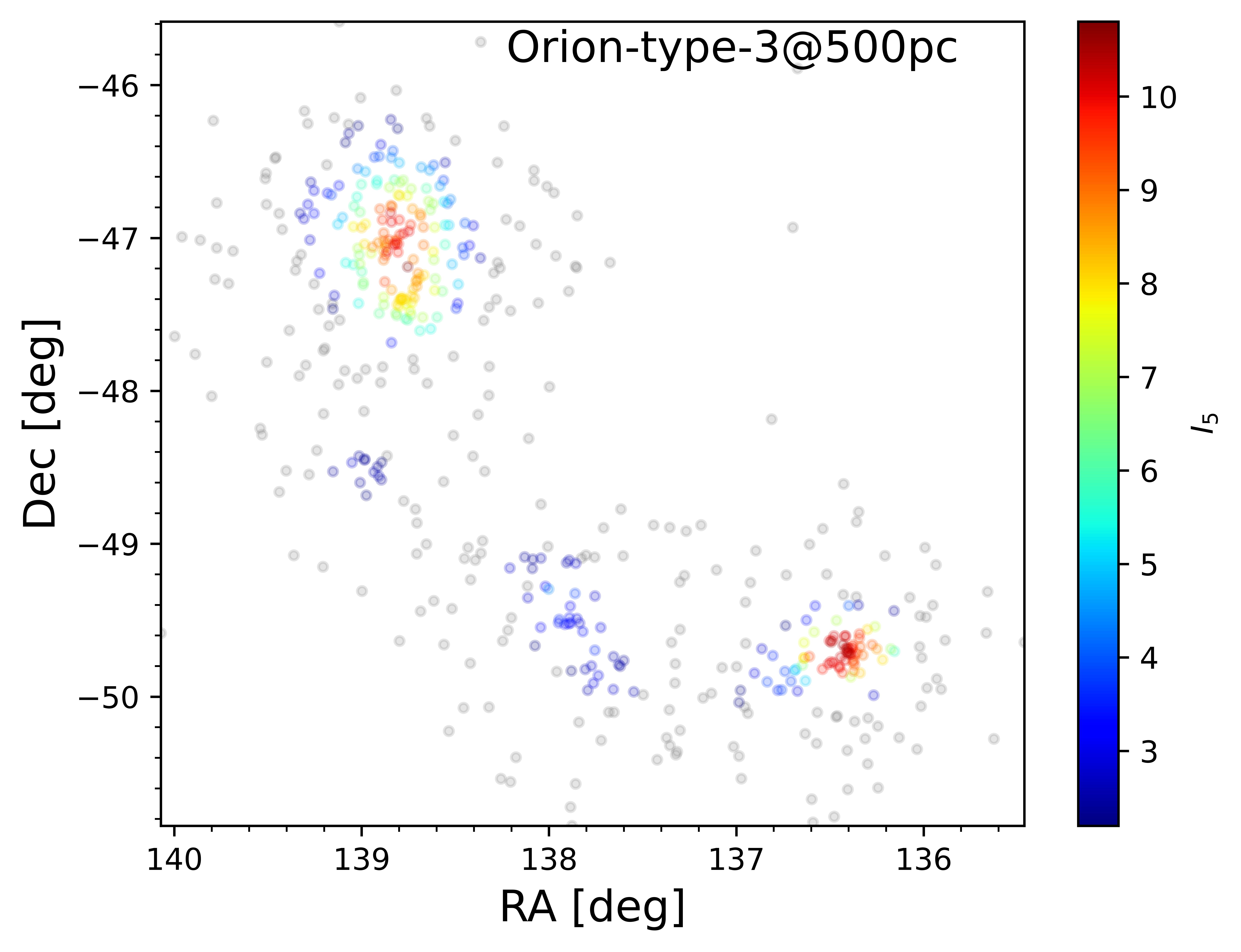}
     \includegraphics[width=0.32\textwidth]{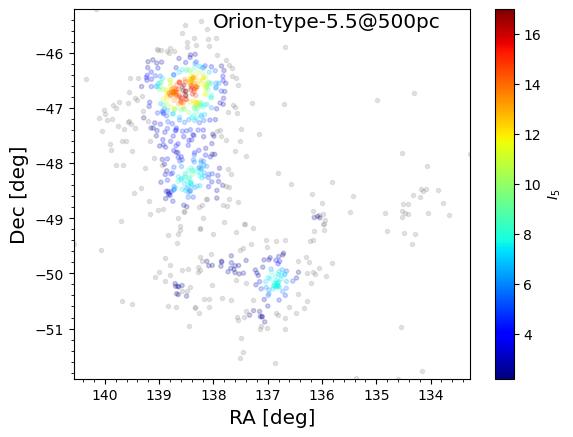}
     \includegraphics[width=0.32\textwidth]{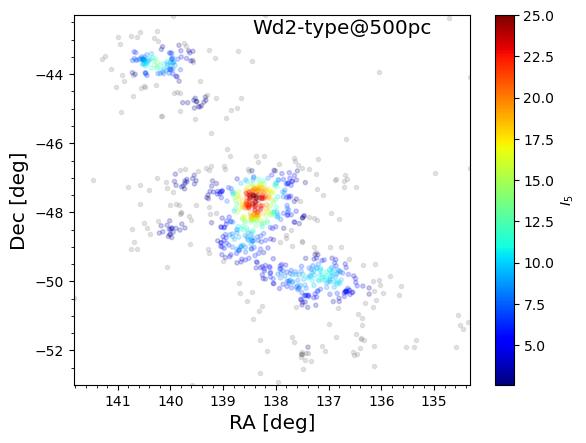}

    \includegraphics[width=0.32\textwidth]{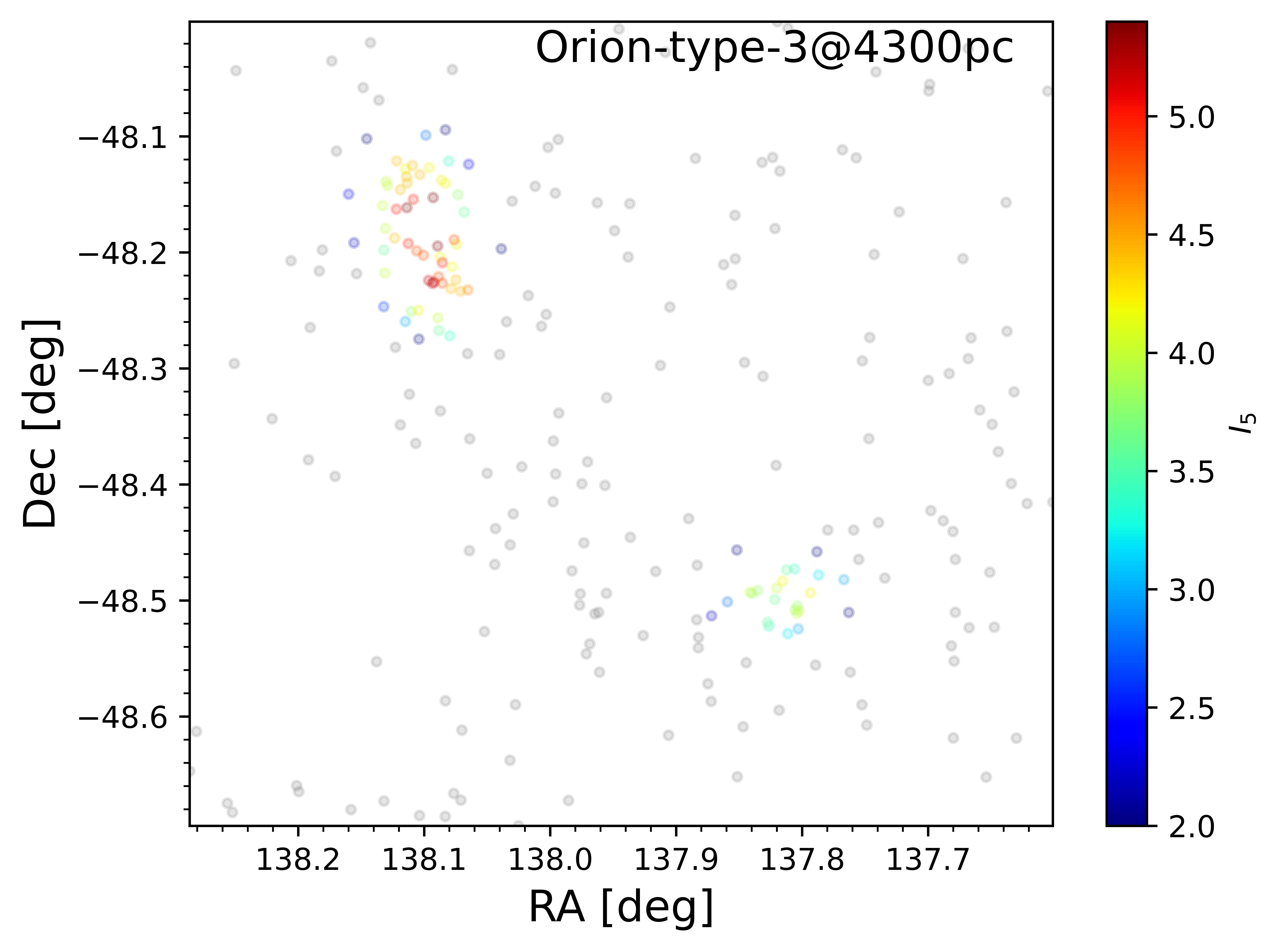}
     \includegraphics[width=0.32\textwidth]{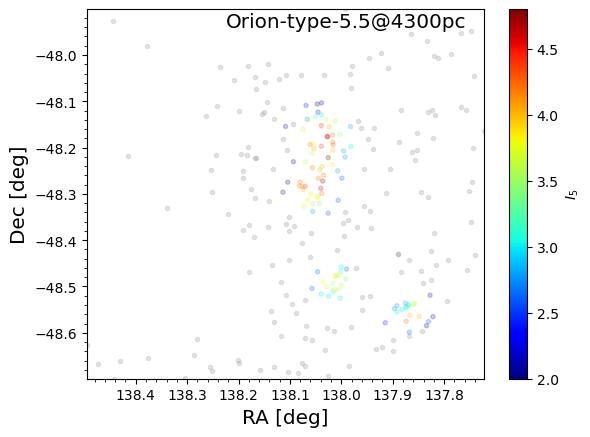}
     \includegraphics[width=0.32\textwidth]{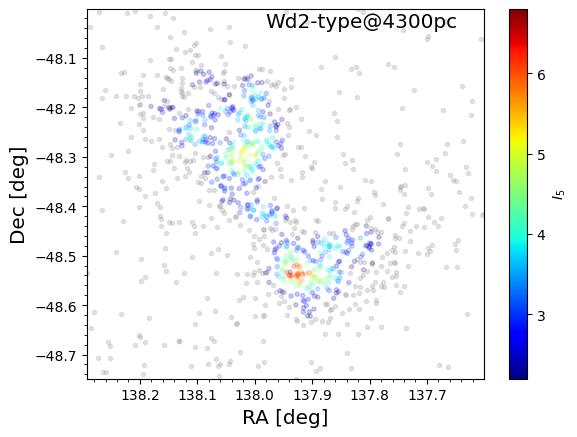}

    \caption{Index values assigned to cluster members by INDICATE for the (Top row:) simulations and observed by {\it Gaia} at (Middle row:) 500\,pc and (Bottom row:) 4300\,pc (see text for details). Grey points indicates stars which are not clustered. The effects of crowding is apparent in bottom panels where there are missing stars 'holes' in the regions with the highest index values.}
    \label{Fig_sims_indicate} 
\end{figure*}

\section{Results}
\label{sec:results}
In this section we discuss the 2D spatial and kinematic properties of the three cluster simulations, then compare these to those derived for our cluster catalogues to assess if these appear qualitatively similar, or whether the biases of {\it Gaia} derived membership lists are significant. We show two different measures of the spatial (Fig.\,\ref{Fig_sims_hdbscan} and Fig.\,\ref{Fig_sims_indicate}), and the kinematic (Fig.\,\ref{Fig_sims_vout}), characteristics of the stars in the original dataset, and placed in the {\it Gaia} catalogue at distances of 500 and 4300\,pc, the most extreme distances we used. Properties for the full catalogue are available in Tables\,\ref{Tab_Q}-\ref{Tab_subclusters3}, and the original simulation in Table\,\ref{Tab_subclusters}.

\begin{table*}
\caption{Properties of subclusters found in the simulations.  \label{Tab_subclusters} 
} \centering                                      
\begin{tabular}{c | c | c| c| c}          
\hline\hline                        
Cluster & Name & Total Members   &$\tilde{V}_{\text{out}}$ &$\tilde{I}_{5}$ \\
&&& [km\,s$^{-1}$] &  \\
\hline                                   
Orion-type-3 & A & 41  &  0.86 & 3.8\\
 & B & 1918  & -0.21 & 11.0 \\
 & C & 541  & 0.71  & 6.8 \\
 & D & 524  & -0.72  & 8.6 \\
 & E & 113  &  2.07 & 9.2 \\
\hline                                   
Orion-type-5.5 & A &565 & 0.60& 8.2 \\
 & B & 452&  0.71 & 4.8 \\
 & C & 604 &  0.09 & 8.4 \\
 & D & 1038&  -0.48 & 10.7 \\

\hline                                   
Wd2-type & A & 33001  & 1.28 & 19.4 \\
 & B & 12379  & -4.04 & 9.6 \\
\hline                                             
\end{tabular}
\end{table*} 

\subsection{Spatial properties of the stars}\label{sect_spatial}

HDBSCAN*  does a reasonable job identifying the rough substructure in our clusters, even  at 4300\,pc, despite relatively few stars being observable (Fig.\,\ref{Fig_sims_hdbscan}). For the more massive Wd2-type region, field overdensities are erroneously detected as additional smaller clusters, only  at distances less than 1.5\,kpc. For the Orion-type-5.5 region, we see that two additional subclusters are picked out at 500\,pc, and one additional cluster is found at 3000\,pc, compared to the original dataset  At the remaining distances typically only 3 clusters are found, and clusters C and D are merged in all observations.  In  the Orion-type-3 cluster larger subclusters B, C and D are most commonly identified, with the smaller A and E clusters less commonly found and additional subclusters (asterims) are rare. By 4300\,pc, we see that only the largest cluster B and D are identifiable.

Fig.\,\ref{Fig_sims_indicate} shows the INDICATE  $I_5$ parameter which indicates the degree of spatial association for each star with its neighbours. All panels show the presence of spatially clustered stellar population in our clusters.  The values of $I_5$ are highest for the simulations of the Wd2-type cluster and for the Orion-type region at the earlier time compared to the later time. Similarly the proportion of member stars which are found to be clustered (rather than dispersed) is highest for the Wd2-type, and Orion-type-3 simulations. For the Orion-type cluster, this suggests that initially the majority of stars in these simulations are tightly spatially clustered, then some dispersal of stars occurs between the two time frames, with those still clustered in the later snapshot now more loosely so. 

These trends for each cluster are mostly preserved when seen with {\it Gaia} at the two distances, though the relative intensities between subclusters are not necessarily preserved due to confusion selecting their true members by HDBSCAN*. The main differences with increasing distance in the Orion-type clusters are (i) an overall decrease in values of $I_5$ ii) a loss of resolution of fine structures, due to missing data, and (iii) the dispersal of stars between the two time frames is not observed at all distances. An example of (ii) is shown in Fig.\,\ref{Fig_sims_indicate}, in the region of the Orion-type-3 simulation at $(x,y)\approx(15,15)$, which has the greatest degree of association of the cluster and a complex substructure consisting of 3 smaller radial concentrations. When observed at both 500\,pc and 4300\,pc, this region is still perceived to have the greatest degree of association, but the complex substructure is lost. The Westerlund2-type cluster does not contain fine structures so we are unable to confirm the universality of this result. A similar overall decrease in values of $I_5$ is observed in this cluster as with the Orion-types. Its two subclusters A and B are both consistently correctly identified as spatially clustered and their relative degrees of association ascertained (with the exception of 4300\,pc where they appear to be the same; Table\,\ref{Tab_subclusters3}).

Regarding the global structure, the highest Q parameter is found for the Wd2-type simulation ($\mathcal{Q}=0.9$). In the Orion-type cluster there is ab increase with time from 3\,Myr ($\mathcal{Q}=0.6$) to 5.5\,Myr ($\mathcal{Q}=0.7$). . Similar behaviour is seen both in our observed catalogue, but typically the Orion-type cluster appears to be less substructured and differences between the two times are sometimes absent (Table\,\ref{Tab_Q}). The largest change is observed for the Wd2-type cluster at 500\,pc, which coupled with the HDBSCAN* analysis, may result in the cluster perceived to be more heavily sub-structured than reality.

The most obvious difference apparent in  Fig.\ref{Fig_sims_hdbscan} and Fig\,\ref{Fig_sims_indicate} when observed is simply that there are far fewer stars. Fig\,\ref{Fig_seensim} indicates that typically fewer than 20\% of the stars are recovered. Our results indicate that despite this, measures of the spatial structure of the stars are fairly robust. 

\subsection{Kinematic properties of the stars}
We calculate the overall $v_{\rm{out}}$ for each simulated region, and also for each subcluster within the region, with their properties listed in Table~\ref{Tab_subclusters}. Overall, the Orion-type region is undergoing expansion at both ages (3 and 5.5 Myr), with the overall expansion increasing from 0.01 km\,s$^{-1}$ at the earlier time to 0.19 km\,s$^{-1}$ later. 
 The synthetic observations (Table\,\ref{Tab_Q}) of both Orion-type clusters give mixed results,
 with some appearing to contract and others expand. At some distances the expansion increase is observed, others the expansion appears to be slowing or an increased rate of contraction is occurring. On the other hand, the Westerlund2-type simulation is undergoing contraction at -0.65 km\,s$^{-1}$m which is correctly identified within an order of magnitude in the observations >1\,kpc, and two orders of magnitude at 500\,pc.

\begin{figure*}
\centering

    \includegraphics[width=0.33\textwidth]{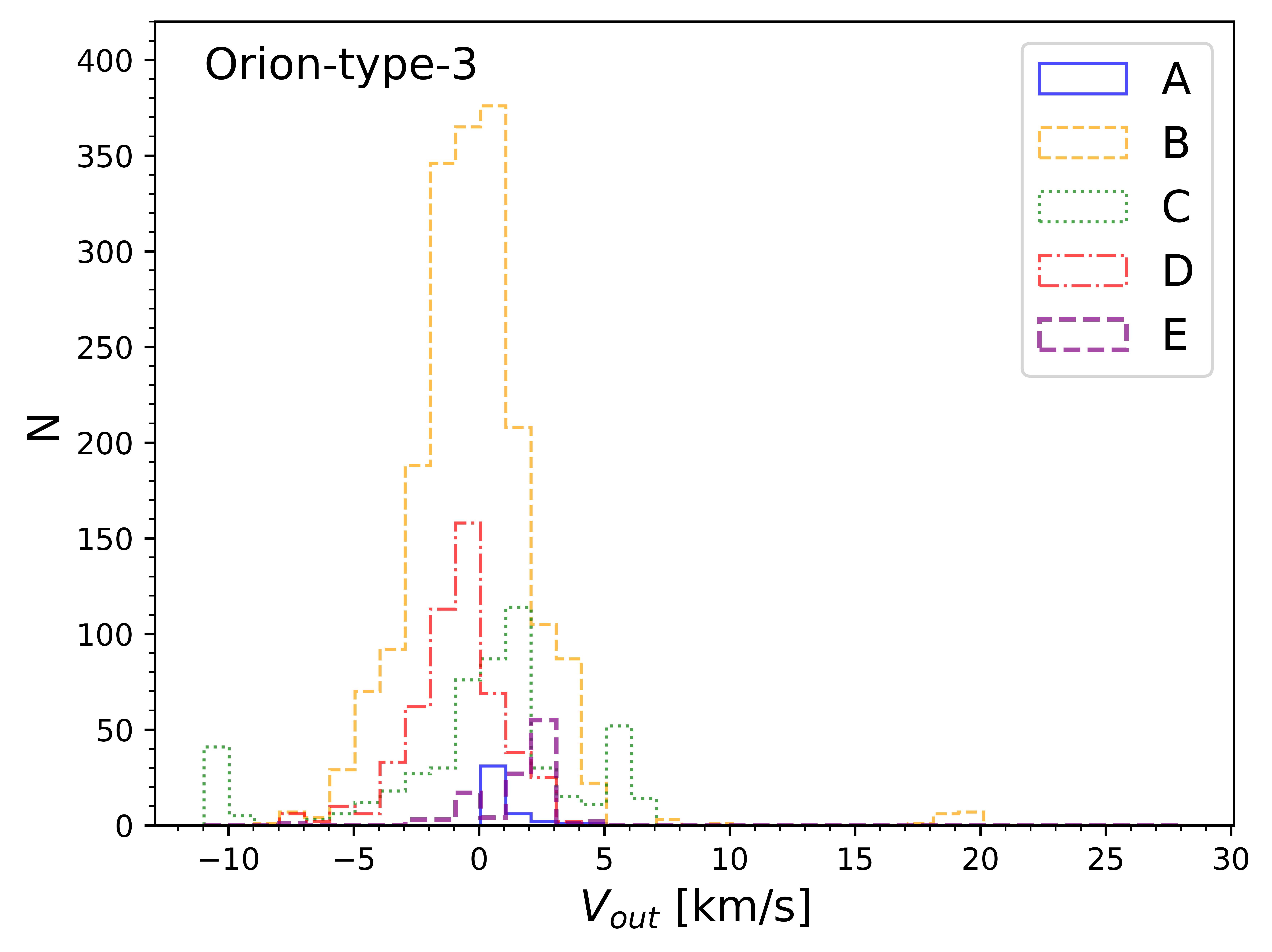}
    \includegraphics[width=0.33\textwidth]{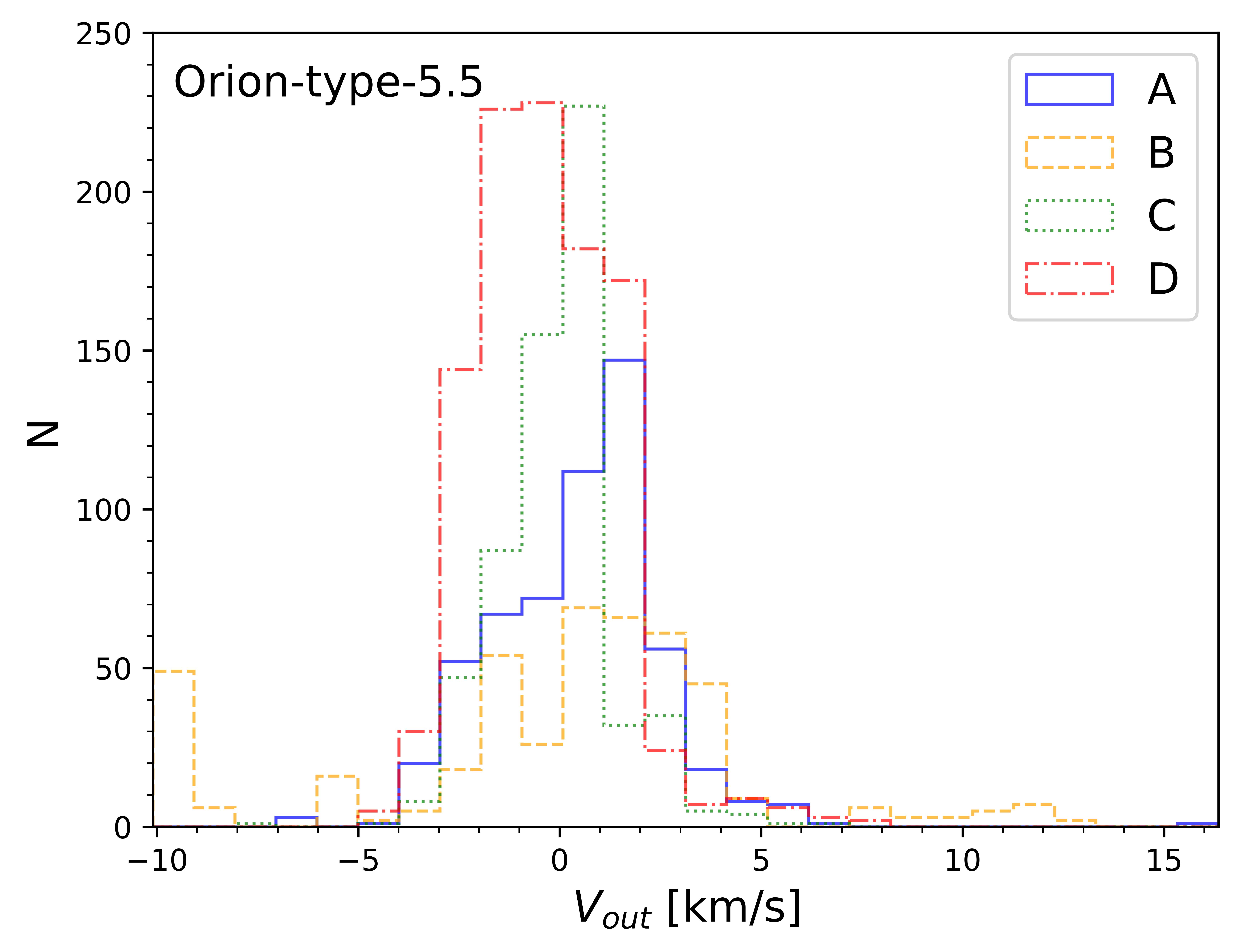}
    \includegraphics[width=0.33\textwidth]{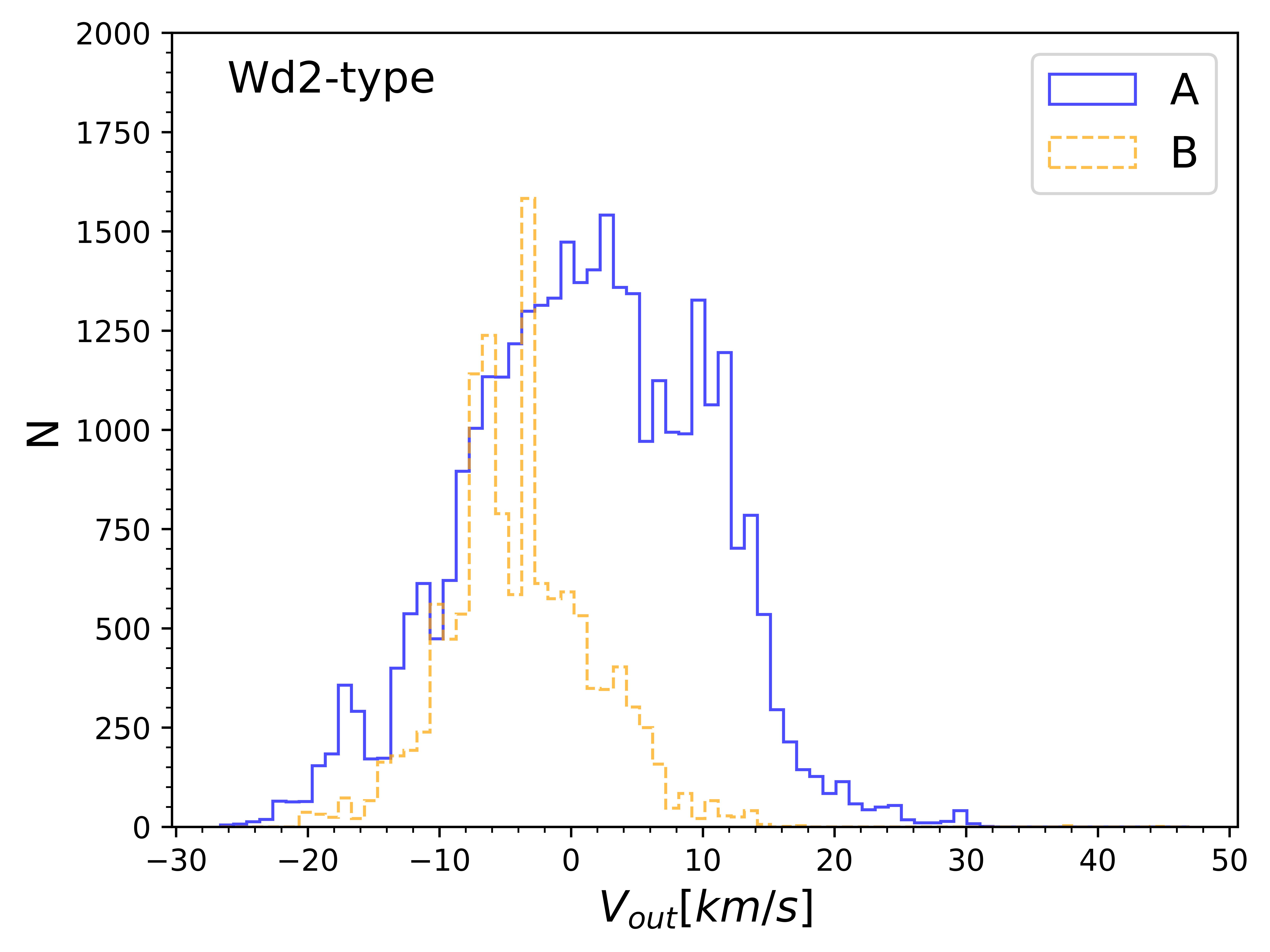}

    \includegraphics[width=0.33\textwidth]{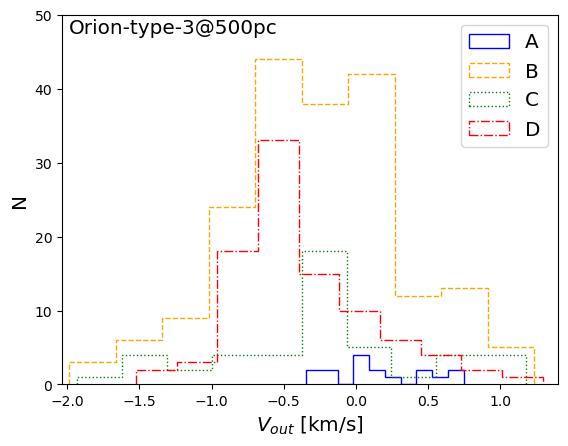} 
    \includegraphics[width=0.33\textwidth]{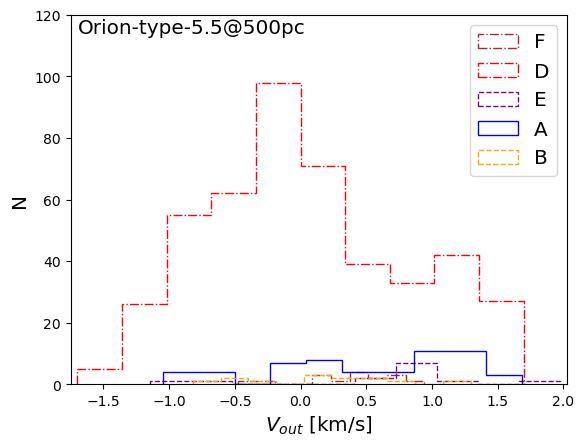}
    \includegraphics[width=0.33\textwidth]{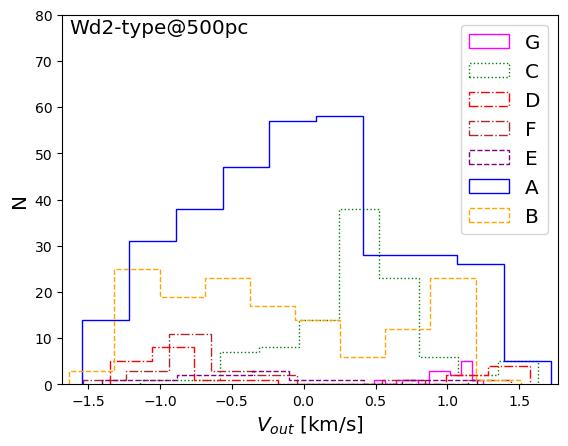}

    \includegraphics[width=0.33\textwidth]{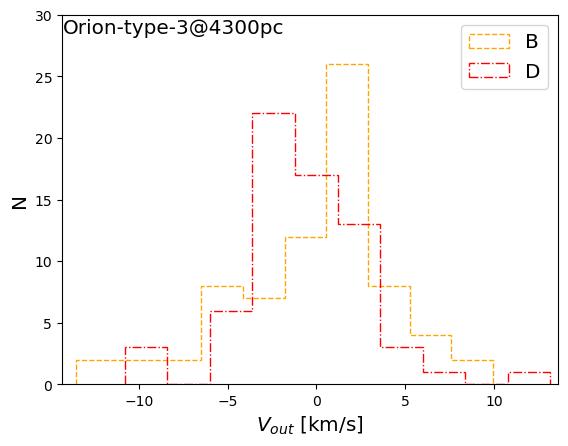} 
    \includegraphics[width=0.33\textwidth]{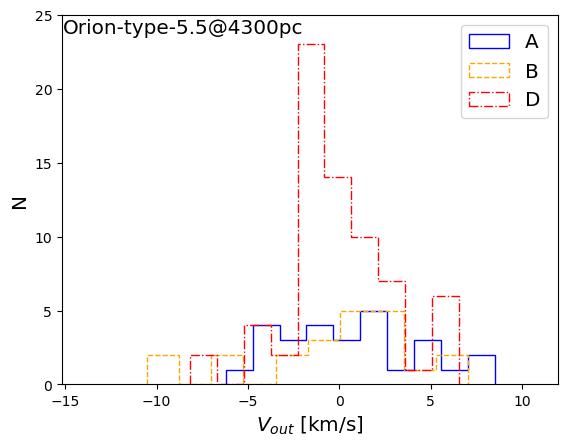}
    \includegraphics[width=0.33\textwidth]{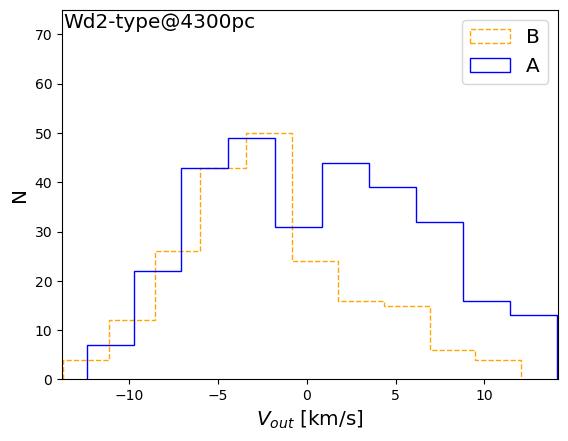}
    
   \caption{Directional 2D velocity histograms (w.r.t. the system center) for the stellar subclusters found in the simulations and observations (smallest and greatest distance) using HDBSCAN* defined in Fig.\,\ref{Fig_sims_hdbscan}.}  \label{Fig_sims_vout} 
\end{figure*}

The net expansion or contraction of the subclusters identified in Sect.\,\ref{sect_spatial} can be misidentified. The distributions of the velocities, simulated and observed, are shown in Fig.\,\ref{Fig_sims_vout} and properties of the latter are detailed in Tables\,\ref{Tab_subclusters1}-\ref{Tab_subclusters3}. Generally the simulated velocities of the clusters have normal distributions suggesting they are true kinematic structures rather than asterisms. The lower panels of Fig.\,\ref{Fig_sims_vout} show the 500 and 4300\,pc observations. For the Orion-type-3 region, the subclusters still tend to have normal distributions, but these become flatter with distance and some (such as A and E) may be mistaken as not real kinematic structures 
The presence of outliers is clear at 4300\,pc, and is an issue in clusters at distances $>$2500\,pc. Their large values suggest they represent stars which have been mistakenly identified as members (Fig.\,\ref{Fig_sims_vout}), which is true in some cases, but in others are real members with either inaccurate velocity measurements, appear more prominent due to missing data in the distribution, and/or assigned to the incorrect subcluster. 
Similar results are found for the Orion-type-5.5 region, and asterisms E and F are clearly identifiable from their velocity distributions, which are flat.

The Wd2-type region consists of two subclusters, one of which is expanding and the other contracting. Again the velocities follow roughly normal distributions. The mean kinematic behaviour of A and B is correctly identified at distances $>1500$\,pc, most likely as they contain many stellar members and exhibit overall broad velocity distributions which are merely narrowed by the selection criteria. Most of the asterisms found at 500\,pc are clearly identifiable as such from their velocity distrbutions, but the larger subcluster C  found at both 500\,pc and 1000\,pc may be mistaken as a true structure.

\section{Discussion and Conclusions}

Despite more stars having full 3D spatial and kinematic information than ever before, these remain a minority with most stars lacking good radial positions within the cluster and radial velocities. 
 {\it Gaia} provides radial velocities only for brighter stars and even the new generation of wide-field high-multiplex fibre-fed spectrographs can only obtain $\sim$$10^3$ radial velocities in an hour.
 Parallaxes are more readily available, but even {\it Gaia} parallaxes still cannot pinpoint the position of a star within a star-forming region at 500\,pc accurately enough to yield a true 3D view. 
 Hence in Paper I we explored the impact of perspective effects on the accurate derivation of cluster properties, finding that while it is possible to obtain the correct qualitative conclusions, quantitative conclusions are likely to be inaccurate.
 However, observations of real clusters also suffer from observational biases such as {\it Gaia}’s ability to detect and resolve stars, the effects of extinction and the ability of observers to differentiate between true members and foreground or background stars.
 Therefore in this paper we have addressed the effects of these problems on the results of 2D spatial-kinematic star cluster analyses.

We have taken the results of two simulations which produce two quite different types of object, and placed them, as close as possible to do so, in the real sky, using the {\it Gaia} catalogue of Milky Way stars. One simulation produces a stellar association similar to the Orion OB1 association, which we study at two different time frames, or ages. The other simulation produces two dense clusters, not dissimilar to Westerlund 2. Thus we can make reasonable predictions about our ability to observe Orion-type associations, and young massive clusters like Westerlund 1 and 2, at different distances from the Sun. 

Unsurprisingly, given its mass and density, the Westerlund 2 type cluster is easy to observe at all the distances we tested, up to 4.3\,kpc. This is true even though this region is relatively young and still contains gas. We were able to identify that the cluster consists of two components, or subclusters, and the correct expansion / contraction of each at large distances. In fact the Wd2-type cluster is harder to correctly observe at smaller distances from the Sun, due to the large spread of velocities, and the application of proper motion cuts in order to select members of the cluster.

The Orion-type association is more typical of regions closer to the Sun, and the types of objects that will be harder to study at larger distances. However even though the number of stars detected at large distances in some cases comprises less than 2\% of the original member list, we still recover the basic properties of the region, including the identification of 3 or 4 clusters. As the distance to the association increases however, the ability to identify substructures decreases, and we see examples of clusters that are missed, or merged together into a single feature (compared to the original simulation data), but ultimately the cluster still appears to have some substructure. Likewise the degree of clustering, as indicated by the INDICATE analysis decreases, but the majority of trends are preserved even when the effects of crowding cause 'holes' to appear in the densest regions (in agreement with \citealt{2022A&A...659A..72B}).

Generally the spatial characteristics of the clusters are significantly more robust than the kinematics. With a relatively high contamination to member ratio and less than $20\%$ of real members identified as such we, perhaps unsurprisingly, find it can be difficult to correctly determine whether subclusters are expanding or contracting, compared to the original simulation data, and in the case of the Orion-type cluster the region as a whole as well. We also see that the velocity distributions become flatter with distance, which could lead to clusters mistakenly being identified as asterisms. Reassuringly, we find most asterisms misidentified as subclusters in the spatial analysis are typically easily identifiable as such when their velocity distributions are examined. 

All the results above assume that the `tight criteria’, based on the criteria suggested by \cite{Cantat-Gaudin2020}, with additional position-based trimming of the membership list, are used to identify the larger scale cluster or association. Our analysis strongly supports using this criteria, rather than the minimal or standard spatial and kinematic criteria for young clusters in the galactic mid-plane and/or regions with a high stellar density. With the standard criteria, the contamination of the membership lists was significant, particularly for our smaller simulations at larger distances, with between $\sim\,7- 93\%$ of ‘members’ actually being fore or background to the cluster. For the minimal criteria, the contamination of the membership lists was between $\sim\,51-98\%$, highlighting the importance of proper motion cuts during membership selection.

Overall, we find that observational biases do play a role in the accuracy of perceived 2D properties of clusters. Qualitative conclusions can reasonably be obtained for the spatial properties of young massive clusters and large Orion-scale associations up to distances of 4.3\,kpc using {\it Gaia}, but observers should take care when interpreting the significance of any given result. While it is possible to obtain reasonable kinematic properties, these can be unreliable and lead to the wrong conclusions of contraction/expansion. Our findings here compound the results from Paper I meaning that, although found properties should be mostly indicative of true behaviour, the majority are unlikely to be quantitatively reliable. 

In this paper, we have considered two regions that would be relatively easy to observe in our Galaxy, due to their size and age. In further work we will consider clusters or associations which are more challenging to observe, and see whether such objects could be detected with {\it Gaia}, or indeed other instruments.

\section*{Data Availability}

The final stage Field-of-View files, from which stellar members are selected, at 500pc, 2500pc and 4300pc for all three clusters are available to download from the Zenodo repository at \url{https://zenodo.org/records/10053996}. Additional distances and analysis data underlying this article will be shared on reasonable request to the corresponding author.

The simulation data underlying this article was provided by CLD by permission, which will shared on request to the corresponding author with permission of CLD.

\section*{Acknowledgements}
AB, CLD and TN are funded by the European Research Council H2020-EU.1.1 ICYBOB project (Grant No. 818940). SR acknowledges funding from the European Research Council Horizon 2020 research and innovation programme (Grant No. 833925, project STAREX). This work has made use of data from the European Space Agency (ESA) mission
{\it Gaia} (\url{https://www.cosmos.esa.int/gaia}), processed by the {\it Gaia}
Data Processing and Analysis Consortium (DPAC,
\url{https://www.cosmos.esa.int/web/gaia/dpac/consortium}). Funding for the DPAC
has been provided by national institutions, in particular the institutions
participaTting in the {\it Gaia} Multilateral Agreement.

\bibliographystyle{mnras}
\bibliography{ABuckner_OBYMCC_P2_final}
\bsp

\section*{Appendix A}

\begin{table*}
\caption{Table of observed $\mathcal{Q}$ and $V_{\rm out}$ values for our cluster catalogue.  \label{Tab_Q} } \centering                                      
\begin{tabular}{c | c | c| c}          
\hline\hline                        
Distance & Cluster & $\mathcal{Q}$ & $V_{\rm out}$ \\
{[pc]} & & & [km\,s$^{-1}$]\\
\hline

500 & Orion-type-3 & 0.6 & -0.21\\
 & Orion-type-5.5 & 0.7 & 0.13\\
 & Wd2-type & 0.6 & -0.01\\
\hline
1000 & Orion-type-3 & 0.7 & -0.11\\
 & Orion-type-5.5 & 0.9 & 0.06\\
 & Wd2-type & 0.8 & -0.09\\
\hline
1500 & Orion-type-3 & 0.7 & -0.09\\
 & Orion-type-5.5 & 0.8 & -0.05\\
 & Wd2-type & 0.8 &  -0.24\\
\hline
2000 & Orion-type-3 & 0.7 & 0.03\\
 & Orion-type-5.5 & 0.8 & 0.01\\
 & Wd2-type & 0.8 & -0.31\\
\hline
2500 & Orion-type-3 & 0.7 & 0.18\\
 & Orion-type-5.5 & 0.7 & 0.31\\
 & Wd2-type & 0.8 & -0.51\\
\hline
3000 & Orion-type-3 & 0.7 & 0.11\\
 & Orion-type-5.5 & 0.7 & 0.12\\
 & Wd2-type & 0.8 & -0.41\\
\hline
3500 & Orion-type-3 & 0.7 & -0.14\\
 & Orion-type-5.5 & 0.7 & 0.36\\
 & Wd2-type & 0.8 & -0.56\\
\hline
4000 & Orion-type-3 & 0.7 & -0.09\\
 & Orion-type-5.5 & 0.8 & -0.18\\
 & Wd2-type & 0.8 & -0.31\\
\hline
4300 & Orion-type-3 & 0.7 & -0.46\\
 & Orion-type-5.5 & 0.8 & -0.82\\
 & Wd2-type & 0.8 & -0.77\\
\hline                                   

\hline                                             
\end{tabular}
\end{table*} 

\begin{table*}
\caption{Table of subcluster properties found in the observations of Orion-type-3.  \label{Tab_subclusters1} 
} \centering                                      
\begin{tabular}{c | c | c| c| c}          
\hline\hline                        
Distance & Cluster & Total Members & $\tilde{V}_{\rm out}$ &$\tilde{I}_{5}$ \\ 
&&&[km\,s$^{-1}$] & \\
\hline                                   
500 & A &16 &0.07 &2.4 \\
 & B &196 &-0.28 &5.9 \\
 & C &47 &-0.16 &2.8\\
 & D &94 &-0.50 &7.2 \\
 & E & - & - & - \\
1000 & A & - & - & - \\
 &B &407 &-0.22 &10.4  \\
 & C &118 &0.40 &4.4 \\
 & D &121 &-0.60 &7.0  \\
 & E & - & - & - \\
1500  & A & - & - & - \\
 &B &418 &-0.25 &5.8  \\
 & C &81 &1.02 &3.0  \\
 & D &105 &-0.76 &7.6  \\
 & E &17 &1.76 &2.8  \\
2000  & A & - & - & - \\
 &B &330 &-0.22 &4.6  \\
 & C &185 &0.76 &1.2  \\
 & D &64 &-1.15 &7.5  \\
  & E &30 &1.22 &3.1  \\
2500  & A & - & - & - \\
 &B &238 &0.22 &4.5  \\
 & C &95 &0.61 &1.6  \\
 & D &58 &-1.69 &6.3  \\
 & E &30 &1.34 &2.9 \\
 & F &31 &-0.04 &1.8  \\
3000  & A & - & - & - \\
 &B &164 &-0.12 &4.6  \\
 & C &48 &1.08 &1.9  \\
 & D &69 &-1.24 &2.6  \\
 & E & - & - & - \\
3500  & A & - & - & - \\
 &B &116 &0.04 &4.2  \\
 & C &32 &-0.13 &1.5  \\
 & D &51 &-1.51 &4.8 \\
 & E & - & - & - \\
4000  & A & - & - & - \\
 &B &111 &-0.04 &5.2  \\
 & C &19 &1.37 &1.8  \\
 & D &37 &-1.40 &4.4  \\
 & E & - & - & - \\
4300  & A &- & - & - \\
 &B &73 &0.75 &4.0  \\
 & C& & & \\
 & D &66 &-1.00 &1.5  \\
 & E & - & - & - \\

\hline                                             
\end{tabular}
\end{table*} 

\begin{table*}
\caption{Table of subcluster properties found in the observations of Orion-type-5.5.  \label{Tab_subclusters2} 
} \centering                                      
\begin{tabular}{c | c | c| c| c}          
\hline\hline                        
Distance & Cluster & Total Members & $\tilde{V}_{\rm out}$ &$\tilde{I}_{5}$ \\  
&&&[km\,s$^{-1}$] & \\
\hline   
500 & A &56 &0.66 &6.7 \\
 & B &13 &0.15 &2.6 \\
 & C & - & - & - \\
 & D &458 &-0.07 &5.8 \\
 & E &13 &0.79 &2.4 \\
 &F &16 &0.47 &1.0 \\
1000 &A &160 &0.77 &3.6 \\
 & B &97 &-0.10 &8.4 \\
 & C & - & - & - \\
 & D &364 &-0.39 &13.6 \\
1500 &A &383 &0.37 &1.8 \\
  & B & - & - & - \\
  & C & - & - & - \\
  & D &500 &-0.26 &2.8 \\
2000 & A &107 &0.03 &4.6 \\
  & B &103 &0.60 &3.0 \\
  & C & - & - & - \\
  &D &360 &-0.32 &3.4 \\
2500 & A &88 &0.15 &3.4 \\
  & B &91 &0.82 &2.0 \\
  & C & - & - & - \\
  &D &251 &0.18 &3.0 \\
3000 & A &64 &0.33 &2.2 \\
 & B &69 &0.69 &1.6 \\
 & C & - & - & - \\
 &D &152 &-0.11 &3.2 \\
 & E &23 &-1.42 &1.4 \\
3500 &A &49 &0.93 &2.4 \\
 & B &59 &1.17 &2.0 \\
 & C & - & - & - \\
 & D &120 &0.08 &3.5 \\
4000 &A &35 &-0.23 &3.0 \\
 & B &31 &1.56 &2.2 \\
 & C & - & - & - \\
 & D &89 &-0.45 &3.4 \\
4300 &A &27 &0.09 &2.8 \\
 & B &22 &1.10 &3.2 \\
 & C & - & - & - \\
 & D &69 &-0.50 &3.6 \\
\hline                                             
\end{tabular}
\end{table*} 

\begin{table*}
\caption{Table of subcluster properties found in the observations of Westerlund2-type.  \label{Tab_subclusters3} 
} \centering                                      
\begin{tabular}{c | c | c| c| c}          
\hline\hline                        
Distance & Cluster & Total Members & $\tilde{V}_{\rm out}$ &$\tilde{I}_{5}$ \\
&&&[km\,s$^{-1}$] & \\
\hline                                   
500 & A &332 &-0.01 &14.3 \\
 & B &143 &-0.36 &7.8 \\
 & C &105 &0.42 &7.6 \\
 & D &21 &-0.95 &2.0 \\
 & E &13 &-0.27 &3.2 \\
 & F &22 &-0.73 &3.4 \\
 & G &17 &1.01 &2.6 \\
1000 & A &721 &0.08 &11.0 \\
  & B &302 &-0.85 &7.0 \\
  & C &99 &0.92 &3.2 \\
1500 & A &1083 &-0.02 &9.2 \\
 & B &665 &-1.38 &6.8 \\
2000 & A &1139 &0.04 &9.4 \\
 & B &686 &-1.82 &8.2 \\
2500 &A &981 &0.19 &5.2 \\
 & B &608 &-2.20 &5.0 \\
3000 &A &987 &0.31 &5.2 \\
 & B &570 &-2.45 &4.2 \\
3500 & A &656 &0.50 &8.6 \\
 & B &512 &-2.94 &6.6 \\
4000 &A &465 &0.89 &6.8 \\
 & B &302 &-2.76 &6.4 \\
4300 & A &296 &0.76 &3.2 \\
 & B &200 &-2.59 &3.2 \\
\hline                                             
\end{tabular}
\end{table*}

\label{lastpage}
\end{document}